\documentclass[a4paper,fleqn,usenatbib,useAMS]{mnras}

\usepackage{etoolbox}
\makeatletter
\patchcmd\@combinedblfloats{\box\@outputbox}{\unvbox\@outputbox}{}
\makeatother
\usepackage{graphicx}	
\usepackage{amsmath}	
\usepackage{amssymb}	
\usepackage{multicol}        
\usepackage{bm}		
\usepackage{pdflscape}	
\usepackage{subfig}

\usepackage[T1]{fontenc}
\usepackage{ae,aecompl}

\usepackage{newtxtext,newtxmath}

\title{Upper limits on the escape fraction of ionizing radiation from galaxies at $2\lesssim z < 6$}

\author[U. Me\v{s}tri\'{c} et al.]{
U. Me\v{s}tri\'{c},$^{1,2,3}$\thanks{E-mail: uros.mestric@inaf.it and umestric@yandex.ru},
E. V. Ryan-Weber,$^{1,2}$
J. Cooke,$^{1,2}$
R. Bassett,$^{1,2}$
L. J. Prichard,$^{4}$ \newauthor
M. Rafelski,$^{4,5}$
\\
$^{1}$Centre for Astrophysics and Supercomputing, Swinburne University of Technology, Hawthorn, VIC 3122, Australia\\
$^{2}$ARC Centre of Excellence for All Sky Astrophysics in 3 Dimensions (ASTRO 3D), Australia\\
$^{3}$INAF -- OAS, Osservatorio di Astrofisica e Scienza dello Spazio di Bologna, via Gobetti 93/3, I-40129 Bologna, Italy\\
$^{4}$Space Telescope Science Institute, 3700 San Martin Drive, Baltimore, MD 21218, USA\\
$^{5}$Department of Physics \& Astronomy, Johns Hopkins University, Baltimore, MD 21218, USA\\
}

\date{}

\pubyear{2020}

\begin{document}
\label{firstpage}
\pagerange{\pageref{firstpage}--\pageref{lastpage}}
\maketitle

\begin{abstract}
In this work, we investigate upper limits on the global escape fraction of ionizing photons ($f_{\rm esc/global}^{\rm abs}$) from a sample of galaxies probed for Lyman-continuum (LyC) emission characterized as non-LyC and LyC leakers. 
We present a sample of 9 clean non-contaminated (by low redshift interlopers, CCD problems and internal reflections of the instrument) galaxies which do not show significant ($>$ $3\sigma$) LyC flux between 880\AA\ $<\lambda_{rest}<$ 910\AA.
The 9 galaxy stacked spectrum reveals no significant LyC flux with an upper limit of $f_{\rm esc}^{\rm abs} \leq 0.06$. 
In the next step of our analysis, we join all estimates of $f_{\rm esc}^{\rm abs}$ upper limits derived from different samples of $2\lesssim z < 6$ galaxies from the literature reported in last $\sim$20 years and include the sample presented in this work.
We find the $f_{\rm esc}^{\rm abs}$ upper limit $\leq$ 0.084 for the galaxies recognized as non-LyC leakers.
After including all known detections from literature $f_{\rm esc/global}^{\rm abs}$ upper limit $\leq$ 0.088 for all galaxies examined for LyC flux.
Furthermore, $f_{\rm esc}^{\rm abs}$ upper limits for different groups of galaxies indicate that the strongest LyC emitters could be galaxies classified as Lyman alpha emitters.
We also discuss the possible existence of a correlation among the observed flux density ratio $(F_{\nu}^{LyC}/F_{\nu}^{UV})_{\rm obs}$ and Lyman alpha equivalent width EW(Ly$\alpha)$, where we confirm the existence of moderately significant correlation among galaxies classified as non-LyC leakers.
\end{abstract}




\section{Introduction}

A thorough understanding of the history and nature of sources responsible for reionization is still unclear. 
Even though a vast amount of the observational time is spent in providing detailed answers, we still lack conclusions based on statistically large samples.
Summarizing results from different observational research we can state with confidence that the Epoch of Reionization (EoR) endpoint lies in the $5.7\lesssim z \lesssim7$ range \citep[e.g.,][]{FAN2006, Kashikawa2006, Ouchi2010, Pentericci2011, Becker2015, Greig2017, Mason2018, Bosman2018, Eilers2018}.
Moreover, it is well known that reionization of hydrogen in the intergalactic medium (IGM) is triggered by Lyman continuum (LyC) radiation mostly generated by massive stars (e.g., Population III stars, O type stars) and active galactic nuclei (AGN).
Complications arise when we try to understand the contribution of various sources to the total amount of LyC radiation that escapes into the IGM.
An additional complexity to this question arises from the fact that the contribution to the ultraviolet background (UVB) from different sources is not constant over cosmic time, rather it changes as the different population of the objects evolve \citep[e.g.,][]{Wyithe2011,BECKER2013,Kakiichi2018}. 

\raggedbottom

Detecting and studying galaxies that emit LyC radiation (Lyman continuum galaxies, LCGs) at $z<6$ provides an opportunity to understand their role during EoR.
Moreover, detecting LyC radiation from galaxies will allow us to test indirect methods for selecting LyC leakers, predicting the escape fraction ($f_{\rm esc}$) of ionizing radiation into the IGM and estimating the average $f_{\rm esc}$ from galaxies at $z>6$.
It is essential to establish methods to indirectly recognize LCGs beyond $z>6$ because direct detection of the LyC flux is almost impossible, due to the rapid increase in density of neutral hydrogen in the IGM  at the higher redshifts \citep{INOUE2008,Inoue2014}.
In the last $\sim20$ years less than 20 secure LCGs have been reported in the literature at $2\lesssim z \lesssim4$ \citep{Vanzella2012, Vanzella2015,Shapley2016, Vanzella2016, deBARROS2016, Bian2017,Vanzella2018, Steidel2018, Fletcher2019, Ji2020}.
For these reasons, it is rational to suspect that the previous studies which targeted Lyman break galaxies as a potential LyC leakers were biased toward weak and non-LyC leakers \citep{COOKE2014}.
Another possible explanation for the low LCG detection rate is that faint low-mass galaxies contribute most of the LyC radiation \citep[e.g.,][]{Wise2009,Robertson2013,Paardekooper2015, BOUWENS2015} and at the moment they are beyond the detection limit of current instruments. 
On the other hand, the vast number of galaxies studied in the past is characterized as non-LyC leakers which may not be true for two reasons: they are too faint for current instruments or they leak LyC into the IGM but the escaped LyC radiation is absorbed on its way to the observer.

\raggedbottom

This work aims to study upper limits on the absolute escape fraction ($f_{\rm esc}^{\rm abs}$) of LyC photons from galaxies probed for LyC radiation and classified as non-LyC leakers.
Although they do not show LyC radiation, combined they can provide insight into limits on LyC leaking properties of the majority of galaxies studied to date. 
In this work, we present and analyze our sample of 9 non-contaminated galaxies classified as non-detection that was probed spectroscopically for LyC radiation with Keck Low-Resolution Imaging Spectrometer (LRIS). 
Moreover, we study the absolute escape fraction upper limits from our sample and samples of galaxies classified as non-detections in the literature.
For the purpose of this work a sample of galaxies is generated from available LyC non-detection samples in the literature reported in the last $\sim 20$ years at $2\lesssim z < 6$ interval.

This paper is organised as follows: the observations and sample selection is presented in Section \ref{2}, the reduction process and analysis of our data is presented in Section \ref{3}, the analysis of all upper limits gathered from literature and the discussion is described in Section \ref{4} and we summarize or conclusions in Section \ref{5}.

\section{Sample, observation and reduction}\label{2}

Here we summarize multiple observation runs with Keck Low-Resolution Imaging Spectrometer \citep[LRIS,][]{Oke1995, STEIDEL2004} in multi-slit mode, carried out in the 2015, 2016, and 2020. 
The common goal of all these runs was to identify $z\sim3-5$ galaxies with potential ionizing flux at $\lambda_{rest}<912$\AA\ and investigate their potential as analogues to $z>6$ sources of reionization.
Here, we focus on the galaxies in our sample with no significant escaping LyC flux. 
Galaxies in our sample with candidate LyC flux detections for these, and related observations, will be discussed in forthcoming papers.

Although the scientific goal was the same (detecting Lyman continuum radiation and confirming redshifts), the galaxy selection criteria evolved over the three runs, as updated multi-band photometry, additional deep u-band photometry, and deep UV imaging from our Hubble Space Telescope ($HST$) program  (GO 15100; PI Cooke) were acquired. 
The selection criteria are outlined for each run in the sub-sections below.
The sample of galaxies is selected from the FourStar galaxy evolution survey \citep[ZFOURGE,][]{Straatman2016} footprint located in the Cosmic Evolution Survey \citep[COSMOS,][]{Scoville2007} field.
The ZFOURGE COSMOS footprint was chosen because of the available $\sim30$ band photometry, multiple-band $HST$ coverage, and fitted spectral energy distributions (SEDs).  These data provide reliable $z_{\rm phot}$ estimates ($\sim$2\% accuracy), SFRs, masses, and flags with galaxy type (star-forming, quiescent or dusty). 

\subsection{2015 sample and observation}

The targets were observed on the 20 March 2015.  Because we were limited to 5.3 h of observing time, this pilot study chose to focus on 8 primary targets that are relatively bright in the Canada France Hawaii Telescope (CFHT) $r$ band $23.15<r<24.95$ with photometric redshifts in the range $3<z<3.2$ and located within the $\sim$5$^\prime$ $\times$ 7$^\prime$ area of a single LRIS  multi-object slitmask.
Although fainter and lower mass galaxies are likely to have higher escape fractions of ionizing photons, this conservative strategy was adopted for the initial detection experiment.

The 8 galaxy sample was selected by visually inspecting CFHT $us$-band photometry and $\sim$30-band SEDs from ZFOURGE, enabling us to avoid the selection of low redshift galaxies. 
In addition, we selected the targets based on their morphology and potential contaminating sources from the available $HST$ images, and included galaxies with potential merger signatures to explore LyC detection as a result of galaxy interaction.
We also paid attention that galaxies selected for this observation fall in redshift range where we can observe their optical lines with Multi-Object Spectrometer for Infra-Red Exploration (MOSFIRE) (results from MOSFIRE observation are presented in the work \cite{Bassett2019}, where in the Section 2. extensive description about sample selection is provided).

Though we used information from the CFHT $us$ filter to aid in target selection, we took into consideration that the filter suffers from a red leak at $\sim5000$\AA\ and that its redder wavelength coverage compared to typical $u$-band filters (e.g., Sloan-like $u$-band filters) results in a fraction of the flux from the Ly$\alpha$ forest contributing to the source magnitude for $z<3.4$ galaxies.
We modelled the Ly$\alpha$ forest contribution and also examined the expected $us,g,r,i$ colours of the galaxies on colour-colour diagrams and their expectation for LyC flux using the method described in \citet{COOKE2014} for galaxies within and outside of the conventional Lyman break galaxy colour selection criteria.  
It should be noted that all our targets are limited to galaxies identified in the ZFOURGE sample, which are K-band selected.  
As a result, our sample likely does not include the bluest, youngest, and least dusty $z\sim3-5$ galaxies, and those expected to have higher LyC flux emission.

The observations were performed using the D560 dichroic to redirect all flux shorter than $\sim$5600\AA\ to the 400 lines/mm grism blazed at 3400\AA\ on the blue arm, while flux at longer wavelengths is sent to the 400 lines/mm diffraction grating blazed at 8500\AA\ on the red arm.  
Science exposures were 1200\,s on the blue side and 1130\,s on the red side, due to different CCD readout times.
The total observing time was 19,200\,s and 18,080\,s on the blue and red arms, respectively, with seeing FWHM $\sim$$0.7^{\prime\prime}$--$1.1^{\prime\prime}$.

\subsection{2016 sample and observation}

Observations were performed on 10 and 11 February 2016. 
The targets were selected similarly to the 2015 sample with the following modifications.  
Targets were selected as single `clean' sources (i.e., with no evidence in the $HST$ images of interaction and ensured that there were no visible companions within $\gtrsim$1$^{\prime\prime}$).
We used updated photometry and SEDs from the ZFOURGE survey and assigned higher priority to galaxies that had lower overall photometric uncertainties resulting in more robust fits to the SEDs and those that are located outside the standard Lyman break selection region on colour-colour diagrams.
Two multi-slit masks were observed with a total of 24 LCG candidates.
The LCG candidate magnitudes range from $i \sim$ 24--26.5.  
Those having $i$ $<$ 25.5 were assigned the highest priority in order to obtain redshifts to inform the deep imaging and to search for potential spectroscopic detection of Ly$\alpha$ emission.  
The fainter LCG candidates were assigned as 'fillers' to search for Ly$\alpha$ emission as redshift estimates to help confirm the multi-band photometric redshifts and for potential stacking use.
The remaining area on the slitmasks was used for related science. 

For this run (on both nights) we were using the D500 dichroic beam splitter to redirect all flux shorter than $\sim$ 5000\AA\ to the blue LRIS channel, while flux at the longer wavelengths is sent to the LRIS red channel.  
Two dispersion elements were used to disperse light on the blue and red channel, 400 lines/mm grism blazed at 3400\AA\ and 400 lines/mm diffraction grating blazed at 8500\AA\, respectively.
The CCD binning was set to 2 $\times$ 2 for both nights.
Similar exposure times were acquired per blue/red integration as performed in the 2015 observations.  
Total times of 20,400\,s and 21,258\,s were acquired for the blue and red arm exposures, respectively, on the first night and first mask and total times of 18,000\,s and 18,400\,s for the blue and red arm exposures, respectively, on the second night and second mask under seeing FWHM $\sim$1$^{\prime\prime}$ conditions.

\subsection{2020 sample and observation}

The sample was observed during three half-nights on the 21, 22 and 23 January 2020.
Targets were selected similarly to the 2016 observations in the COSMOS ZFOURGE footprint with, importantly, added deep imaging information from two F336W $\sim$2$^{\prime} \times$ 2$^{\prime}$ fields and three F435W $\sim$2$^{\prime} \times$ 2$^{\prime}$ fields from our $HST$ program.  
The main goals of these observations included confirming redshifts for galaxies with detected LyC flux in the $HST$ images, a search for spectroscopic evidence of LyC flux down to rest-frame $\sim$700\,\AA, depending on target redshift, and an analysis of the LCG ISM features.
Results on the HST F336W and F435W photometry and details on candidate selection criteria are presented in Prichard et al.\ (submitted).
During the three half nights, we observed two multi-slit masks with 17 LCG candidates having a variety of priorities, with some duplication of targets from previous runs (for added depth) and additional science objects as slitmask `filler' sources.  
Similar to 2016 observations, the majority of the LCG targets have $i\sim$ 24--26.5, with the fainter galaxies targeted to search for detectable Ly$\alpha$ emission to help confirm the photometric redshifts.

The observations were performed with the same instrument set up as the previous runs, but with the exposures taken with 1 $\times$ 1 binning.
The total observing time for the first multi-object slitmask was 28,800\,s on the blue arm and 25,080\,s on the red arm, and the total integration on the second slitmask was 16,800\,s and 14,630\,s for the blue and red arms, respectively.  The seeing FWHM for the three half-nights was $\sim$1--1.2$^{\prime\prime}$.

\begin{figure*}
  \centering
  \includegraphics[width=12cm, height=11.13cm]{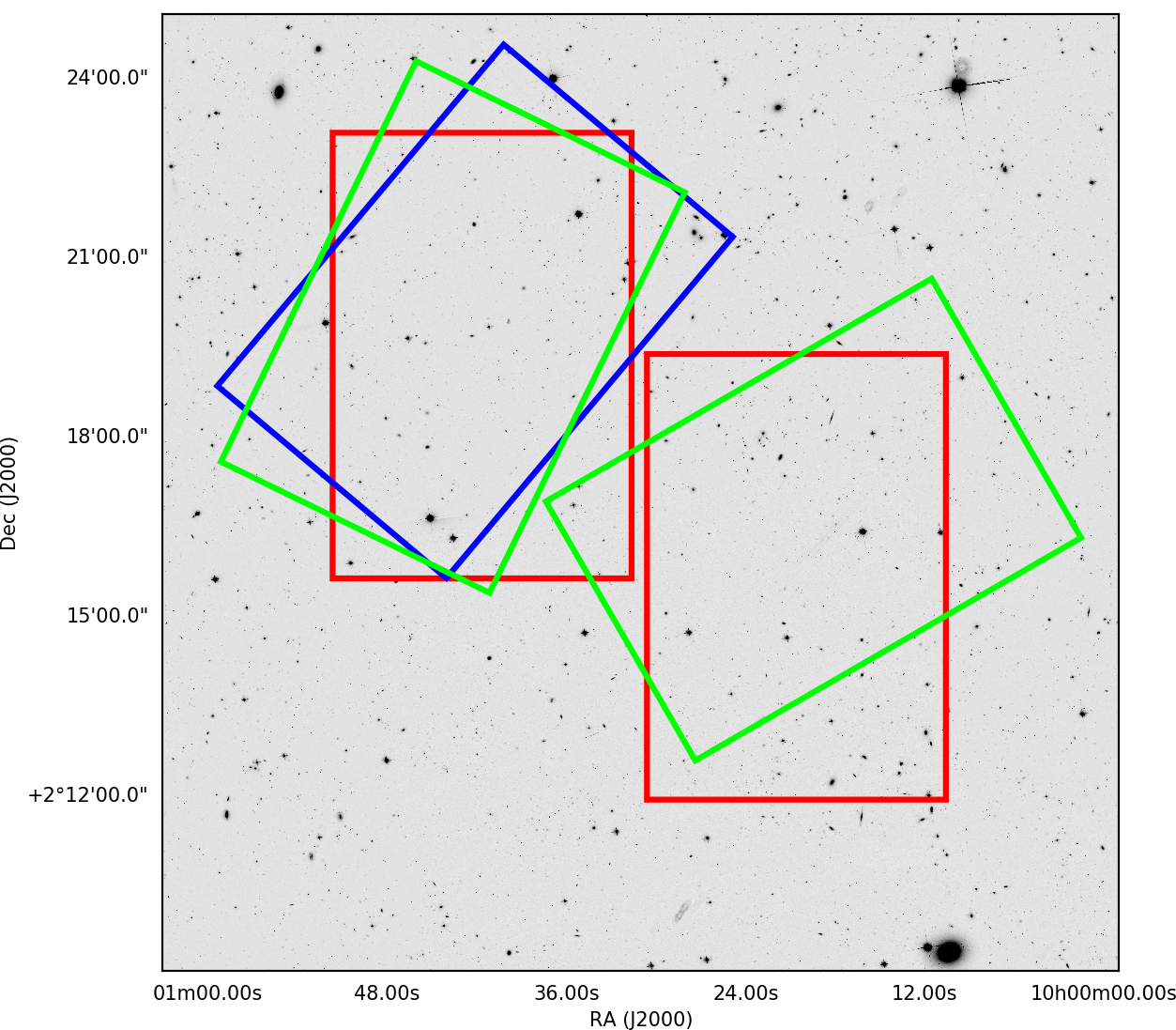}
  \caption{Cut out of the ZFOURGE footprint inside the COSMOS field. Data are taken with HST F814W filter, binned by a factor of 10 in both x and y, created by Anton Koekemoer at STScI using cycle 12+13 data \citep{Koekemoer2007, Massey2010}. The regions shown are different positions of the LRIS masks for different observation runs (2015. blue, 2016. red and 2020. green).}
  \label{HSTcoverage}
\end{figure*}

\subsection{Data reduction}

Data reduction was carried out with the Image Reduction and Analysis Facility software \citep[\textsc{IRAF},][]{Tody1986}.
Spectroscopic data were reduced following standard reduction procedures for multi-object slit spectroscopy.
Before reduction, all light and calibration frames were properly aligned, stacked, and trimmed. 
The reduction procedure was organized in the conventional manner, including the following steps:
\begin{itemize}
\item Flat fielding - Here, we use `twilight sky' flat fields acquired during evening or morning on the same nights as the data and the IRAF \textit{response} function to normalize our flats. 
For this purpose we are using 3 twilight flats observed on the same night and those twilight flats are averaged together to create master flat field.
\item 1D spectra extraction - IRAF \textit{apall} function was used for background subtraction, tracing, and extraction of the 1D spectra. 
During this process, since our objects are faint and LyC flux is hard to detect, we set \textit{apall} parameter \textit{t-{nsum}} to 100. 
This means that we are summing 100 dispersion lines, which will enable us to properly trace faint dispersion trace from our candidate particularly in the LyC part if detected.
\item Wavelength calibration - Spectra from Hg, Ne, Ar, Cd and Zn calibration lamps were used to perform the wavelength calibration.  These spectra were taken on sky with the observations to minimize the effects from instrument flexure.
\item Flux calibration - Spectra of the suitable blue spectrophotometric stars from \cite{Oke1990} were obtained during the same nights as the data for flux calibration. 
\end{itemize}

\section{Evaluating escape fraction of LyC radiaton}\label{3}

Probing the LyC flux with spectroscopy is the most direct, secure and accurate way.
It provides us with the opportunity to measure LyC in the most plausible range 880\AA\ $<\lambda_{rest}<$ 910\AA\
and to examine how LyC flux density changes at shorter wavelengths beyond the mentioned region if detected.

To evaluate escaping fraction ($f_{\rm esc}$) of LyC radiation from our candidates we adopt the standard relation for relative escape fraction ($f_{\rm esc}^{\rm rel}$) introduced in \cite{STEIDEL2001}:

\begin{equation}  
\label{eq:1}
f_{\rm esc}^{\rm rel}=\frac{(F_{\nu}^{LyC}/F_{\nu}^{UV})_{\rm obs}}{(L_{\nu}^{LyC}/L_{\nu}^{UV})_{\rm int}}\exp{}(\tau_{IGM}^{LyC}),
\end{equation}

\noindent where $(F_{\nu}^{LyC}/F_{\nu}^{UV})_{\rm obs}$ is the observed restframe LyC to UV flux density ratio. 
The $(L_{\nu}^{LyC}/L_{\nu}^{UV})_{\rm int}$ is the model-dependent intrinsic ratio of the galactic ionizing to non-ionizing luminosity density, and  $\tau_{IGM}^{LyC}$ is the redshift-dependent attenuation of LyC photons due to hydrogen in the intergalactic medium along the line of sight.
Moreover, as recomended in \cite{Inoue2005, Siana2007} after multiplying $f_{\rm esc}^{\rm rel}$ with dust attenuation $A_{\lambda}=k_{1500}E(B-V)$ we can get the absolute escape fraction ($f_{\rm esc}^{\rm abs}$) or the fraction of LyC photons that escape from the galaxy into the IGM.

\begin{equation}
\label{eq:2}
f_{\rm esc}^{\rm abs}=f_{\rm esc}^{\rm rel}\times 10^{-0.4(k_{1500}E(B-V))},
\end{equation}

\noindent here $k_{1500}=10.33$ for a Calzetti reddening law \citep{Calzetti1997} and $E(B-V)$ is total dust attenuation for the studied galaxy.
For our case we adopted $E(B-V)$ coefficients from \cite{Laigle2016} and they are presented in the Table \ref{tab:final_sample}.

\subsection{Probing LyC and UV flux from spectra}

To create the cleanest possible sample, we chose only galaxies whose spectra are not contaminated by the signal from low-redshift interlopers, bright neighbouring galaxies, flux from adjacent alignment stars bleeding into the galaxy spectrum, instrument internal reflections, and/or bad columns on the CCD.
The latter effect had a particularly negative effect on the 2016 2 $\times$ 2 binned data.
Moreover, we only include the spectra of the galaxies if we were able to confirm their redshift and whose dispersion was traceable blueward from the Ly$\alpha$ line.
Finally, we do not include galaxies from the sample that exhibit candidate LyC emission and/or are under further investigation.
The final sample consists of 9 objects meeting the above conditions from the LCG targets that included varied priorities and are galaxies among the brightest of the sample, with $i$ $<$ 25.1. 
The 2D HST imaging in three bands and 2D LRIS spectra of all 9 clean candidates are presented in APPENDIX A and APPENDIX B, respectively.

The spectra of each galaxy in the final sample is shifted to the rest frame and searched for LyC flux between 880\AA\ $<\lambda_{rest}<$ 910\AA, with the non-ionizing UV flux measured at 1450\AA\ $\lesssim \lambda_{rest}\lesssim$ 1500\AA.
It is important to emphasize, in this work we are not discussing the existence of potential signal blueward from $\lambda_{rest}<880$\AA\ and we are not presenting a few high priority galaxies with potential LyC flux.
All findings and results relating to detected LyC signal in the final sample and full sample will be presented in a separate study. 

After measuring the mean flux density per pixel in region 880\AA\ $<\lambda_{rest}<$ 910\AA\ of the 9 galaxies, we did not find any statistically significant LyC signal.
We report the mean flux density and signal-to-noise ratio (SNR) in Table \ref{tab:final_sample}.
The uncertainties used for the SNR calculation are 1$\sigma$ uncertainties per pixel derived by IRAF after spectrum extraction and flux calibration.

In this work we considered all measured LyC signal below three sigma as non-detection, three to five sigma as possible detection and above five sigma secure detection. 
To probe spectra for LyC in more detail we perform resampling of the spectra in the LyC 880\AA\ $<\lambda_{rest}<$ 910\AA\ interval.

\begin{figure*}
  \centering
  \includegraphics[width=18cm, height=22.4cm]{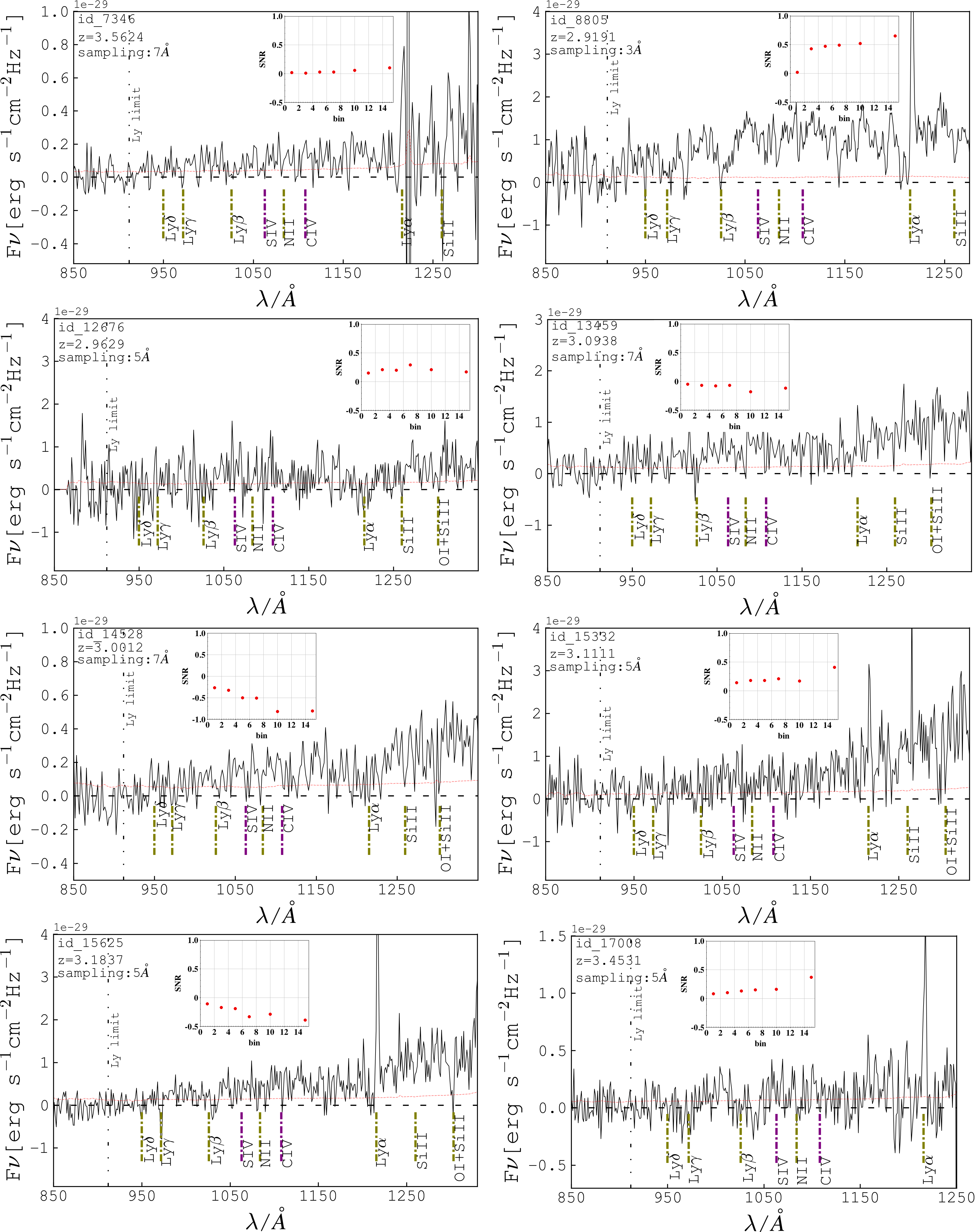}
  \caption{1D spectra of the clean sample observed with Keck/LRIS, shifted to their rest-frame wavelengths. For each spectrum, the ZFOURGE id, redshift, and sampling factor is listed in the upper left corner of the plot.  The $1\sigma$ per-pixel uncertainty (red) is overlaid on each spectrum (black) and the Lyman limit is marked as a dash-dot dot line. The inset plot at the top of each spectrum shows the 880\AA$<\lambda<910$\AA\ LyC flux SNR vs.\ binning factor. No statistically significant LyC flux is measured. }
  \label{fig:sample}
\end{figure*}

\renewcommand{\thefigure}{\arabic{figure} (Continued)}
\addtocounter{figure}{-1}

\begin{figure}
  \centering
  \includegraphics[width=\columnwidth]{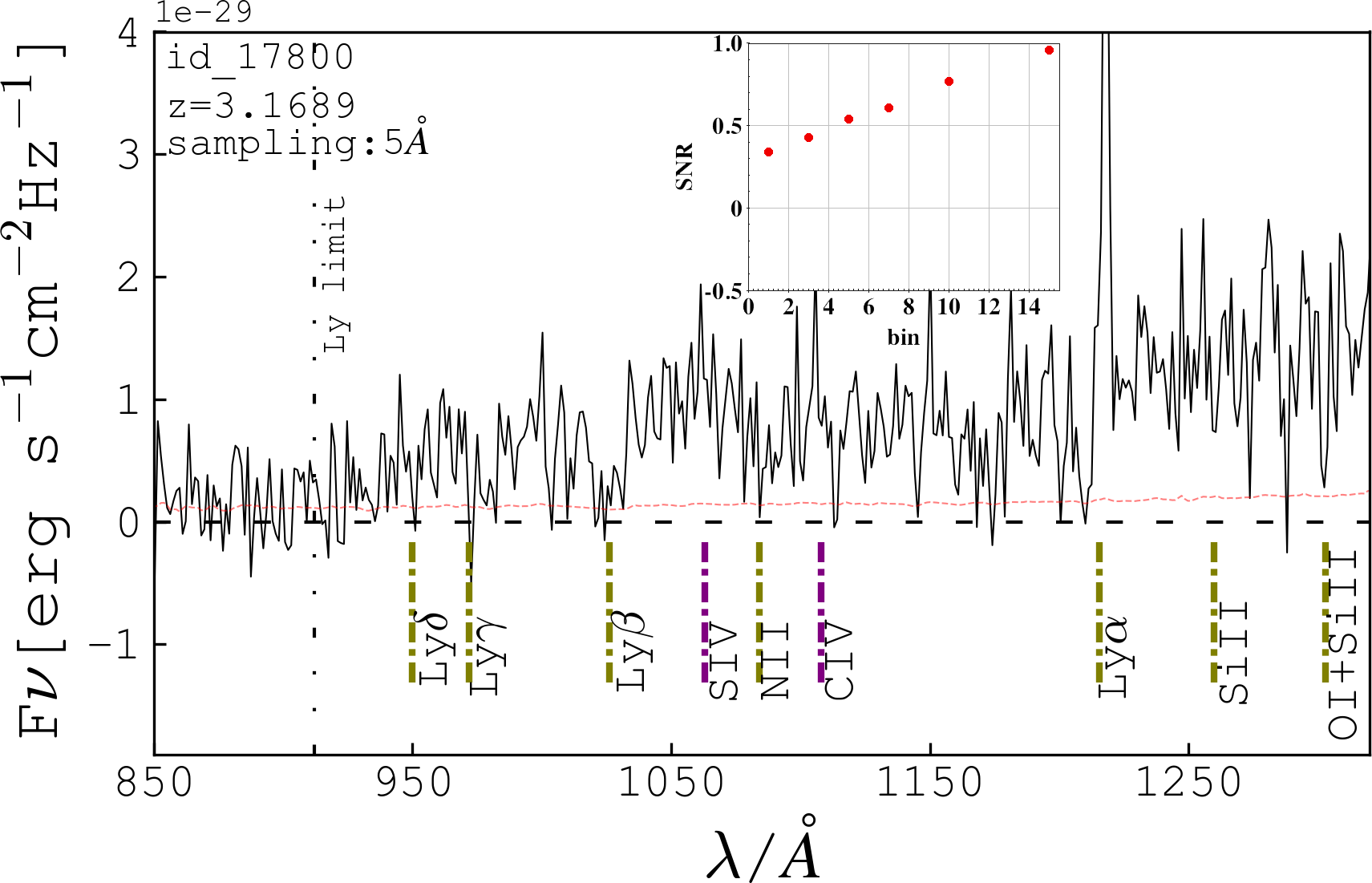}
  \caption{Same as Figure \ref{fig:sample}.}
  \label{sample_cont}
\end{figure}
\renewcommand{\thefigure}{\arabic{figure}}

The resampling of the LyC portion of the spectra is done by Python tool for resampling \textsc{SpectRes} \citep{Carnall2017} which resamples the flux densities and their uncertainties onto the required wavelength grids and preserves the integrated flux.
For all 9 candidates from the final sample the LyC flux and uncertainty elements are resampled in five different binning factors 3\AA, 5\AA, 7\AA, 10\AA\ and 15\AA.
Binning the spectra in the LyC range will increase the signal to noise ratio (SNR) if any statistically significant LyC signal is presented in the spectra of our candidates.
It is expected that with increasing the binning factor SNR will increase by $\sqrt{binning factor}$. 
In the case where no signal is detected in the probed wavelength frame, the SNR will stay same no matter how large binning factor we are using in the probed wavelength frame.
It is important to emphasise that this holds only if the photon counting is our primary source of noise.

The spectra of all 9 candidates are shown in Figure \ref{fig:sample}. 
On each 1D spectrum in the upper left corner the object id and measured redshift are noted.
The redshift is measured from stellar and interstellar absorption features marked as purple and olive dash dotted lines and the red line in the spectrum is the $1\sigma$ uncertainty.
Additionally we overlay Lyman break galaxy composite spectra from \cite{Shapley2003} to confirm our estimated redshifts (not shown in Figure \ref{fig:sample}). 
Moreover, for each 1D spectrum the inset plot is provided which shows the SNR as a function of the binning factor for the probed LyC interval.
This plot shows how the SNR of measured flux density changes with changing binning factor in the region 880\AA\ $<\lambda_{rest}<$ 910\AA.
As we can see from the inset plots in Figure \ref{fig:sample}, the SNR for most candidates remains constant below $\rm SNR=1$ which indicates that no statistically significant LyC flux is detected from the presented sample of galaxies.

\subsection{Probing LyC from composite spectra}

\begin{figure*}
  \centering
  \includegraphics[width=15cm, height=10cm]{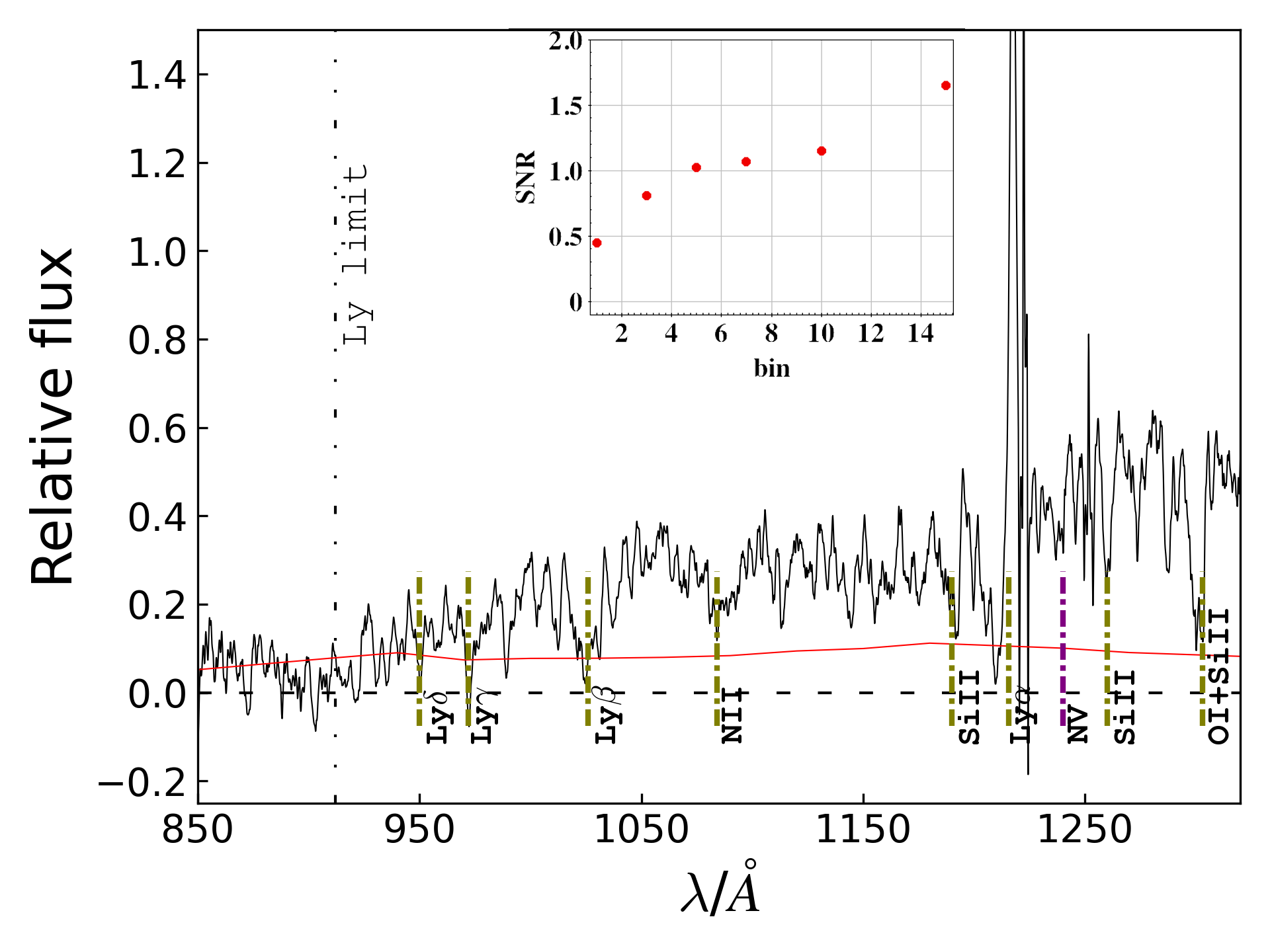}
  \caption{Composite 1D spectrum of the clean non-detection sample (9 candidates). The black dash-dot line represents the LyC limit and red line represents one sigma uncertainty of the stacked spectrum. The olive coloured dash-dotted lines are ISM absorption features and with purple dash-dotted lines stellar absorption features are marked. In the upper right corner plot SNR vs bin factor is presented for the composite spectra where binning is done in the 880\AA$<\lambda<910$\AA\ interval. No statistically significant LyC flux is measured.}
  \label{composite}
\end{figure*}

Significant ($\rm SNR>3$) LyC flux in the region 880\AA\ $<\lambda_{rest}<$ 910\AA\ is not detected in any of our candidates from the final sample. 
As a next step we analyze the composite spectrum created by stacking the 1D spectra of the 9 galaxies.
To create the 1D composite spectrum first each spectrum was shifted to rest-frame using \textsc{IRAF} task \textit{dopcor}.
In the next step we normalized each spectra by the mean value in its wavelength range 1450\AA\ $\lesssim \lambda_{rest}\lesssim$ 1500\AA.
Presented composite spectrum Figure \ref{composite} is created as a weighted average of all 9 1D spectra presented in this work.
We assign weights to each spectrum based on their SNR in the 880\AA\ $<\lambda_{rest}<$ 910\AA\ range.
Creating composite spectrum this way we avoid giving more importance to the objects with greater SNR in other parts of the spectrum but without LyC signal.
The reason for choosing this approach is motivated by our main goal to detect LyC radiation from the resulting composite spectrum.
The stacked 1D spectrum shows clear interstellar absorption features (olive colour dash-dotted lines) and a weak stellar absorption feature NV (purple dash-dotted lines) confirming that the measured redshifts for the observed galaxies are correct.
As with the single 1D spectra, we resampled the resulting 1D composite spectrum by using five different binning factors (3\AA, 5\AA, 7\AA, 10\AA\ and 15\AA).
The SNR of the stacked spectrum was $\sim0.5$ for non binned spectra an rise up to $\sim1.6$ for the biggest bin, inset plot Figure \ref{composite}.
From the resulting composite spectrum we measured a flux density ratio of $(F_{\nu}^{LyC}/F_{\nu}^{UV})_{\rm obs} = 0.027$.

Although no statistically significant LyC detection is reported we proceed with estimating $f_{\rm esc}^{\rm rel}$ and $f_{\rm esc}^{\rm abs}$ upper limits from composite spectra.
First, we evaluated $\exp{}(\tau_{IGM}^{LyC})$ for each candidate separately by using the results from the updated analytic model for transmission presented by \cite{Inoue2014}.
Then we averaged transmission factors $\exp{}(\tau_{IGM}^{LyC})$ from all candidates included in the final composite spectrum.
The final estimated averaged mean for the probed sample is $\exp{}(\tau_{IGM}^{LyC}) = 0.567$. Furthermore, the average $E(B-V)$ is found to be 0.1 for our stacked sample.
The intrinsic luminosity ratio $(L_{\nu}^{LyC}/L_{\nu}^{UV})_{\rm int}=0.333$ is adopted from literature \citep[e.g][]{Grazian2016, Marchi2017} and this ratio describes our assumptions that observed galaxies are mostly young star-forming galaxies, and it would be easier to compare our results with other studies.
The computed upper limits for relative and absolute escape fraction from the stacked 1D spectrum reveal $f_{\rm esc}^{\rm rel}=0.14$ and $f_{\rm esc}^{\rm abs}=0.06$.

\begin{table*}
\caption{Final sample of 9 clean galaxies probed for LyC flux}
\label{tab:final_sample}
\begin{tabular}{cccccccccc}
\hline \hline
id$^{1}$ & \begin{tabular}[c]{@{}c@{}}RA\\ $[\rm deg]$\end{tabular} & \begin{tabular}[c]{@{}c@{}}DEC\\ $[\rm deg]$\end{tabular} & \begin{tabular}[c]{@{}c@{}}$r$\\ $[\rm mag]$\end{tabular} & $E(B-V)$ & \begin{tabular}[c]{@{}c@{}}$F_{\nu}^{\rm LyC}$$^{2}$\\ $[10^{-30}]$\end{tabular} & \begin{tabular}[c]{@{}c@{}}SNR\\ $[F_{\rm LyC}]$\end{tabular} & \begin{tabular}[c]{@{}c@{}}$F_{\nu}^{\rm UV}$$^{2}$\\ $[10^{-29}]$\end{tabular} & $z_{spec}$ & \begin{tabular}[c]{@{}c@{}}EW($Ly\alpha$)\\ \AA \end{tabular} \\ \hline
7346 & 150.09767 & 2.25563 & 25.11$\pm0.02$  & 0.1 & 0.02 & 0.02 & 0.3 & 3.5624 & 13$\pm5$ \\
8805 & 150.12250 & 2.26981 & 24.36$\pm0.01$  & 0.1 & 2.08 & 0.02 & 1.41 & 2.9191 & 15$\pm5$ \\
12676 & 150.21806 & 2.31344 & 24.71$\pm0.04$ & 0.1 & 1.26 & 0.15 & 1.14 & 2.9629 & $\lesssim$ 0 \\
13459 & 150.20192 & 2.32179 & 24.72$\pm0.03$ & 0.1 & -0.24 & -0.05 & 3.86 & 3.0938 & 6$\pm5$ \\
14528 & 150.15553 & 2.33388 & 24.75$\pm0.04$ & 0.1 & -0.46 & -0.27 & 1.21 & 3.0012 & $\lesssim$ 0 \\
15332 & 150.16920 & 2.34218 & 24.95$\pm0.06$ & 0.1 & 0.74 & 0.08 & 1.19 & 3.1111 & 8$\pm5$ \\
15625 & 150.13919 & 2.34531 & 24.82$\pm0.04$ & 0.1 & -0.39 & -0.11 & 1.52 & 3.1837 & 20$\pm5$ \\
17008 & 150.16879 & 2.35901 & 24.99$\pm0.02$ & 0.1 & 0.16 & 0.08 & 0.29 & 3.4531 & 25$\pm10$ \\
17800 & 150.17387 & 2.36797 & 24.26$\pm0.03$ & 0.1 & 1.74 & 0.54 & 1.70 & 3.1689 & 15$\pm5$ \\  \hline \hline
\end{tabular}

\begin{flushleft}
Notes:\\$^{1}$ ZFOURGE id's.\\
$^{2}$ [erg s$^{-1}$ \rm cm$^{-2}$ Hz$^{-1}$]\\
\end{flushleft}
\end{table*}

\section{Analysis and Discussion}\label{4}
\raggedbottom

Many studies over the past 20 years have tried to detect LyC flux and estimate the escape fraction of ionizing radiation from various samples of galaxies at $z>2$ (most of these are summarized in Table \ref{Literature upper limits}).
Only a few have been successful in detecting LyC flux from a single galactic source and report the most probable $f_{\rm esc}^{\rm abs}$.
The selection methods and strategies to create samples of galaxies which shows LyC flux vary between studies and, in most cases probed galaxies are classified as Lyman alpha emitters (LAEs), Lyman break galaxies (LBGs), star-forming galaxies (SFGs), galaxies with strong restframe optical nebular emission lines, or randomly selected galaxies (e.g., magnitude limited searches) at $2\lesssim z < 6$.
The methods of probing LyC flux vary and involve spectroscopic and photometric (broadband and narrowband) space and ground-based observations.

\begin{table*}
\caption{Reported upper limits on $f_{\rm esc}$ from literature}
\label{Literature upper limits}
\begin{tabular}{c|c|c|c|c|c|c|c|c|c|}
\hline \hline
$\#$ & Author & \begin{tabular}[c]{@{}c@{}} $f_{\rm esc/upp}^{\rm abs}$ $^1$ \\ (adopted)\end{tabular} & \begin{tabular}[c]{@{}c@{}} $f_{\rm esc/upp}^{\rm abs}$ $^2$ \\ (reported)\end{tabular} & \begin{tabular}[c]{@{}c@{}} $f_{\rm esc/upp}^{\rm rel}$ $^3$ \\ (reported)\end{tabular} & z range $^4$ & \begin{tabular}[c]{@{}c@{}} Sample\\ type\end{tabular} & Sample size & observation & HST$^5$ \\ \hline
1 &  \cite{STEIDEL2001} & 0.60 & 0.60$\pm0.13$ $^6$ & 0.76$\pm0.16$ $^6$ & 3.31-3.49 & LBG & 29 & spectroscopy & - \\
2 &  \cite{Giallongo2002} & 0.047$^*$ & - & 0.16 & 2.96-3.32 & LBG & 2 & spectroscopy & - \\
3 &  \cite{Fernandez} & 0.17 & 0.17 & - & 1.95-2.85 & random & 14 & broadband & yes \\
4 &  \cite{Fernandez} & 0.004 & 0.004 & - & 2.85-3.50 & random & 13 & broadband & yes \\
5 &  \cite{Inoue2005} & 0.169 & 0.169 & 0.72 & 3.20 & LBG & 1 & narrowband & yes \\
6 &  \cite{Inoue2005} & 0.38 & 0.38 & 2.16 & 3.20 & LBG & 1 & narrowband & yes \\
7 &  \cite{Shapley2006} & 0.049$^{**}$ & - & 0.14 & 2.88-3.29 & LBG & 14 & spectroscopy & yes \\
8 &  \cite{Chen2007} & 0.075 & 0.075 & - & 2.03-3.12 & GRB & 28 & spectroscopy & - \\
9 &  \cite{Iwata2009} & 0.20 & 0.20 & 0.83 & 3.10 & LBG & 7 & narrowband & yes \\
10 & \cite{Fynbo2009} & 0.075 & 0.075 & - & 2.04-4.05 & GRB &33 & spectroscopy & - \\
11 & \cite{Vanzella2010} & 0.03 & 0.03 & 0.05 & 3.40-4.50 & LBG & 102 & broadband & yes \\
12 & \cite{Boutsia2011} & 0.015$^*$ & - & 0.05 & 3.27-3.35 & LBG & 11 & broadband & yes \\
13 & \cite{Nestor2011} & 0.1 & 0.1 & 0.5 & 3.06-3.09 & LBG & 20 & narrowband & yes \\
14 & \cite{Nestor2013} & 0.07 & 0.07 & - & 3.06-3.29 & LBG & 38 & narrowband & yes \\
15 & \cite{Nestor2013} & 0.30 & 0.30 & - & 3.04-3.11 & LAE & 88 & narrowband & yes \\
16 & \cite{Mostardi2013}$^7$& 0.02 & 0.02 & 0.08 & 2.81-3.41 & LBG & 49 & narrowband & partially \\
17 & \cite{Mostardi2013}$^8$ & 0.15 & 0.15 & 0.49 & 2.83-2.93 & LAE & 91 & narrowband & partially \\
18 & \cite{Amorin2014} & 0.23 & 0.23 & - & 3.42 & random & 1 & broadband & yes \\
19 & \cite{Micheva2016} & 0.15 & 0.15 & 0.69 & 3.05-3.15 & LAE & 138 & narrowband & yes \\
20 & \cite{Micheva2016} & 0.01 & 0.01 & 0.18 & 3.0-4.0 & LBG & 127 & narrowband & yes \\
21 & \cite{Guiata2016}$^9$ & 0.046$^{***}$ & - & 0.12 & 3.11-3.52 & SFG/LAE & 86 & narrowband & yes \\
22 & \cite{Grazian2016} & 0.008$^{***}$ & - & 0.02 & 3.27-3.40 & SFG & 37 & broadband & yes \\
23 & \cite{Vasei2016}  & 0.02 & 0.02 & 0.08 & 2.38 & random & 1 & broadband & yes \\
24 & \cite{Matthee2017} & 0.064 & 0.064 & 231 & 2.2 & HAE & 191 & broadband & yes \\
25 & \cite{Japelj2017}$^{10}$ & 0.102$^{***}$ & - & - & 3.0-4.0 & SFG & 145 & broadband & yes \\
26 & \cite{Marchi2017} & 0.035$^{***}$ & - & 0.09 & 3.51-4.42 & random & 33 & spectroscopy & yes \\
27 & \cite{Grazian2017} & 0.007$^{***}$ & - & 0.017 & 3.27-3.40 & SFG & 69 & broadband & yes \\
28 & \cite{Rutkowski2017} & 0.056 & 0.056 & 0.07 & 2.38-2.90 & SFG & 208 & spectroscopy & yes \\
29 & \cite{Rutkowski2017} & 0.067 & 0.067 & 0.078 & 2.31-2.25 & ELG & 41 & broadband & yes \\
30 & \cite{Smith2018} & 0.22 & 0.22$^{+0.44}_{-0.22}$ & - & 2.28-2.45 & SFG & 17 & broadband & yes \\
31 & \cite{Smith2018} & 0.53 & 0.53 & - & 2.57-3.08 & SFG & 7 & broadband & yes \\
32 & \cite{Smith2018} & 0.55 & 0.55 & - & 3.10-4.20 & SFG & 10 & broadband & yes \\
33 & \cite{Naidu2018} & 0.024$^{***}$ & - & 0.063 & 3.27-3.58 & SELG/EELG & 73 & broadband & yes \\
34 & \cite{Kakiichi2018}  & 0.08 & 0.08 & - & 5.8 & LBG & 6 & spectroscopy & yes \\
35 & \cite{Fletcher2019} & 0.005 & 0.005 & 0.006 & 3.10 & LAE & 42 & broadband & yes \\
36 & \cite{Tanvir2019}  & 0.015 & 0.015 & - & 1.6-6.7 & GRB & 140 & spectroscopy & - \\
37 & \cite{Ji2020} & 0.006 & 0.0063 & 0.029 & 3.40-3.95 & LBG & 107 & broadband & yes \\
38 & \cite{Uros2020} & 0.006 & 0.006 & - & 3.80-6.0 & random & 39 & broadband & yes \\
39 & \cite{Vielfaure2020}$^{11}$  & 0.35 & 0.35 & 0.43 & 3.5 & GRB & 1 & spectroscopy & yes \\
40 & \cite{Bosman2020}  & 0.01 & 0.01 & - & 5.8 & LAE & 1 & spectroscopy & yes \\
41 & \cite{Bian2020}  & 0.42 & 0.14-0.32 & - & 3.1 & LAE & 54 & spectroscopy & yes \\
42 & \cite{Pahl2021}  & 0.05 & 0.05 & - & 2.87-3.23 & LBG & 107 & spectroscopy & yes \\
43 & This work & 0.06 & 0.06 & 0.14 & 3.00-3.56  & SFG & 9 & spectroscopy & yes \\ \hline \hline
\end{tabular}  
\begin{flushleft}
Notes:\\$^{1}$ Adopted $f_{\rm esc}^{\rm abs}$ upper limits and used through entire this work.\\
$^{2}$ Reported $f_{\rm esc}^{\rm abs}$ upper limit.\\
$^{3}$ Reported $f_{\rm esc}^{\rm rel}$ upper limit.\\
$^{4}$ Probed redshift range.\\
$^{5}$ HST coverage, if (-) not available or we were unable to confirm. \\
$^{6}$ Values for $f_{\rm esc}^{\rm abs}$ and $f_{\rm esc}^{\rm rel}$ upper limits taken from \citep{Inoue2005} .\\
$^{7}$ Reported $f_{\rm esc}^{\rm abs}=0.1-0.2$ and $f_{\rm esc}^{\rm rel}=0.5-0.8$ upper limits.\\
$^{8}$ Reported $f_{\rm esc}^{\rm abs}=0.5-0.15$ and $f_{\rm esc}^{\rm rel}=0.18-0.49$ upper limits.\\
$^{9}$ Sample contains 67 SFGs and 19LAEs.\\
$^{10}$ Reported $f_{\rm esc}^{\rm rel}$ < 0.07, 0.2 and 0.6 upper limits at $L \sim L_{\rm z=3}^{*}, 0.5L_{\rm z=3}^{*}$ and $0.1L_{\rm z=3}^{*}$, respectively. We adopt mean value of these three estimates.\\
$^{11}$ Two presented GRBs are omitted since they are part of the sample presented in \citep{Tanvir2019}\\
$^*$ To evaluate $f_{\rm esc/upp}^{\rm abs}$ for LBGs we adopt $E(B-V)=0.13$ from \citep{Shapley2003}.    \\
$^{**}$ To evaluate $f_{\rm esc/upp}^{\rm abs}$ we adopt $E(B-V)=0.11$ from \citep{Shapley2006}. \\
$^{***}$ For all other cases where $f_{\rm esc/upp}^{\rm abs}$ was not reported we are using $E(B-V)=0.1$ to evaluate $f_{\rm esc/upp}^{\rm abs}$. \\

\end{flushleft}
\end{table*}

\subsection{The global $f_{\rm esc}^{\rm abs}$ upper limits advantages and limitations} 

Nearly all $2<z<6$ galaxies searched for LyC flux are classified as non-detections.
This implies that cases where LyC flux is detected from individual galaxies is extremely rare, with only $\sim$20 detections reported at $2\lesssim z \lesssim 4$.
The single galaxies with detected LyC flux demonstrate the possibility that a significant fraction ($20\%$ and higher) of LyC photons can escape from a galaxy into the IGM, e.g., Ion2 \citep{Vanzella2016,deBARROS2016}, Ion3 \citep{Vanzella2018}, uncontaminated KLCS sample \cite{Pahl2021}, LACES gold sample \citep{Fletcher2019}.
These findings also imply that detected LyC leakers are observed through lucky lines of sights with lower than mean HI column densities \cite{Bassett2021}.
In addition to this we should keep in mind that galaxy which leak LyC radiation into IGM it can be undetected due to the detection limit of the instrument or dense IGM (toward observed line of sight) which can result in complete absorption of escaped LyC radiation (in the observer direction).
Thus, if galaxies are the major drivers of reionization, the majority of the LyC photons are produced by galaxies for which we are unable to detect LyC radiation (i.e., non-detections as classified in most past studies).

The common approach of the research listed in Table \ref{Literature upper limits} when not detecting statistically significant LyC flux from individual galaxies is to probe LyC flux from a stacked sample.
This outcome is usually achieved by creating an average or weighted average data set from multiple 1D spectra or 2D photometric images.
If no significant LyC flux is detected, the results are reported as upper limits on $f_{\rm esc}^{\rm abs}$ or $f_{\rm esc}^{\rm rel}$.
Here we summarize much of the work published in the last $\sim$20 years reporting upper limits on escape fraction of ionizing radiation (including the sample from this work).
We note that LyC escape fractions are model dependent with different authors applying different models and assumptions. 
Therefore, it can be difficult to compare or to join results on LyC detections from different research, particularly when $f_{\rm esc}^{\rm abs}$ or $f_{\rm esc}^{\rm rel}$ are presented as a range.
However, it is less complicated to combine results on reported $f_{\rm esc}$ upper limits from different work because all results are evaluated over the large samples containing tens and sometimes hundreds of galaxies where modelled and physical properties are averaged.
As a result, an evaluation of the $f_{\rm esc}^{\rm abs}$ upper limits from the galaxy samples compiled here can provide a global LyC escape fraction ($f_{\rm esc/global}^{\rm abs}$) upper limit for the majority of galaxies probed for LyC radiation to date after adding to this sample confirmed LyC detections.

\subsection{Can galaxies reionize the Universe based on current observational limits on global $f_{\rm esc}^{\rm abs}$~?}

To investigate the upper limits on the global escape fraction we divide listed samples from Table \ref{Literature upper limits} into six different groups based on sample type.
The sample type is the name of the investigated population of galaxies probed for LyC flux during the particular research.
We ended up with six different groups of galaxies:
\begin{itemize}
\item Random - galaxies that are randomly selected or the author did not specify the type of the sample.
\item Star-Forming Galaxies (SFG)
\item Extreme Line Galaxies/Extreme Emitting Line Galaxies/ Strong Emitting Line Galaxies/Hydrogen Alpha Emitters (ELG/EELG/SELG/HAE)
\item Gama Ray Bursts (GRB) - GRB hosts are probed for LyC flux
\item Lyman Alpha Emitters (LAE)
\item Lyman Break Galaxies (LBG)
\end{itemize}

\begin{figure*}
  \centering
  \includegraphics[width=15cm, height=9.35cm]{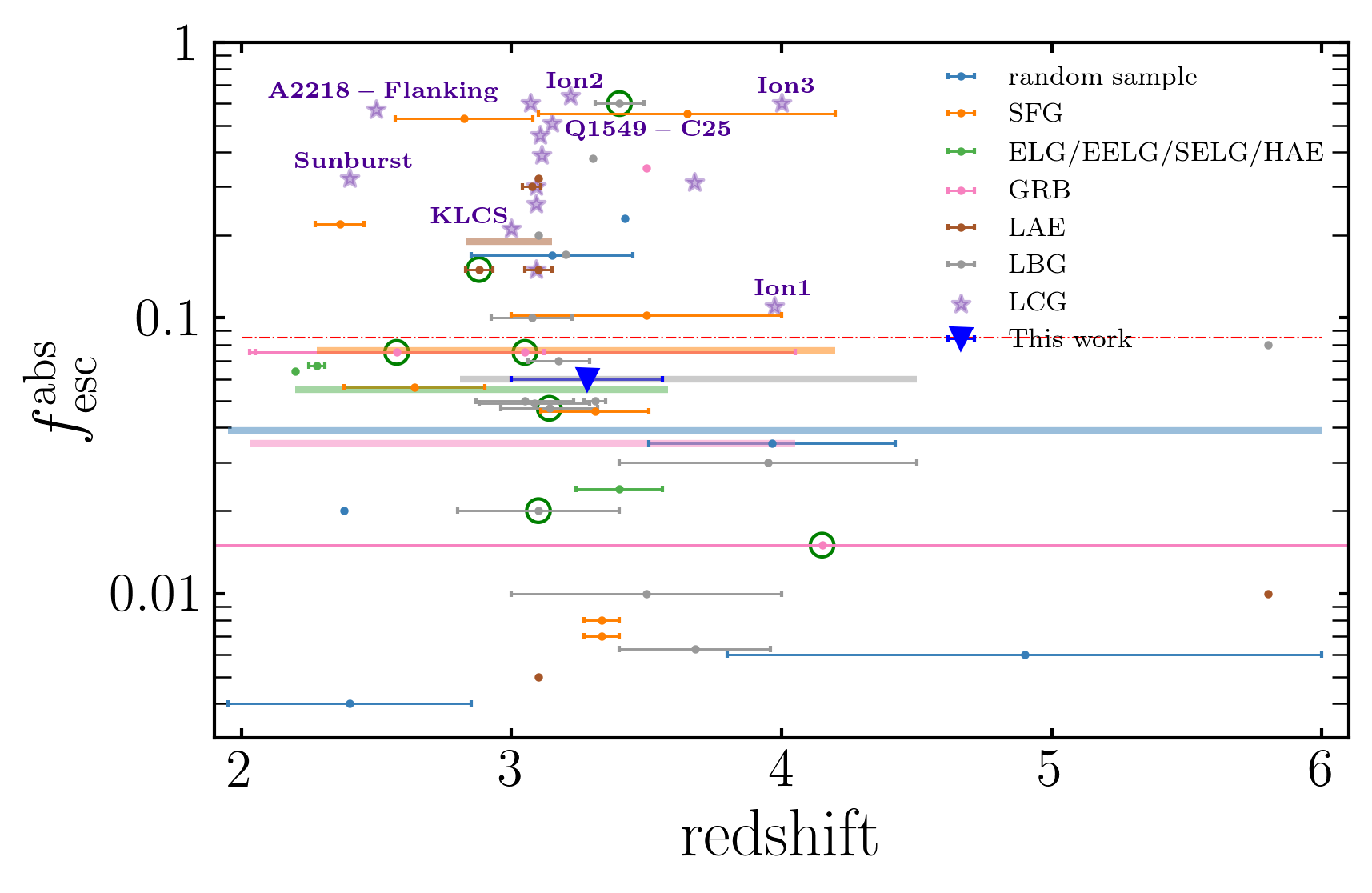}
  \caption{Reported upper limits on escape fraction of ionizing photons as a function of redshift for 43 samples of galaxies divided into 6 groups. Each group of galaxies is marked in different colours where horizontal error bars represent the probed redshift range. The purple stars are single galaxies with detected LyC flux and reported $f_{\rm esc}^{\rm abs}$ called as Lyman Continuum Galaxies. Thick horizontal lines are weighted mean upper limits on $f_{\rm esc}^{\rm abs}$ for a specific group of galaxies and dash-dotted red thin line is the upper limit for the escape fraction of ionizing photons at $z= 2-6$ redshift interval estimated from all groups of galaxies (see text for details). Samples of galaxies for which we are unable to confirm existence of the HST imaging are marked with the green circles. }
  \label{upper_limits_plot}
\end{figure*}

In Figure \ref{upper_limits_plot} we present the LyC absolute escape fraction upper limits for galaxy samples listed in Table \ref{Literature upper limits} as a function of redshift.
The upper limit for the sample presented in this work is marked with a blue triangle.
Upper limits for different groups from literature are represented as blue (random), orange (SFG), green (ELG/EELG/SELG/HAE), pink (GRB), brown (LAE) and grey (LBG) points.
Horizontal error bars represent the probed redshift range of the sample.
It is important to note that several samples are additionally marked with green circles.
For those samples, we were unable to confirm the existence of HST imaging what can imply that those samples can be contaminated by low redshift interlopers.
The purple stars are known Lyman continuum galaxies from the literature \citep{Vanzella2016,Shapley2016,deBARROS2016,Bian2017,Vanzella2018,Ji2020, Pahl2021}. 
Seven purple stars without names are the LACES gold sample from \cite{Fletcher2019}. 
The non-detection samples of galaxies for which $f_{\rm esc}^{\rm abs}$ are not reported appropriate $E(B-V)$ are adopted and we evaluated $f_{\rm esc}^{\rm abs}$ for those samples.
In the Table \ref{Literature upper limits} those samples are marked with single, double and triple asterisk signs.
The $f_{\rm esc}^{\rm abs}$ weighted mean upper limits were calculated using the number of galaxies in each literature sample to provide a values for each of the 6 different galaxy types.
The $f_{\rm esc}^{\rm abs}$ weighted mean upper limits for each group are marked in Figure \ref{upper_limits_plot} as a horizontal thick line where the length of the line is redshift coverage of the group.

After dividing samples in two redshift bins ($2\lesssim z \leq 3$ and $3\leq z \leq 4$) we evaluate weighted average $f_{\rm esc}^{\rm abs}$ upper limits where sample size is used as weights.
In this process, 14 samples were dropped since their redshift range was more than $z=0.1$ outside two defined redshift bins.
Finally, using weighted means of $f_{\rm esc}^{\rm abs}$ for each group we evaluate weighted mean escape fraction of ionizing photons ($f_{\rm esc}^{\rm abs}$) upper limit for galaxies where group sizes are used as weights ($2\lesssim z < 6$ range).
Resulting weighted mean $f_{\rm esc}^{\rm abs}$ upper limit for all LyC non-detections is plotted as horizontal red dash-dotted line Figure \ref{upper_limits_plot} and it is 0.084 (8.4\%).
The computed weighted mean $f_{\rm esc}^{\rm abs}$ upper limits for each group and group size are summarized in Table \ref{tab:Group sample stat}, where $f_{\rm esc}^{\rm abs}$ upper limits at $2\lesssim z \leq 3$, $3\leq z \leq 4$ and $2\lesssim z < 6$ range ($f_{\rm esc}^{\rm abs}$) are summarized in Table \ref{tab:z sample stat}.

Evaluated $f_{\rm esc}^{\rm abs}$ upper limits for a different group of galaxies indicates that on average LAEs and SFGs can potentially have the highest escape fractions of LyC radiation, $0.189$ and $0.076$  respectively.
Where upper limits on escape fractions of LyC radiation for other groups as LBGs, random, ELG/EELG/SELG/HAE and GRBs are $f_{\rm esc}^{\rm abs} = 0.039 - 0.076$.
In the case of LAEs and LBGs this is to some extent in agreement with the results derived from three-dimensional radiative transfer models for primordial galaxies where $f_{\rm esc}^{\rm abs}=0.07-0.47$ for LAEs and $f_{\rm esc}^{\rm abs}=0.06-0.17$ for LBGs is estimated \citep[]{Yajima2009}.
This is also consistent with the higher LyC detection rate of LAEs vs LBGs when comparing \cite{Fletcher2019} and \cite{Steidel2018}, shown in the \cite{Bassett2021}. 

Finally to evaluate global escape fraction $f_{\rm esc/global}^{\rm abs}$ from all galaxies probed for LyC radiation (in $2\lesssim z < 6$ range) we are including result from well known LCGs presented in the literature.
In total there are 26 confirmed LCGs, 13 galaxies from uncontaminated KLCS sample \citep{Pahl2021}, 7 galaxies from LACES \citep[gold sample,][]{Fletcher2019}, Ion1 \citep{Ji2020}, Ion2 \citep{Vanzella2016}, Ion3 \citep{Vanzella2018}, Sunburst \citep{Rivera2019}, A2218-Flanking \citep{Bian2017} and Q1549-C25 \citep{Shapley2016}.
We add all their reported $f_{\rm esc}^{\rm abs}$ estimates and divide by number of galaxies (26), after this we combine those results with estimated $f_{\rm esc}^{\rm abs}$ from all non-detections by evaluating weighted average. 
At the end resulting upper limits from all galaxies probed for LyC radiation in in $2\lesssim z < 6$ range is \bm{$f_{\rm esc/global}^{\rm abs}\sim0.088$} ($\sim 8.8\%$).

At the moment it is not clear which $f_{\rm esc}^{\rm abs}$ is required on average for galaxies to reionize the Universe.
For example, \cite{Finkelstein2019} present a semi-empirical model for reionization that prefers a model with increasing $f_{\rm esc}$ at high redshift and fainter magnitudes, specifically a global $f_{\rm esc}^{\rm abs}>0.1$ at $z\sim15$ and which drops to $f_{\rm esc}^{\rm abs}<0.05$ at $z<10$. 
On the other hand \cite{Khaire2016} suggest $f_{\rm esc}^{\rm abs}=0.14-0.22$ and \cite{Ouchi2010} suggest that galaxies at $z>7$ should have $f_{\rm esc}^{\rm abs}>0.2$ if galaxies alone are sufficient to support reionization.
Furthermore, by combining results from the 2012 Hubble Ultra Deep Field \citep[UDF12][]{Ellis2013,Koekemoer2013} and the observations of cosmic microwave background, \cite{Robertson2013} conclude under the assumption of $f_{\rm esc}^{\rm abs} = 0.2$ and clumping factor of $C_{\rm HII} \approx 3$ that galaxies probed at the depths of the high redshift surveys as UDF12 are not sufficient to reionize the Universe by $z\sim6$.
Either way, the presented $f_{\rm esc/global}^{\rm abs}<0.088$ upper limits in this work, can point in three directions (based on different assumptions) as to whether galaxies are efficient enough to reionize the IGM:

\begin{itemize}
\item First, there is a chance that galaxies can support reionization if the required $f_{\rm esc}^{\rm abs} < 0.05$ as reported in \cite{Finkelstein2019}. This is possible if escaped LyC radiation is below the current detection limit for galaxies classified as non-detections.
Since estimated upper limits indicate $f_{\rm esc/global}^{\rm abs} < 0.088$. 
\item Second, galaxies can not support reionization if $f_{\rm esc}^{\rm abs} \gtrsim 0.2$ is required \citep{Ouchi2010, Robertson2013, Khaire2016}. But there are indications that faint star-forming galaxies contributing most to the ionizing emissivity \citep[e.g.,][]{BOUWENS2015}, which leave galaxies in the race for the title of a dominant driver of the reionization.
\item Third, the line of sight or $f_{\rm esc}^{\rm abs}$ upper limits are not a direct measure of a true $f_{\rm esc}^{\rm abs}$ if LyC escape is anisotropic and if some LyC leakers are not detected due to high IGM opacity. Therefore $f_{\rm esc}^{\rm abs}$ estimates can not be interpreted as the total amount of ionizing radiation that escape from galaxies into IGM.
\end{itemize}
All these statements relying on the assumption that galaxies at $z<6$ have similar or the same $f_{\rm esc}^{\rm abs}$ properties as their $z>6$ counterparts.
Although there are promising observational  results that suggest detection of high $f_{\rm esc}^{\rm abs}\gtrsim 0.42$ from galaxies at $z>6$ \citep[e.g.,][]{Romain2020, Jeon2020}, too few sources have been detected to infer any trends.

The evaluated upper limits on $f_{\rm esc/global}^{\rm abs}$ indicates that the majority of galaxies on average do not show LyC escape fraction greater than $\sim8.8\%$ in the $2\lesssim z < 6$ range.
Due to the manner in which the data are presented in the literature and lack of large enough samples, it is difficult to divide data into smaller redshift bins and investigate how the $f_{\rm esc}^{\rm abs}$ upper limits evolves.
Because of this we only manage to divide sample into two redshift bins ($2\lesssim z \leq 3$ and $3\leq z \leq 4$) Table \ref{tab:z sample stat}, which enables to keep decent sample size in each redshift bin.  
Results from cosmological radiative transfer simulations presented by \cite{Khaire2016} suggest that $f_{\rm esc}^{\rm abs}=0-0.05$ from galaxies is required to keep Universe ionized at $z<3.5$.
It is hard to build any solid conclusion from $f_{\rm esc}^{\rm abs}$ upper limits estimated for galaxies in $2\lesssim z \leq 3$ and $3\leq z \leq 4$ range, 0.086 and 0.105 respectively, but this scenario can not be ruled out yet.

As a bottom line, from the presented upper limits on $f_{\rm esc}^{\rm abs}$ from galaxies classified as non-LyC leakers, and under different assumptions (previously discussed) we can state that our results imply that galaxies can stay in the game as a main, but not necessary single, contributor of ionizing photons during and after EoR.

\begin{table}
\caption{Mean upper limits on $f_{\rm esc}^{\rm abs}$ for each group}
\label{tab:Group sample stat}
\begin{tabular}{ccc}
\hline \hline
Group & \begin{tabular}[c]{@{}c@{}} $f_{\rm esc}^{\rm abs}$ upper limit\\ (weighted mean)\end{tabular} & Group size \\ \hline
random & 0.039 & 110 \\
SFG & 0.076 & 579 \\
ELG/EELG/SELG/HAE & 0.055 & 305 \\
GRB & 0.035 & 202 \\
LAE & 0.189 & 414 \\
LBG & 0.060 & 621 \\ \hline \hline
\end{tabular}
\end{table}

\begin{table}
\caption{Mean upper limits on $f_{\rm esc}^{\rm abs}$ from non-detections at particular redshift range}
\label{tab:z sample stat}
\begin{tabular}{ccc}
\hline \hline
Redshift range & \begin{tabular}[c]{@{}c@{}} $f_{\rm esc}^{\rm abs}$ upper limit\\ (weighted mean)\end{tabular} & Sample size \\ \hline
$2\lesssim z \leq 3$ & 0.086 & 570 \\
$3\leq z \leq 4$ & 0.105 & 1084 \\
$2\lesssim z < 6$ & 0.084 & 2231 \\ \hline \hline
\end{tabular}
\begin{flushleft}

\end{flushleft}
\end{table}

\subsection{Strength of EW(Ly$\alpha)$ as an indicator of LyC leakage?}

Understanding the relationship between the properties of the Ly$\alpha$ line, the structure of regions and leaking mechanisms from where LyC radiation efficiently escapes into the IGM at lower redshifts can provide insight into the LyC properties of galaxies beyond $z>6$ \citep[e.g.,][]{Izotov2018A}. 
Numerous theoretical efforts are focused on trying to understand processes related to leakage of Ly$\alpha$ and LyC radiation through the ISM and CGM and how they are correlated \citep{Dijkstra2016,Kakiichi2019,Kimm2019}. 
Mostly they agree that Ly$\alpha$ photons leave the galaxy more efficiently than LyC radiation and that they positively correlate.
Furthermore, results from simulations \citep{Verhamme2015,Kakiichi2018} show that multiple peaked Ly$\alpha$ line and their properties (peak separation) can indicate the existence of LyC leakage.
This is also confirmed with observations where some of the LCGs at low redshift \citep{Verhamme2017} and high redshift \citep{Vanzella2020} have multiple peaked Ly$\alpha$ line profiles.

\begin{figure*}
  \centering
  \includegraphics[width=16cm, height=10.5cm]{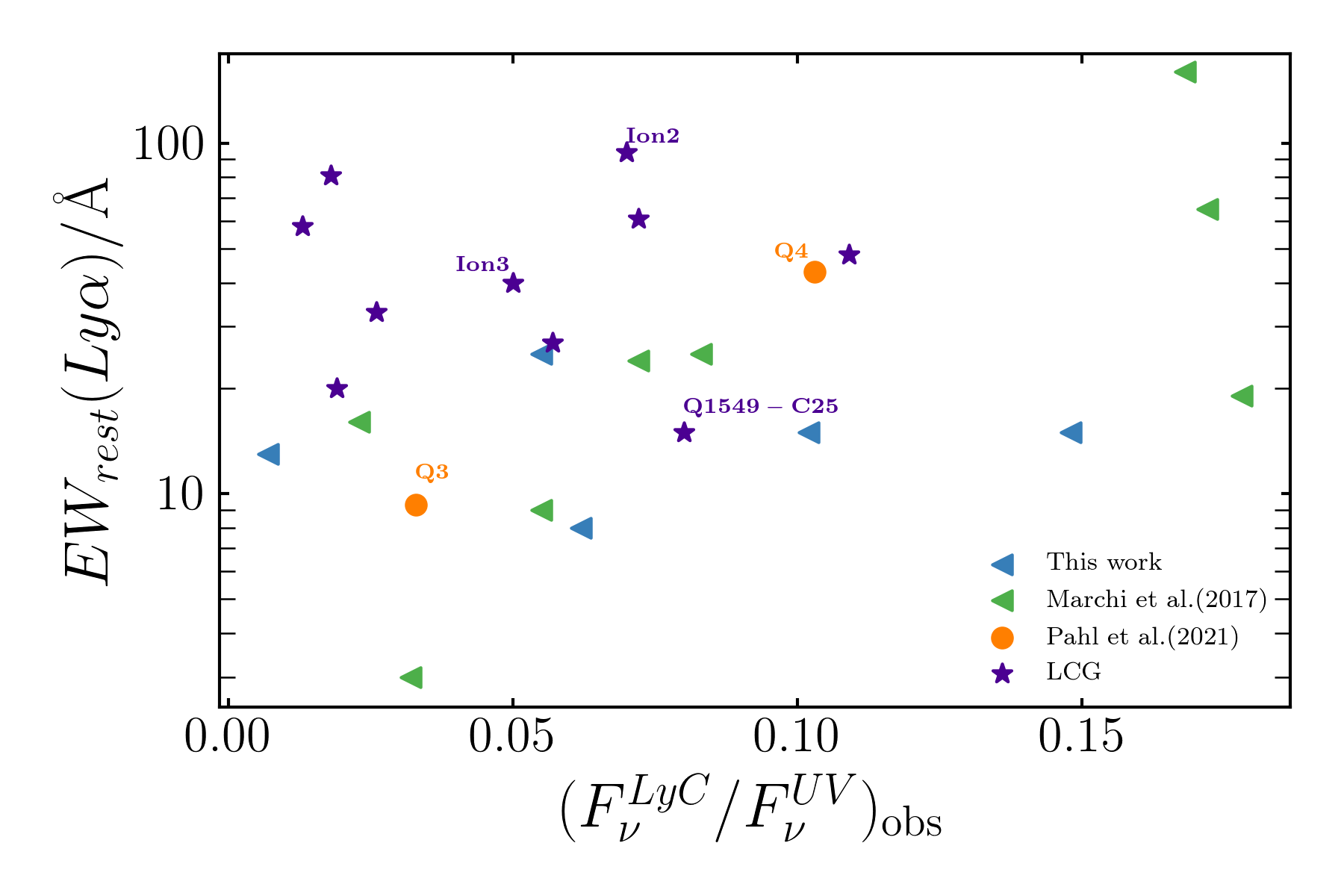}
  \caption{Rest frame EW(Ly$\alpha)>0$ is plotted as a function of observed flux density ratio $(F_{\nu}^{LyC}/F_{\nu}^{UV})_{\rm obs}$ for the sample of galaxies with non-detected LyC from the literature as green triangles and orange circles. The sample from this work is marked with blue triangles and confirmed LCGs from the literature are marked with purple stars.  It is important to note that the sample from \protect\cite{Pahl2021} has reported net EW(Ly$\alpha)$ (i.e., the full Ly$\alpha$ line, including the Ly$\alpha$ absorption component) leading to lower equivalent widths when compared with other plotted data. The orange circles are mean values for the Q3 and Q4 samples. }
  \label{ew_vs_obs}
\end{figure*}

Here we investigate a possible connection between reported upper limits on the observed flux density ratio $(F_{\nu}^{LyC}/F_{\nu}^{UV})_{\rm obs}$ and EW(Ly$\alpha)$ using our sample and including an available sample of galaxies from the literature (only positive $(F_{\nu}^{LyC}/F_{\nu}^{UV})_{\rm obs}$ values are considered for analysis).
In Figure \ref{ew_vs_obs} rest frame EW(Ly$\alpha)$ is shown as a function of $(F_{\nu}^{LyC}/F_{\nu}^{UV})_{\rm obs}$ and only galaxies with EW(Ly$\alpha)>0$ are included.
We would like to emphasize the fact that EW(Ly$\alpha)$ is sometimes measured in a different way among different studies, due to this the measurements can differ up to $50\%$.
To minimize these discrepancies, for our sample, we adopt the same method of measuring EW(Ly$\alpha)$ described in \cite{Cassata2015} that was also used by \cite{Marchi2017} where EW(Ly$\alpha$) is measured from the emission part of the Ly$\alpha$ line only.
Data in Figure \ref{ew_vs_obs} are presented in the following way, \cite{Marchi2017} green triangles, this work blue triangles, the LCG sample is presented as purple stars where LCGs from \cite{Fletcher2019} are plotted without their names.
In addition to the single measurements we add mean results for the sample of Q3 (LBG galaxies with 0\AA$\lesssim \rm EW(Ly\alpha) \lesssim$9.3\AA) and Q4 (LBG galaxies with 9.3\AA$\lesssim \rm EW(Ly\alpha) \lesssim$43\AA) galaxies \citep[orange circles,][]{Shapley2003}.
The EW(Ly$\alpha)$ for sample marked with orange circles is reported as net EW(Ly$\alpha)$ which means that the absorption part of the Ly$\alpha$ line is taken into consideration when equivalent width is measured. 
This can result in an underestimated EW (Ly$\alpha$) when compared with results from \cite{Marchi2017} and our work.

To check presented samples of galaxies with non-detected LyC flux (samples marked with triangles only) for the existence of any correlation we calculated Pearson's and Spearman's correlation coefficients $r=0.56$ and $r_{s}=0.55$ respectively.
The given result suggests the existence of a moderate statistically significant positive correlation among EW(Ly$\alpha$) and $(F_{\nu}^{LyC}/F_{\nu}^{UV})_{\rm obs}$.
Confirmation of this kind of correlation is in agreement with previous results from \cite{Marchi2017,Marchi2018,Steidel2018}, indicating that same mechanisms can be responsible for LyC and Ly$\alpha$ leakage as stated in their work.
After including the confirmed LCGs in the statistics, the Pearson's and Spearmans's coefficients reduced to 0.31 and 0.13, respectively, meaning there is a positive but weak correlation among EW(Ly$\alpha$) and $(F_{\nu}^{LyC}/F_{\nu}^{UV})_{\rm obs}$.
One possible explanation for decreasing in correlation after including confirmed LCGs is that the strong LyC emitters ($f_{\rm esc}^{\rm abs}\gtrsim20\%$) do not follow the same correlation as galaxies with low or non-escape fraction of the ionizing photons.
The same results (weak correlation, $r=0.33$ and $r_{s}=0.20$) are also found after including in analysis uncontaminated non-detected samples from \cite{Pahl2021} even if they use a different method of estimating EW(Ly$\alpha$).
It is important to note that observed trend (moderate or weak) can be false since data sets presented in the Figure \ref{5} are at different absolute magnitudes where different upper limits could reflect different source luminosity.

With the currently available sample, we can see that Ly$\alpha$ line in emission is presented with all confirmed LyC detection, except Ion1 who has the lowest estimated $f_{\rm esc}^{\rm abs}$.
Implying that Ly$\alpha$ in emission, in most cases, is the characteristic feature in the spectra of LCG galaxies.
This is additionally supported if we look at the confirmed sample of low redshift LCGs reported in \cite{Izotov2016b,Izotov2018} and if we take into account the fact that detection rate from the LAE sample \citep{Fletcher2019} is higher than detection rate from LBG selected sample presented by \cite{Steidel2018}.

\section{Summary and Conclusion}\label{5}

We have presented a sample of 9 non-contaminated galaxies from our larger Lyman continuum galaxy sample targeted for LyC flux in which no significant flux is detected.
The galaxies were observed on three different years and the selection criteria for each run was modified as new data for the fields were acquired.
After stacking the individual non-detection galaxies, no significant LyC flux above 3$\sigma$ was detected in the composite spectrum.
The upper limit on $f_{\rm esc}^{\rm abs}$ from the composite spectrum was evaluated at 0.06 (6$\%$).

The next step of the analysis involved extracting from literature (as many as can be found in the past $\sim20$ years) reported upper limits on $f_{\rm esc}^{\rm abs}$ and $f_{\rm esc}^{\rm rel}$ from samples of differing galaxy selection criteria.
After combining all the reported results, we evaluate the upper limits on $f_{\rm esc}^{\rm abs}$ for different galaxies types and estimate $f_{\rm esc}^{\rm abs}$ upper limits based on different redshift bins ($2\lesssim z \leq 3$ and $3\leq z \leq 4$). 
We estimate a global upper limit on escape fraction of LyC radiation including all 2231 galaxies classified as non-detections and 26 confirmed LCGs in total 2257 galaxies at  $2\lesssim z < 6$. 

Finally, we gather a large sample of galaxies with reported $(F_{\nu}^{LyC}/F_{\nu}^{UV})_{\rm obs}$ and EW(Ly$\alpha)$, including our sample, and test for the existence of a correlation between EW(Ly$\alpha)$ and escaping LyC flux reported in past theoretical and observational studies.

The most important results of our work are summarized below.

\begin{itemize}
\item Based on the sample of 2257 galaxies ($2\lesssim z < 6$) we estimate the global upper limit on escape fraction of ionizing photons $f_{\rm esc/global}^{\rm abs}<0.088$ or $< 8.8\%$.
\item Depending on which assumption from simulations we adopt ( $f_{\rm esc}^{\rm abs}<0.05$ or $f_{\rm esc}^{\rm abs}>0.2$ as a required minimum for galaxies to reionize the Universe) and the assumption that $f_{\rm esc}$ or the ionizing efficiency of galaxies doesn't decrease with redshift, it is possible that that ionizing flux from galaxies is sufficient to reionize the IGM.
\item We evaluate the upper limit on $f_{\rm esc}^{\rm abs}$ for galaxies in the $2\lesssim z \leq 3$ and $3\leq z \leq 4$ range, as 0.086 and 0.105 respectively.
At lower redshifts, $z<3.5$, upper limits on $f_{\rm esc}^{\rm abs}$ for galaxies supports the findings in \cite{Khaire2016} ($f_{\rm esc}^{\rm abs}=0.05$ is required to keep the Universe ionized at $z<3.5$, and no additional contribution from QSOs is required).
However, for galaxies at higher redshifts (between $z>3.5 - 5.5$), $f_{\rm esc}^{\rm abs}$ is required to increase at least a factor of 3 \cite{Khaire2016}. 
In order to keep Universe reionized but this trend is not evident from our results.    
\item The evaluated upper limits for different groups of galaxies imply that the highest $f_{\rm esc}^{\rm abs}$ upper limits are due to galaxies classified as LAEs and SFGs, whose $f_{\rm esc}^{\rm abs}$ upper limits are 0.189 and 0.060 respectively. 
\item  The moderately significant correlation among observed flux density ratio and EW(Ly$\alpha$) exist if we take into consideration only samples from \cite{Marchi2017} and this work with $r=0.56$ and $r_{s}=0.55$, but after including confirmed LCGs seen correlation is recognized as weak, which may suggest the existence of different correlations for strong LCGs.
\item We notice that no LCG with EW(Ly$\alpha)\lesssim25$\AA\ are reported in the literature, in contrast galaxies classified as non-LyC leakers have EW(Ly$\alpha$) that are higher and lower than 25\AA.   
\end{itemize}

In the future, we are planning to expand this work by doing a more detailed analysis of the presented samples. 
We are planning to include in the analysis other properties, absolute magnitudes, E(B-V) etc. by extracting them from the literature or evaluating them, in case if not available. 
In addition to this, we are planning to include in our sample population of AGNs at studied redshift since their contribution to hydrogen ionization, most likely, becomes more important as we are approaching lower redshifts.

The author of this paper would like to apologize if some samples from the literature relevant to this work have been inadvertently omitted from this analysis.
We attempted to include all available results.

\section*{Acknowledgements}

We wish to thank the referee for a careful reading of our manuscript, providing comments and corrections that improve and clarify our work.
We would like also to thank Eros Vanzella for useful discussion and opinions related to this work. 
This research was conducted by the Australian Research Council Centre of Excellence for All Sky Astrophysics in 3 Dimensions (ASTRO 3D), through project number CE170100013.
Some of the data presented herein were obtained at the W. M. Keck Observatory, which is operated as a scientific partnership among the California Institute of Technology, the University of California and the National Aeronautics and Space Administration. The Observatory was made possible by the generous financial support of the W. M. Keck Foundation.
The authors wish to recognize and acknowledge the very significant cultural role and reverence that the summit of Maunakea has always had within the indigenous Hawaiian community.  We are most fortunate to have the opportunity to conduct observations from this mountain.

\section*{DATA AVAILABILITY}

The Keck/LRIS data underlying this article will be shared on reasonable request to the corresponding author.




\bibliographystyle{mnras}
\bibliography{bib} 

\begin{thebibliography}{}
\makeatletter
\relax
\def\mn@urlcharsother{\let\do\@makeother \do\$\do\&\do\#\do\^\do\_\do\%\do\~}
\def\mn@doi{\begingroup\mn@urlcharsother \@ifnextchar [ {\mn@doi@}
  {\mn@doi@[]}}
\def\mn@doi@[#1]#2{\def\@tempa{#1}\ifx\@tempa\@empty \href
  {http://dx.doi.org/#2} {doi:#2}\else \href {http://dx.doi.org/#2} {#1}\fi
  \endgroup}
\def\mn@eprint#1#2{\mn@eprint@#1:#2::\@nil}
\def\mn@eprint@arXiv#1{\href {http://arxiv.org/abs/#1} {{\tt arXiv:#1}}}
\def\mn@eprint@dblp#1{\href {http://dblp.uni-trier.de/rec/bibtex/#1.xml}
  {dblp:#1}}
\def\mn@eprint@#1:#2:#3:#4\@nil{\def\@tempa {#1}\def\@tempb {#2}\def\@tempc
  {#3}\ifx \@tempc \@empty \let \@tempc \@tempb \let \@tempb \@tempa \fi \ifx
  \@tempb \@empty \def\@tempb {arXiv}\fi \@ifundefined
  {mn@eprint@\@tempb}{\@tempb:\@tempc}{\expandafter \expandafter \csname
  mn@eprint@\@tempb\endcsname \expandafter{\@tempc}}}

\bibitem[\protect\citeauthoryear{{Amor{\'\i}n} et~al.,}{{Amor{\'\i}n}
  et~al.}{2014}]{Amorin2014}
{Amor{\'\i}n} R.,  et~al., 2014, \mn@doi [\apjl] {10.1088/2041-8205/788/1/L4},
  \href {https://ui.adsabs.harvard.edu/abs/2014ApJ...788L...4A} {788, L4}

\bibitem[\protect\citeauthoryear{{Bassett} et~al.,}{{Bassett}
  et~al.}{2019}]{Bassett2019}
{Bassett} R.,  et~al., 2019, \mn@doi [\mnras] {10.1093/mnras/sty3320}, \href
  {http://adsabs.harvard.edu/abs/2019MNRAS.483.5223B} {483, 5223}

\bibitem[\protect\citeauthoryear{{Bassett}, {Ryan-Weber}, {Cooke},
  {Me{\v{s}}tri{\'c}}, {Kakiichi}, {Prichard}  \& {Rafelski}}{{Bassett}
  et~al.}{2021}]{Bassett2021}
{Bassett} R.,  {Ryan-Weber} E.~V.,  {Cooke} J.,  {Me{\v{s}}tri{\'c}} U.,
  {Kakiichi} K.,  {Prichard} L.,   {Rafelski} M.,  2021, \mn@doi [\mnras]
  {10.1093/mnras/stab070}, \href
  {https://ui.adsabs.harvard.edu/abs/2021MNRAS.502..108B} {502, 108}

\bibitem[\protect\citeauthoryear{{Becker} \& {Bolton}}{{Becker} \&
  {Bolton}}{2013}]{BECKER2013}
{Becker} G.~D.,  {Bolton} J.~S.,  2013, \mn@doi [\mnras]
  {10.1093/mnras/stt1610}, \href
  {http://adsabs.harvard.edu/abs/2013MNRAS.436.1023B} {436, 1023}

\bibitem[\protect\citeauthoryear{{Becker}, {Bolton}  \& {Lidz}}{{Becker}
  et~al.}{2015}]{Becker2015}
{Becker} G.~D.,  {Bolton} J.~S.,   {Lidz} A.,  2015, \mn@doi [\mnras]
  {10.1017/pasa.2015.45}, \href
  {http://adsabs.harvard.edu/abs/2015PASA...32...45B} {32, e045}

\bibitem[\protect\citeauthoryear{{Bian} \& {Fan}}{{Bian} \&
  {Fan}}{2020}]{Bian2020}
{Bian} F.,  {Fan} X.,  2020, \mn@doi [\mnras] {10.1093/mnrasl/slaa007}, \href
  {https://ui.adsabs.harvard.edu/abs/2020MNRAS.493L..65B} {493, L65}

\bibitem[\protect\citeauthoryear{{Bian}, {Fan}, {McGreer}, {Cai}  \&
  {Jiang}}{{Bian} et~al.}{2017}]{Bian2017}
{Bian} F.,  {Fan} X.,  {McGreer} I.,  {Cai} Z.,   {Jiang} L.,  2017, \mn@doi
  [\apjl] {10.3847/2041-8213/aa5ff7}, \href
  {https://ui.adsabs.harvard.edu/abs/2017ApJ...837L..12B} {837, L12}

\bibitem[\protect\citeauthoryear{{Bosman}, {Fan}, {Jiang}, {Reed}, {Matsuoka},
  {Becker}  \& {Haehnelt}}{{Bosman} et~al.}{2018}]{Bosman2018}
{Bosman} S.~E.~I.,  {Fan} X.,  {Jiang} L.,  {Reed} S.,  {Matsuoka} Y.,
  {Becker} G.,   {Haehnelt} M.,  2018, \mn@doi [\mnras]
  {10.1093/mnras/sty1344}, \href
  {http://adsabs.harvard.edu/abs/2018MNRAS.479.1055B} {479, 1055}

\bibitem[\protect\citeauthoryear{{Bosman}, {Kakiichi}, {Meyer}, {Gronke},
  {Laporte}  \& {Ellis}}{{Bosman} et~al.}{2020}]{Bosman2020}
{Bosman} S. E.~I.,  {Kakiichi} K.,  {Meyer} R.~A.,  {Gronke} M.,  {Laporte} N.,
    {Ellis} R.~S.,  2020, \mn@doi [\apj] {10.3847/1538-4357/ab85cd}, \href
  {https://ui.adsabs.harvard.edu/abs/2020ApJ...896...49B} {896, 49}

\bibitem[\protect\citeauthoryear{{Boutsia} et~al.,}{{Boutsia}
  et~al.}{2011}]{Boutsia2011}
{Boutsia} K.,  et~al., 2011, \mn@doi [\apj] {10.1088/0004-637X/736/1/41}, \href
  {https://ui.adsabs.harvard.edu/abs/2011ApJ...736...41B} {736, 41}

\bibitem[\protect\citeauthoryear{{Bouwens}, {Illingworth}, {Oesch}, {Caruana},
  {Holwerda}, {Smit}  \& {Wilkins}}{{Bouwens} et~al.}{2015}]{BOUWENS2015}
{Bouwens} R.~J.,  {Illingworth} G.~D.,  {Oesch} P.~A.,  {Caruana} J.,
  {Holwerda} B.,  {Smit} R.,   {Wilkins} S.,  2015, \mn@doi [\apj]
  {10.1088/0004-637X/811/2/140}, \href
  {http://adsabs.harvard.edu/abs/2015ApJ...811..140B} {811, 140}

\bibitem[\protect\citeauthoryear{{Calzetti}}{{Calzetti}}{1997}]{Calzetti1997}
{Calzetti} D.,  1997, \mn@doi [\aj] {10.1086/118242}, \href
  {http://adsabs.harvard.edu/abs/1997AJ....113..162C} {113, 162}

\bibitem[\protect\citeauthoryear{{Carnall}}{{Carnall}}{2017}]{Carnall2017}
{Carnall} A.~C.,  2017, arXiv e-prints, \href
  {https://ui.adsabs.harvard.edu/abs/2017arXiv170505165C} {p. arXiv:1705.05165}

\bibitem[\protect\citeauthoryear{{Cassata} et~al.,}{{Cassata}
  et~al.}{2015}]{Cassata2015}
{Cassata} P.,  et~al., 2015, \mn@doi [\aap] {10.1051/0004-6361/201423824},
  \href {https://ui.adsabs.harvard.edu/abs/2015A&A...573A..24C} {573, A24}

\bibitem[\protect\citeauthoryear{{Chen}, {Prochaska}  \& {Gnedin}}{{Chen}
  et~al.}{2007}]{Chen2007}
{Chen} H.-W.,  {Prochaska} J.~X.,   {Gnedin} N.~Y.,  2007, \mn@doi [\apjl]
  {10.1086/522306}, \href
  {https://ui.adsabs.harvard.edu/abs/2007ApJ...667L.125C} {667, L125}

\bibitem[\protect\citeauthoryear{{Cooke}, {Ryan-Weber}, {Garel}  \&
  {D{\'{\i}}az}}{{Cooke} et~al.}{2014}]{COOKE2014}
{Cooke} J.,  {Ryan-Weber} E.~V.,  {Garel} T.,   {D{\'{\i}}az} C.~G.,  2014,
  \mn@doi [\mnras] {10.1093/mnras/stu635}, \href
  {http://adsabs.harvard.edu/abs/2014MNRAS.441..837C} {441, 837}

\bibitem[\protect\citeauthoryear{{Dijkstra}, {Gronke}  \&
  {Venkatesan}}{{Dijkstra} et~al.}{2016}]{Dijkstra2016}
{Dijkstra} M.,  {Gronke} M.,   {Venkatesan} A.,  2016, \mn@doi [\apj]
  {10.3847/0004-637X/828/2/71}, \href
  {https://ui.adsabs.harvard.edu/abs/2016ApJ...828...71D} {828, 71}

\bibitem[\protect\citeauthoryear{{Eilers}, {Davies}  \& {Hennawi}}{{Eilers}
  et~al.}{2018}]{Eilers2018}
{Eilers} A.-C.,  {Davies} F.~B.,   {Hennawi} J.~F.,  2018, \mn@doi [\apj]
  {10.3847/1538-4357/aad4fd}, \href
  {http://adsabs.harvard.edu/abs/2018ApJ...864...53E} {864, 53}

\bibitem[\protect\citeauthoryear{{Ellis} et~al.,}{{Ellis}
  et~al.}{2013}]{Ellis2013}
{Ellis} R.~S.,  et~al., 2013, \mn@doi [\apjl] {10.1088/2041-8205/763/1/L7},
  \href {https://ui.adsabs.harvard.edu/abs/2013ApJ...763L...7E} {763, L7}

\bibitem[\protect\citeauthoryear{{Fan} et~al.,}{{Fan} et~al.}{2006}]{FAN2006}
{Fan} X.,  et~al., 2006, \mn@doi [\aj] {10.1086/504836}, \href
  {http://adsabs.harvard.edu/abs/2006AJ....132..117F} {132, 117}

\bibitem[\protect\citeauthoryear{{Fern{\'a}ndez-Soto}, {Lanzetta}  \&
  {Chen}}{{Fern{\'a}ndez-Soto} et~al.}{2003}]{Fernandez}
{Fern{\'a}ndez-Soto} A.,  {Lanzetta} K.~M.,   {Chen} H.~W.,  2003, \mn@doi
  [\mnras] {10.1046/j.1365-8711.2003.06622.x}, \href
  {https://ui.adsabs.harvard.edu/abs/2003MNRAS.342.1215F} {342, 1215}

\bibitem[\protect\citeauthoryear{{Finkelstein} et~al.,}{{Finkelstein}
  et~al.}{2019}]{Finkelstein2019}
{Finkelstein} S.~L.,  et~al., 2019, arXiv e-prints, \href
  {http://adsabs.harvard.edu/abs/2019arXiv190202792F} {}

\bibitem[\protect\citeauthoryear{{Fletcher}, {Tang}, {Robertson}, {Nakajima},
  {Ellis}, {Stark}  \& {Inoue}}{{Fletcher} et~al.}{2019}]{Fletcher2019}
{Fletcher} T.~J.,  {Tang} M.,  {Robertson} B.~E.,  {Nakajima} K.,  {Ellis}
  R.~S.,  {Stark} D.~P.,   {Inoue} A.,  2019, \mn@doi [\apj]
  {10.3847/1538-4357/ab2045}, \href
  {https://ui.adsabs.harvard.edu/abs/2019ApJ...878...87F} {878, 87}

\bibitem[\protect\citeauthoryear{{Fynbo} et~al.,}{{Fynbo}
  et~al.}{2009}]{Fynbo2009}
{Fynbo} J.~P.~U.,  et~al., 2009, \mn@doi [\apjs] {10.1088/0067-0049/185/2/526},
  \href {https://ui.adsabs.harvard.edu/abs/2009ApJS..185..526F} {185, 526}

\bibitem[\protect\citeauthoryear{{Giallongo}, {Cristiani}, {D'Odorico}  \&
  {Fontana}}{{Giallongo} et~al.}{2002}]{Giallongo2002}
{Giallongo} E.,  {Cristiani} S.,  {D'Odorico} S.,   {Fontana} A.,  2002,
  \mn@doi [\apjl] {10.1086/340254}, \href
  {https://ui.adsabs.harvard.edu/abs/2002ApJ...568L...9G} {568, L9}

\bibitem[\protect\citeauthoryear{{Grazian} et~al.,}{{Grazian}
  et~al.}{2016}]{Grazian2016}
{Grazian} A.,  et~al., 2016, \mn@doi [\aap] {10.1051/0004-6361/201526396},
  \href {http://adsabs.harvard.edu/abs/2016A%26A...585A..48G} {585, A48}

\bibitem[\protect\citeauthoryear{{Grazian} et~al.,}{{Grazian}
  et~al.}{2017}]{Grazian2017}
{Grazian} A.,  et~al., 2017, \mn@doi [\aap] {10.1051/0004-6361/201730447},
  \href {https://ui.adsabs.harvard.edu/abs/2017A&A...602A..18G} {602, A18}

\bibitem[\protect\citeauthoryear{{Greig} \& {Mesinger}}{{Greig} \&
  {Mesinger}}{2017}]{Greig2017}
{Greig} B.,  {Mesinger} A.,  2017, \mn@doi [\mnras] {10.1093/mnras/stx2118},
  \href {https://ui.adsabs.harvard.edu/abs/2017MNRAS.472.2651G} {472, 2651}

\bibitem[\protect\citeauthoryear{{Guaita} et~al.,}{{Guaita}
  et~al.}{2016}]{Guiata2016}
{Guaita} L.,  et~al., 2016, \mn@doi [\aap] {10.1051/0004-6361/201527597}, \href
  {http://adsabs.harvard.edu/abs/2016A%26A...587A.133G} {587, A133}

\bibitem[\protect\citeauthoryear{{Inoue} \& {Iwata}}{{Inoue} \&
  {Iwata}}{2008}]{INOUE2008}
{Inoue} A.~K.,  {Iwata} I.,  2008, \mn@doi [\mnras]
  {10.1111/j.1365-2966.2008.13350.x}, \href
  {http://adsabs.harvard.edu/abs/2008MNRAS.387.1681I} {387, 1681}

\bibitem[\protect\citeauthoryear{{Inoue}, {Iwata}, {Deharveng}, {Buat}  \&
  {Burgarella}}{{Inoue} et~al.}{2005}]{Inoue2005}
{Inoue} A.~K.,  {Iwata} I.,  {Deharveng} J.-M.,  {Buat} V.,   {Burgarella} D.,
  2005, \mn@doi [\aap] {10.1051/0004-6361:20041769}, \href
  {http://adsabs.harvard.edu/abs/2005A%26A...435..471I} {435, 471}

\bibitem[\protect\citeauthoryear{{Inoue}, {Shimizu}, {Iwata}  \&
  {Tanaka}}{{Inoue} et~al.}{2014}]{Inoue2014}
{Inoue} A.~K.,  {Shimizu} I.,  {Iwata} I.,   {Tanaka} M.,  2014, \mn@doi
  [\mnras] {10.1093/mnras/stu936}, \href
  {http://adsabs.harvard.edu/abs/2014MNRAS.442.1805I} {442, 1805}

\bibitem[\protect\citeauthoryear{{Iwata} et~al.,}{{Iwata}
  et~al.}{2009}]{Iwata2009}
{Iwata} I.,  et~al., 2009, \mn@doi [\apj] {10.1088/0004-637X/692/2/1287}, \href
  {http://adsabs.harvard.edu/abs/2009ApJ...692.1287I} {692, 1287}

\bibitem[\protect\citeauthoryear{{Izotov}, {Guseva}, {Fricke}  \&
  {Henkel}}{{Izotov} et~al.}{2016}]{Izotov2016b}
{Izotov} Y.~I.,  {Guseva} N.~G.,  {Fricke} K.~J.,   {Henkel} C.,  2016, \mn@doi
  [\mnras] {10.1093/mnras/stw1973}, \href
  {https://ui.adsabs.harvard.edu/abs/2016MNRAS.462.4427I} {462, 4427}

\bibitem[\protect\citeauthoryear{{Izotov}, {Schaerer}, {Worseck}, {Guseva},
  {Thuan}, {Verhamme}, {Orlitov{\'a}}  \& {Fricke}}{{Izotov}
  et~al.}{2018a}]{Izotov2018}
{Izotov} Y.~I.,  {Schaerer} D.,  {Worseck} G.,  {Guseva} N.~G.,  {Thuan} T.~X.,
   {Verhamme} A.,  {Orlitov{\'a}} I.,   {Fricke} K.~J.,  2018a, \mn@doi
  [\mnras] {10.1093/mnras/stx3115}, \href
  {http://adsabs.harvard.edu/abs/2018MNRAS.474.4514I} {474, 4514}

\bibitem[\protect\citeauthoryear{{Izotov}, {Worseck}, {Schaerer}, {Guseva},
  {Thuan}, {Fricke}  \& {Orlitov{\'a}}}{{Izotov} et~al.}{2018b}]{Izotov2018A}
{Izotov} Y.~I.,  {Worseck} G.,  {Schaerer} D.,  {Guseva} N.~G.,  {Thuan} T.~X.,
   {Fricke} Verhamme A.,   {Orlitov{\'a}} I.,  2018b, \mn@doi [\mnras]
  {10.1093/mnras/sty1378}, \href
  {https://ui.adsabs.harvard.edu/abs/2018MNRAS.478.4851I} {478, 4851}

\bibitem[\protect\citeauthoryear{{Japelj} et~al.,}{{Japelj}
  et~al.}{2017}]{Japelj2017}
{Japelj} J.,  et~al., 2017, \mn@doi [\mnras] {10.1093/mnras/stx477}, \href
  {https://ui.adsabs.harvard.edu/abs/2017MNRAS.468..389J} {468, 389}

\bibitem[\protect\citeauthoryear{{Jeon} et~al.,}{{Jeon}
  et~al.}{2020}]{Jeon2020}
{Jeon} J.,  et~al., 2020, arXiv e-prints, \href
  {https://ui.adsabs.harvard.edu/abs/2020arXiv201105918J} {p. arXiv:2011.05918}

\bibitem[\protect\citeauthoryear{{Ji} et~al.,}{{Ji} et~al.}{2020}]{Ji2020}
{Ji} Z.,  et~al., 2020, \mn@doi [\apj] {10.3847/1538-4357/ab5fdc}, \href
  {https://ui.adsabs.harvard.edu/abs/2020ApJ...888..109J} {888, 109}

\bibitem[\protect\citeauthoryear{{Kakiichi} \& {Gronke}}{{Kakiichi} \&
  {Gronke}}{2019}]{Kakiichi2019}
{Kakiichi} K.,  {Gronke} M.,  2019, arXiv e-prints, \href
  {https://ui.adsabs.harvard.edu/abs/2019arXiv190502480K} {p. arXiv:1905.02480}

\bibitem[\protect\citeauthoryear{{Kakiichi} et~al.,}{{Kakiichi}
  et~al.}{2018}]{Kakiichi2018}
{Kakiichi} K.,  et~al., 2018, \mn@doi [\mnras] {10.1093/mnras/sty1318}, \href
  {https://ui.adsabs.harvard.edu/abs/2018MNRAS.479...43K} {479, 43}

\bibitem[\protect\citeauthoryear{{Kashikawa} et~al.,}{{Kashikawa}
  et~al.}{2006}]{Kashikawa2006}
{Kashikawa} N.,  et~al., 2006, \mn@doi [\apj] {10.1086/504966}, \href
  {http://adsabs.harvard.edu/abs/2006ApJ...648....7K} {648, 7}

\bibitem[\protect\citeauthoryear{{Khaire}, {Srianand}, {Choudhury}  \&
  {Gaikwad}}{{Khaire} et~al.}{2016}]{Khaire2016}
{Khaire} V.,  {Srianand} R.,  {Choudhury} T.~R.,   {Gaikwad} P.,  2016, \mn@doi
  [\mnras] {10.1093/mnras/stw192}, \href
  {https://ui.adsabs.harvard.edu/abs/2016MNRAS.457.4051K} {457, 4051}

\bibitem[\protect\citeauthoryear{{Kimm}, {Blaizot}, {Garel}, {Michel-Dansac},
  {Katz}, {Rosdahl}, {Verhamme}  \& {Haehnelt}}{{Kimm} et~al.}{2019}]{Kimm2019}
{Kimm} T.,  {Blaizot} J.,  {Garel} T.,  {Michel-Dansac} L.,  {Katz} H.,
  {Rosdahl} J.,  {Verhamme} A.,   {Haehnelt} M.,  2019, \mn@doi [\mnras]
  {10.1093/mnras/stz989}, \href
  {https://ui.adsabs.harvard.edu/abs/2019MNRAS.486.2215K} {486, 2215}

\bibitem[\protect\citeauthoryear{{Koekemoer} et~al.,}{{Koekemoer}
  et~al.}{2007}]{Koekemoer2007}
{Koekemoer} A.~M.,  et~al., 2007, \mn@doi [\apj] {10.1086/520086}, \href
  {http://adsabs.harvard.edu/abs/2007ApJS..172..196K} {172, 196}

\bibitem[\protect\citeauthoryear{{Koekemoer} et~al.,}{{Koekemoer}
  et~al.}{2013}]{Koekemoer2013}
{Koekemoer} A.~M.,  et~al., 2013, \mn@doi [\apjs] {10.1088/0067-0049/209/1/3},
  \href {https://ui.adsabs.harvard.edu/abs/2013ApJS..209....3K} {209, 3}

\bibitem[\protect\citeauthoryear{{Laigle} et~al.,}{{Laigle}
  et~al.}{2016}]{Laigle2016}
{Laigle} C.,  et~al., 2016, \mn@doi [\apjs] {10.3847/0067-0049/224/2/24}, \href
  {https://ui.adsabs.harvard.edu/abs/2016ApJS..224...24L} {224, 24}

\bibitem[\protect\citeauthoryear{{Marchi} et~al.,}{{Marchi}
  et~al.}{2017}]{Marchi2017}
{Marchi} F.,  et~al., 2017, \mn@doi [\aap] {10.1051/0004-6361/201630054}, \href
  {http://adsabs.harvard.edu/abs/2017A%26A...601A..73M} {601, A73}

\bibitem[\protect\citeauthoryear{{Marchi} et~al.,}{{Marchi}
  et~al.}{2018}]{Marchi2018}
{Marchi} F.,  et~al., 2018, \mn@doi [\aap] {10.1051/0004-6361/201732133}, \href
  {http://adsabs.harvard.edu/abs/2018A%26A...614A..11M} {614, A11}

\bibitem[\protect\citeauthoryear{{Mason}, {Treu}, {Dijkstra}, {Mesinger},
  {Trenti}, {Pentericci}, {de Barros}  \& {Vanzella}}{{Mason}
  et~al.}{2018}]{Mason2018}
{Mason} C.~A.,  {Treu} T.,  {Dijkstra} M.,  {Mesinger} A.,  {Trenti} M.,
  {Pentericci} L.,  {de Barros} S.,   {Vanzella} E.,  2018, \mn@doi [\apj]
  {10.3847/1538-4357/aab0a7}, \href
  {https://ui.adsabs.harvard.edu/abs/2018ApJ...856....2M} {856, 2}

\bibitem[\protect\citeauthoryear{{Massey}, {Stoughton}, {Leauthaud}, {Rhodes},
  {Koekemoer}, {Ellis}  \& {Shaghoulian}}{{Massey} et~al.}{2010}]{Massey2010}
{Massey} R.,  {Stoughton} C.,  {Leauthaud} A.,  {Rhodes} J.,  {Koekemoer} A.,
  {Ellis} R.,   {Shaghoulian} E.,  2010, \mn@doi [\mnras]
  {10.1111/j.1365-2966.2009.15638.x}, \href
  {http://adsabs.harvard.edu/abs/2010MNRAS.401..371M} {401, 371}

\bibitem[\protect\citeauthoryear{{Matthee}, {Sobral}, {Best}, {Khostovan},
  {Oteo}, {Bouwens}  \& {R{\"o}ttgering}}{{Matthee} et~al.}{2017}]{Matthee2017}
{Matthee} J.,  {Sobral} D.,  {Best} P.,  {Khostovan} A.~A.,  {Oteo} I.,
  {Bouwens} R.,   {R{\"o}ttgering} H.,  2017, \mn@doi [\mnras]
  {10.1093/mnras/stw2973}, \href
  {https://ui.adsabs.harvard.edu/abs/2017MNRAS.465.3637M} {465, 3637}

\bibitem[\protect\citeauthoryear{{Me{\v{s}}tri{\'c}}
  et~al.,}{{Me{\v{s}}tri{\'c}} et~al.}{2020}]{Uros2020}
{Me{\v{s}}tri{\'c}} U.,  et~al., 2020, \mn@doi [\mnras]
  {10.1093/mnras/staa920}, \href
  {https://ui.adsabs.harvard.edu/abs/2020MNRAS.494.4986M} {494, 4986}

\bibitem[\protect\citeauthoryear{{Meyer}, {Laporte}, {Ellis}, {Verhamme}  \&
  {Garel}}{{Meyer} et~al.}{2020}]{Romain2020}
{Meyer} R.~A.,  {Laporte} N.,  {Ellis} R.~S.,  {Verhamme} A.,   {Garel} T.,
  2020, \mn@doi [\mnras] {10.1093/mnras/staa3216}, \href
  {https://ui.adsabs.harvard.edu/abs/2020MNRAS.500..558M} {500, 558}

\bibitem[\protect\citeauthoryear{Micheva, Iwata, Inoue, Matsuda, Yamada  \&
  Hayashino}{Micheva et~al.}{2016}]{Micheva2016}
Micheva G.,  Iwata I.,  Inoue A.~K.,  Matsuda Y.,  Yamada T.,   Hayashino T.,
  2016, \mn@doi [Monthly Notices of the Royal Astronomical Society]
  {10.1093/mnras/stw2700}, 465, 316

\bibitem[\protect\citeauthoryear{{Mostardi}, {Shapley}, {Nestor}, {Steidel},
  {Reddy}  \& {Trainor}}{{Mostardi} et~al.}{2013}]{Mostardi2013}
{Mostardi} R.~E.,  {Shapley} A.~E.,  {Nestor} D.~B.,  {Steidel} C.~C.,  {Reddy}
  N.~A.,   {Trainor} R.~F.,  2013, \mn@doi [\apj] {10.1088/0004-637X/779/1/65},
  \href {http://adsabs.harvard.edu/abs/2013ApJ...779...65M} {779, 65}

\bibitem[\protect\citeauthoryear{{Naidu}, {Forrest}, {Oesch}, {Tran}  \&
  {Holden}}{{Naidu} et~al.}{2018}]{Naidu2018}
{Naidu} R.~P.,  {Forrest} B.,  {Oesch} P.~A.,  {Tran} K.-V.~H.,   {Holden}
  B.~P.,  2018, \mn@doi [\mnras] {10.1093/mnras/sty961}, \href
  {https://ui.adsabs.harvard.edu/abs/2018MNRAS.478..791N} {478, 791}

\bibitem[\protect\citeauthoryear{{Nestor}, {Shapley}, {Steidel}  \&
  {Siana}}{{Nestor} et~al.}{2011}]{Nestor2011}
{Nestor} D.~B.,  {Shapley} A.~E.,  {Steidel} C.~C.,   {Siana} B.,  2011,
  \mn@doi [\apj] {10.1088/0004-637X/736/1/18}, \href
  {http://adsabs.harvard.edu/abs/2011ApJ...736...18N} {736, 18}

\bibitem[\protect\citeauthoryear{{Nestor}, {Shapley}, {Kornei}, {Steidel}  \&
  {Siana}}{{Nestor} et~al.}{2013}]{Nestor2013}
{Nestor} D.~B.,  {Shapley} A.~E.,  {Kornei} K.~A.,  {Steidel} C.~C.,   {Siana}
  B.,  2013, \mn@doi [\apj] {10.1088/0004-637X/765/1/47}, \href
  {https://ui.adsabs.harvard.edu/abs/2013ApJ...765...47N} {765, 47}

\bibitem[\protect\citeauthoryear{{Oke}}{{Oke}}{1990}]{Oke1990}
{Oke} J.~B.,  1990, \mn@doi [\aj] {10.1086/115444}, \href
  {https://ui.adsabs.harvard.edu/abs/1990AJ.....99.1621O} {99, 1621}

\bibitem[\protect\citeauthoryear{{Oke} et~al.,}{{Oke} et~al.}{1995}]{Oke1995}
{Oke} J.~B.,  et~al., 1995, \mn@doi [\pasp] {10.1086/133562}, \href
  {https://ui.adsabs.harvard.edu/abs/1995PASP..107..375O} {107, 375}

\bibitem[\protect\citeauthoryear{{Ouchi} et~al.,}{{Ouchi}
  et~al.}{2010}]{Ouchi2010}
{Ouchi} M.,  et~al., 2010, \mn@doi [\apj] {10.1088/0004-637X/723/1/869}, \href
  {http://adsabs.harvard.edu/abs/2010ApJ...723..869O} {723, 869}

\bibitem[\protect\citeauthoryear{{Paardekooper}, {Khochfar}  \& {Dalla
  Vecchia}}{{Paardekooper} et~al.}{2015}]{Paardekooper2015}
{Paardekooper} J.-P.,  {Khochfar} S.,   {Dalla Vecchia} C.,  2015, \mn@doi
  [\mnras] {10.1093/mnras/stv1114}, \href
  {https://ui.adsabs.harvard.edu/abs/2015MNRAS.451.2544P} {451, 2544}

\bibitem[\protect\citeauthoryear{{Pahl}, {Shapley}, {Steidel}, {Chen}  \&
  {Reddy}}{{Pahl} et~al.}{2021}]{Pahl2021}
{Pahl} A.~J.,  {Shapley} A.,  {Steidel} C.~C.,  {Chen} Y.,   {Reddy} N.~A.,
  2021, \mn@doi [\mnras] {10.1093/mnras/stab1374}, \href
  {https://ui.adsabs.harvard.edu/abs/2021MNRAS.tmp.1487P} {}

\bibitem[\protect\citeauthoryear{{Pentericci} et~al.,}{{Pentericci}
  et~al.}{2011}]{Pentericci2011}
{Pentericci} L.,  et~al., 2011, \mn@doi [\apj] {10.1088/0004-637X/743/2/132},
  \href {https://ui.adsabs.harvard.edu/abs/2011ApJ...743..132P} {743, 132}

\bibitem[\protect\citeauthoryear{{Rivera-Thorsen} et~al.,}{{Rivera-Thorsen}
  et~al.}{2019}]{Rivera2019}
{Rivera-Thorsen} T.~E.,  et~al., 2019, \mn@doi [Science]
  {10.1126/science.aaw0978}, \href
  {https://ui.adsabs.harvard.edu/abs/2019Sci...366..738R} {366, 738}

\bibitem[\protect\citeauthoryear{{Robertson} et~al.,}{{Robertson}
  et~al.}{2013}]{Robertson2013}
{Robertson} B.~E.,  et~al., 2013, \mn@doi [\apj] {10.1088/0004-637X/768/1/71},
  \href {https://ui.adsabs.harvard.edu/abs/2013ApJ...768...71R} {768, 71}

\bibitem[\protect\citeauthoryear{{Rutkowski} et~al.,}{{Rutkowski}
  et~al.}{2017}]{Rutkowski2017}
{Rutkowski} M.~J.,  et~al., 2017, \mn@doi [\apjl] {10.3847/2041-8213/aa733b},
  \href {https://ui.adsabs.harvard.edu/abs/2017ApJ...841L..27R} {841, L27}

\bibitem[\protect\citeauthoryear{{Scoville} et~al.,}{{Scoville}
  et~al.}{2007}]{Scoville2007}
{Scoville} N.,  et~al., 2007, \mn@doi [\apjs] {10.1086/516585}, \href
  {http://adsabs.harvard.edu/abs/2007ApJS..172....1S} {172, 1}

\bibitem[\protect\citeauthoryear{{Shapley}, {Steidel}, {Pettini}  \&
  {Adelberger}}{{Shapley} et~al.}{2003}]{Shapley2003}
{Shapley} A.~E.,  {Steidel} C.~C.,  {Pettini} M.,   {Adelberger} K.~L.,  2003,
  \mn@doi [\apjs] {10.1086/373922}, \href
  {http://adsabs.harvard.edu/abs/2003ApJ...588...65S} {588, 65}

\bibitem[\protect\citeauthoryear{{Shapley}, {Steidel}, {Pettini}, {Adelberger}
  \& {Erb}}{{Shapley} et~al.}{2006}]{Shapley2006}
{Shapley} A.~E.,  {Steidel} C.~C.,  {Pettini} M.,  {Adelberger} K.~L.,   {Erb}
  D.~K.,  2006, \mn@doi [\apj] {10.1086/507511}, \href
  {https://ui.adsabs.harvard.edu/abs/2006ApJ...651..688S} {651, 688}

\bibitem[\protect\citeauthoryear{{Shapley}, {Steidel}, {Strom},
  {Bogosavljevi{\'c}}, {Reddy}, {Siana}, {Mostardi}  \& {Rudie}}{{Shapley}
  et~al.}{2016}]{Shapley2016}
{Shapley} A.~E.,  {Steidel} C.~C.,  {Strom} A.~L.,  {Bogosavljevi{\'c}} M.,
  {Reddy} N.~A.,  {Siana} B.,  {Mostardi} R.~E.,   {Rudie} G.~C.,  2016,
  \mn@doi [\apjl] {10.3847/2041-8205/826/2/L24}, \href
  {http://adsabs.harvard.edu/abs/2016ApJ...826L..24S} {826, L24}

\bibitem[\protect\citeauthoryear{{Siana} et~al.,}{{Siana}
  et~al.}{2007}]{Siana2007}
{Siana} B.,  et~al., 2007, \mn@doi [\apj] {10.1086/521185}, \href
  {http://adsabs.harvard.edu/abs/2007ApJ...668...62S} {668, 62}

\bibitem[\protect\citeauthoryear{{Smith} et~al.,}{{Smith}
  et~al.}{2018}]{Smith2018}
{Smith} B.~M.,  et~al., 2018, \mn@doi [\apj] {10.3847/1538-4357/aaa3dc}, \href
  {https://ui.adsabs.harvard.edu/abs/2018ApJ...853..191S} {853, 191}

\bibitem[\protect\citeauthoryear{{Steidel}, {Pettini}  \&
  {Adelberger}}{{Steidel} et~al.}{2001}]{STEIDEL2001}
{Steidel} C.~C.,  {Pettini} M.,   {Adelberger} K.~L.,  2001, \mn@doi [\apj]
  {10.1086/318323}, \href {http://adsabs.harvard.edu/abs/2001ApJ...546..665S}
  {546, 665}

\bibitem[\protect\citeauthoryear{{Steidel}, {Shapley}, {Pettini}, {Adelberger},
  {Erb}, {Reddy}  \& {Hunt}}{{Steidel} et~al.}{2004}]{STEIDEL2004}
{Steidel} C.~C.,  {Shapley} A.~E.,  {Pettini} M.,  {Adelberger} K.~L.,  {Erb}
  D.~K.,  {Reddy} N.~A.,   {Hunt} M.~P.,  2004, \mn@doi [\apj]
  {10.1086/381960}, \href {http://adsabs.harvard.edu/abs/2004ApJ...604..534S}
  {604, 534}

\bibitem[\protect\citeauthoryear{{Steidel}, {Bogosavljevi{\'c}}, {Shapley},
  {Reddy}, {Rudie}, {Pettini}, {Trainor}  \& {Strom}}{{Steidel}
  et~al.}{2018}]{Steidel2018}
{Steidel} C.~C.,  {Bogosavljevi{\'c}} M.,  {Shapley} A.~E.,  {Reddy} N.~A.,
  {Rudie} G.~C.,  {Pettini} M.,  {Trainor} R.~F.,   {Strom} A.~L.,  2018,
  \mn@doi [\apj] {10.3847/1538-4357/aaed28}, \href
  {http://adsabs.harvard.edu/abs/2018ApJ...869..123S} {869, 123}

\bibitem[\protect\citeauthoryear{{Straatman} et~al.,}{{Straatman}
  et~al.}{2016}]{Straatman2016}
{Straatman} C.~M.~S.,  et~al., 2016, \mn@doi [\apj]
  {10.3847/0004-637X/830/1/51}, \href
  {http://adsabs.harvard.edu/abs/2016ApJ...830...51S} {830, 51}

\bibitem[\protect\citeauthoryear{{Tanvir} et~al.,}{{Tanvir}
  et~al.}{2019}]{Tanvir2019}
{Tanvir} N.~R.,  et~al., 2019, \mn@doi [\mnras] {10.1093/mnras/sty3460}, \href
  {https://ui.adsabs.harvard.edu/abs/2019MNRAS.483.5380T} {483, 5380}

\bibitem[\protect\citeauthoryear{{Tody}}{{Tody}}{1986}]{Tody1986}
{Tody} D.,  1986, in {Crawford} D.~L.,  ed.,  \procspie Vol. 627,
  Instrumentation in astronomy VI. p.~733, \mn@doi{10.1117/12.968154}

\bibitem[\protect\citeauthoryear{{Vanzella}, {Siana}, {Cristiani}  \&
  {Nonino}}{{Vanzella} et~al.}{2010}]{Vanzella2010}
{Vanzella} E.,  {Siana} B.,  {Cristiani} S.,   {Nonino} M.,  2010, \mn@doi
  [\mnras] {10.1111/j.1365-2966.2010.16408.x}, \href
  {http://adsabs.harvard.edu/abs/2010MNRAS.404.1672V} {404, 1672}

\bibitem[\protect\citeauthoryear{{Vanzella} et~al.,}{{Vanzella}
  et~al.}{2012}]{Vanzella2012}
{Vanzella} E.,  et~al., 2012, \mn@doi [\mnras]
  {10.1111/j.1745-3933.2012.01286.x}, \href
  {http://adsabs.harvard.edu/abs/2012MNRAS.424L..54V} {424, L54}

\bibitem[\protect\citeauthoryear{{Vanzella} et~al.,}{{Vanzella}
  et~al.}{2015}]{Vanzella2015}
{Vanzella} E.,  et~al., 2015, \mn@doi [\aap] {10.1051/0004-6361/201525651},
  \href {http://adsabs.harvard.edu/abs/2015A%26A...576A.116V} {576, A116}

\bibitem[\protect\citeauthoryear{{Vanzella} et~al.,}{{Vanzella}
  et~al.}{2016}]{Vanzella2016}
{Vanzella} E.,  et~al., 2016, \mn@doi [\apj] {10.3847/0004-637X/825/1/41},
  \href {http://adsabs.harvard.edu/abs/2016ApJ...825...41V} {825, 41}

\bibitem[\protect\citeauthoryear{{Vanzella} et~al.,}{{Vanzella}
  et~al.}{2018}]{Vanzella2018}
{Vanzella} E.,  et~al., 2018, \mn@doi [\mnras] {10.1093/mnrasl/sly023}, \href
  {https://ui.adsabs.harvard.edu/abs/2018MNRAS.476L..15V} {476, L15}

\bibitem[\protect\citeauthoryear{{Vanzella} et~al.,}{{Vanzella}
  et~al.}{2020}]{Vanzella2020}
{Vanzella} E.,  et~al., 2020, \mn@doi [\mnras] {10.1093/mnras/stz2286}, \href
  {https://ui.adsabs.harvard.edu/abs/2020MNRAS.491.1093V} {491, 1093}

\bibitem[\protect\citeauthoryear{{Vasei} et~al.,}{{Vasei}
  et~al.}{2016}]{Vasei2016}
{Vasei} K.,  et~al., 2016, \mn@doi [\apj] {10.3847/0004-637X/831/1/38}, \href
  {https://ui.adsabs.harvard.edu/abs/2016ApJ...831...38V} {831, 38}

\bibitem[\protect\citeauthoryear{{Verhamme}, {Orlitov{\'a}}, {Schaerer}  \&
  {Hayes}}{{Verhamme} et~al.}{2015}]{Verhamme2015}
{Verhamme} A.,  {Orlitov{\'a}} I.,  {Schaerer} D.,   {Hayes} M.,  2015, \mn@doi
  [\aap] {10.1051/0004-6361/201423978}, \href
  {http://adsabs.harvard.edu/abs/2015A%26A...578A...7V} {578, A7}

\bibitem[\protect\citeauthoryear{{Verhamme}, {Orlitov{\'a}}, {Schaerer},
  {Izotov}, {Worseck}, {Thuan}  \& {Guseva}}{{Verhamme}
  et~al.}{2017}]{Verhamme2017}
{Verhamme} A.,  {Orlitov{\'a}} I.,  {Schaerer} D.,  {Izotov} Y.,  {Worseck} G.,
   {Thuan} T.~X.,   {Guseva} N.,  2017, \mn@doi [\aap]
  {10.1051/0004-6361/201629264}, \href
  {http://adsabs.harvard.edu/abs/2017A%26A...597A..13V} {597, A13}

\bibitem[\protect\citeauthoryear{{Vielfaure} et~al.,}{{Vielfaure}
  et~al.}{2020}]{Vielfaure2020}
{Vielfaure} J.~B.,  et~al., 2020, \mn@doi [\aap] {10.1051/0004-6361/202038316},
  \href {https://ui.adsabs.harvard.edu/abs/2020A&A...641A..30V} {641, A30}

\bibitem[\protect\citeauthoryear{{Wise} \& {Cen}}{{Wise} \&
  {Cen}}{2009}]{Wise2009}
{Wise} J.~H.,  {Cen} R.,  2009, \mn@doi [\apj] {10.1088/0004-637X/693/1/984},
  \href {https://ui.adsabs.harvard.edu/abs/2009ApJ...693..984W} {693, 984}

\bibitem[\protect\citeauthoryear{{Wyithe} \& {Bolton}}{{Wyithe} \&
  {Bolton}}{2011}]{Wyithe2011}
{Wyithe} J.~S.~B.,  {Bolton} J.~S.,  2011, \mn@doi [\mnras]
  {10.1111/j.1365-2966.2010.18030.x}, \href
  {http://adsabs.harvard.edu/abs/2011MNRAS.412.1926W} {412, 1926}

\bibitem[\protect\citeauthoryear{Yajima, Umemura, Mori  \& Nakamoto}{Yajima
  et~al.}{2009}]{Yajima2009}
Yajima H.,  Umemura M.,  Mori M.,   Nakamoto T.,  2009, \mn@doi [Monthly
  Notices of the Royal Astronomical Society]
  {10.1111/j.1365-2966.2009.15195.x}, 398, 715

\bibitem[\protect\citeauthoryear{{de Barros} et~al.,}{{de Barros}
  et~al.}{2016}]{deBARROS2016}
{de Barros} S.,  et~al., 2016, \mn@doi [\aap] {10.1051/0004-6361/201527046},
  \href {http://adsabs.harvard.edu/abs/2016A%26A...585A..51D} {585, A51}

\makeatother
\end{thebibliography}





\begin{appendix}
\section{Images of the 9 LyC candidates IN three HST filters}

\begin{figure*}
\begin{center}
    \subfloat{\includegraphics[width=.25\textwidth]{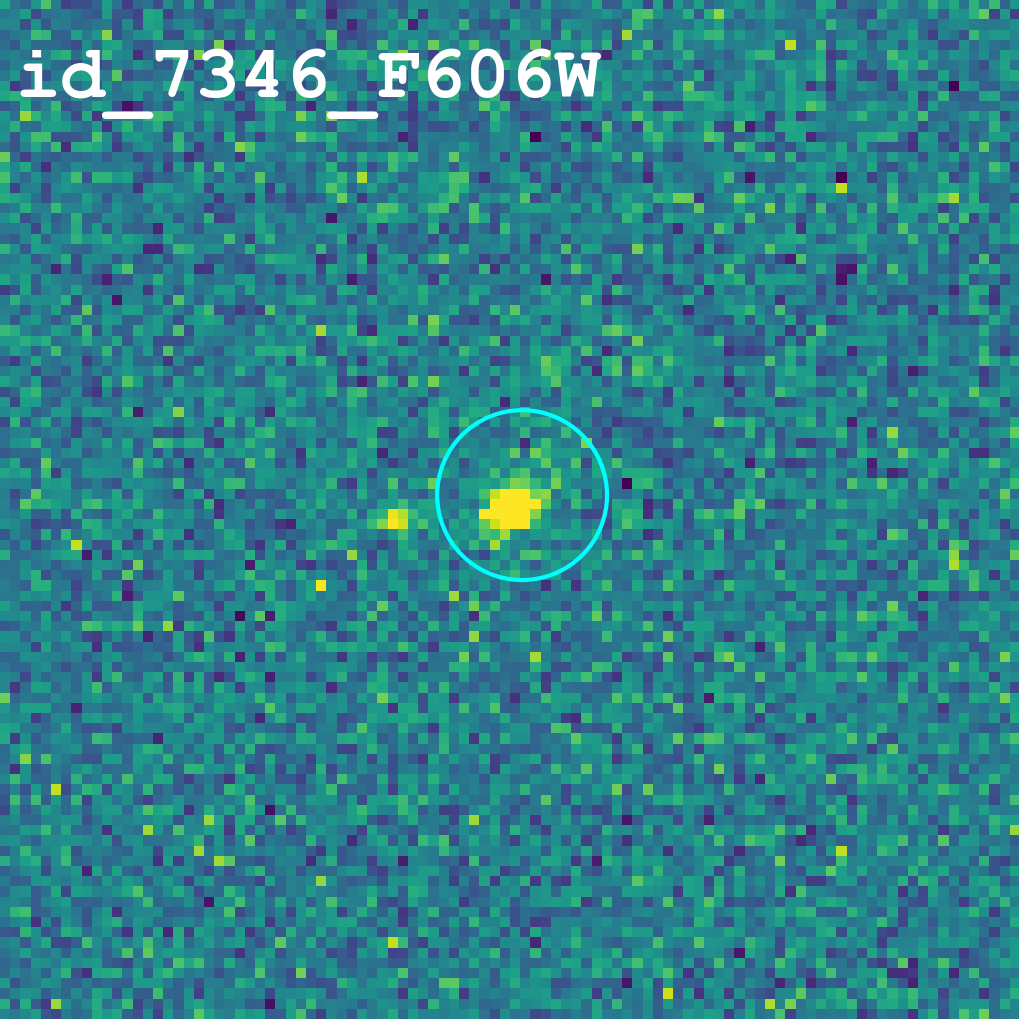}}
    \hspace{-0.01cm}\vspace{-0.05cm}
    \subfloat{\includegraphics[width=.25\textwidth]{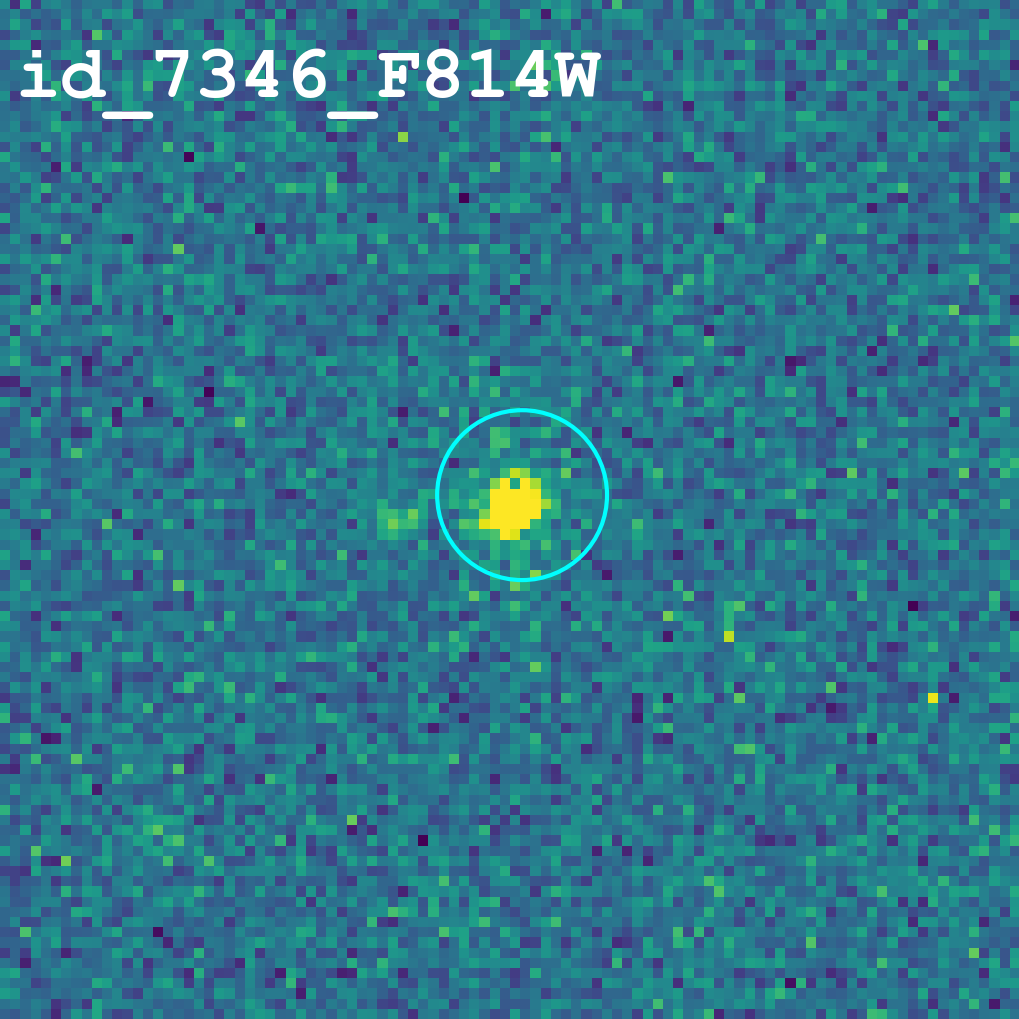}}
    \hspace{-0.01cm}\vspace{-0.05cm}
    \subfloat{\includegraphics[width=.25\textwidth]{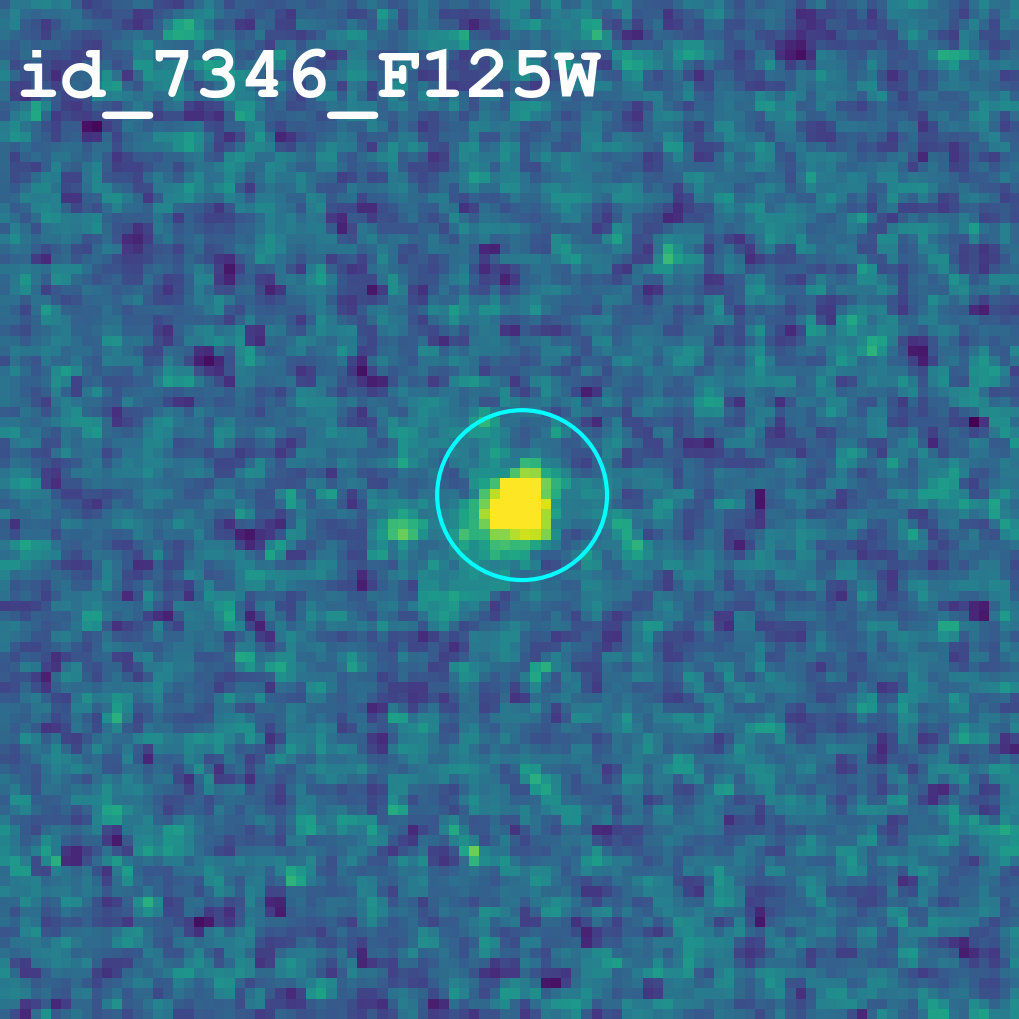}}
    \hspace{-0.01cm}\vspace{-0.05cm}

    \subfloat{\includegraphics[width=.25\textwidth]{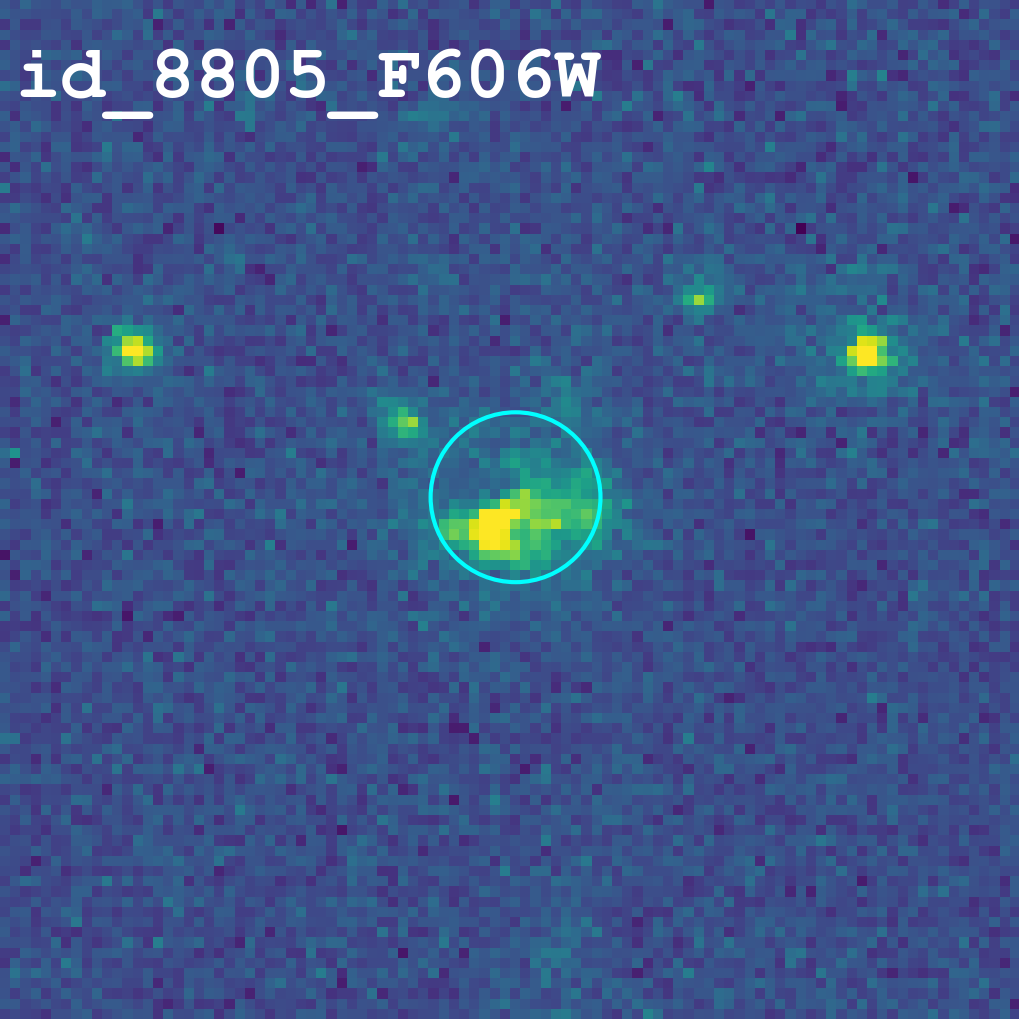}}
    \hspace{-0.01cm}\vspace{-0.05cm}
    \subfloat{\includegraphics[width=.25\textwidth]{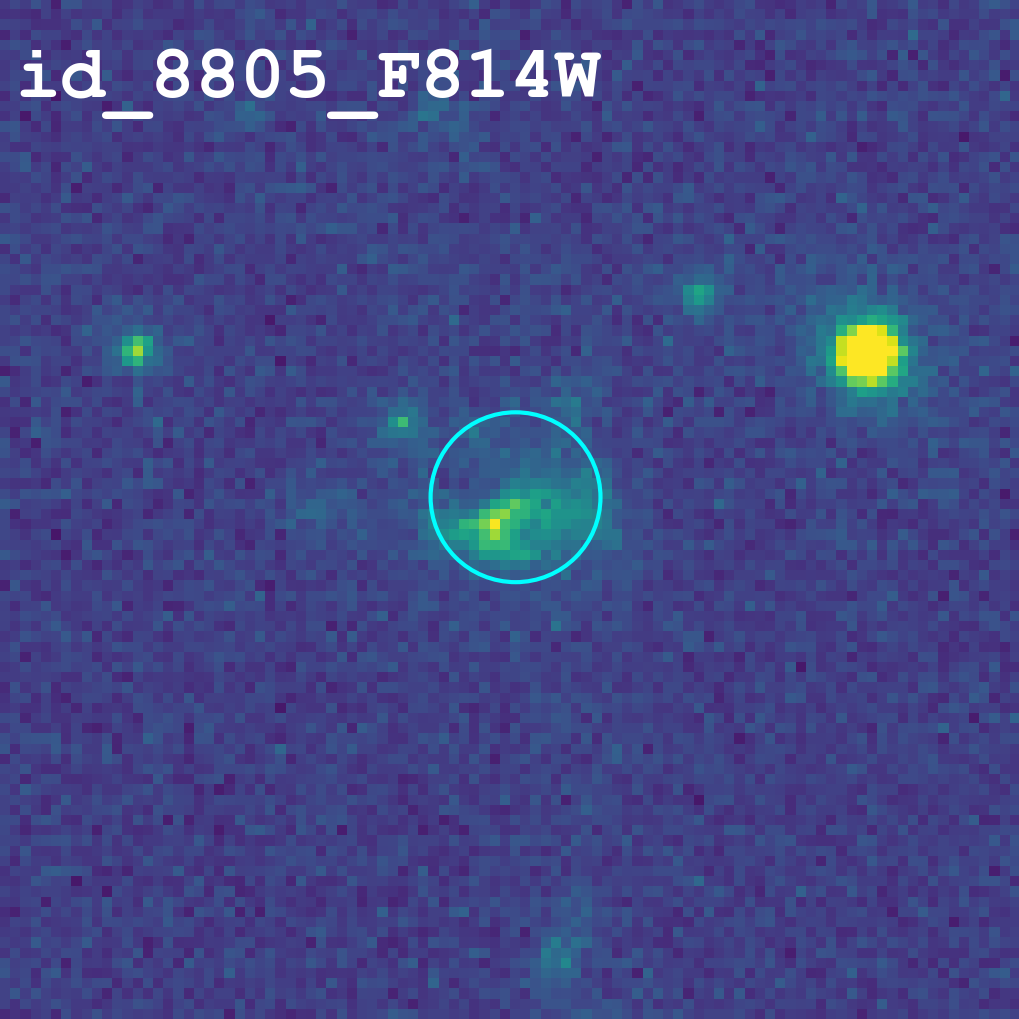}}
    \hspace{-0.01cm}\vspace{-0.05cm}
    \subfloat{\includegraphics[width=.25\textwidth]{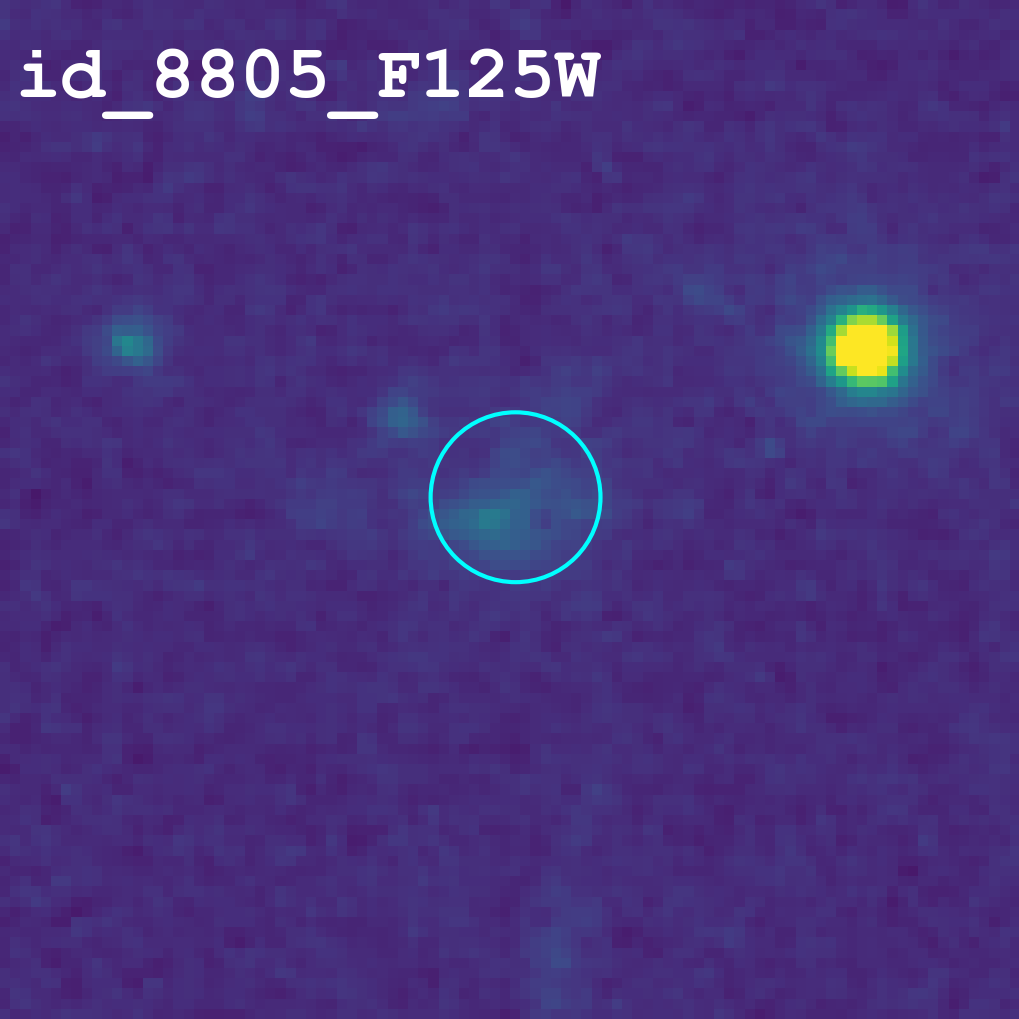}}
    \hspace{-0.01cm}\vspace{-0.05cm}

    \subfloat{\includegraphics[width=.25\textwidth]{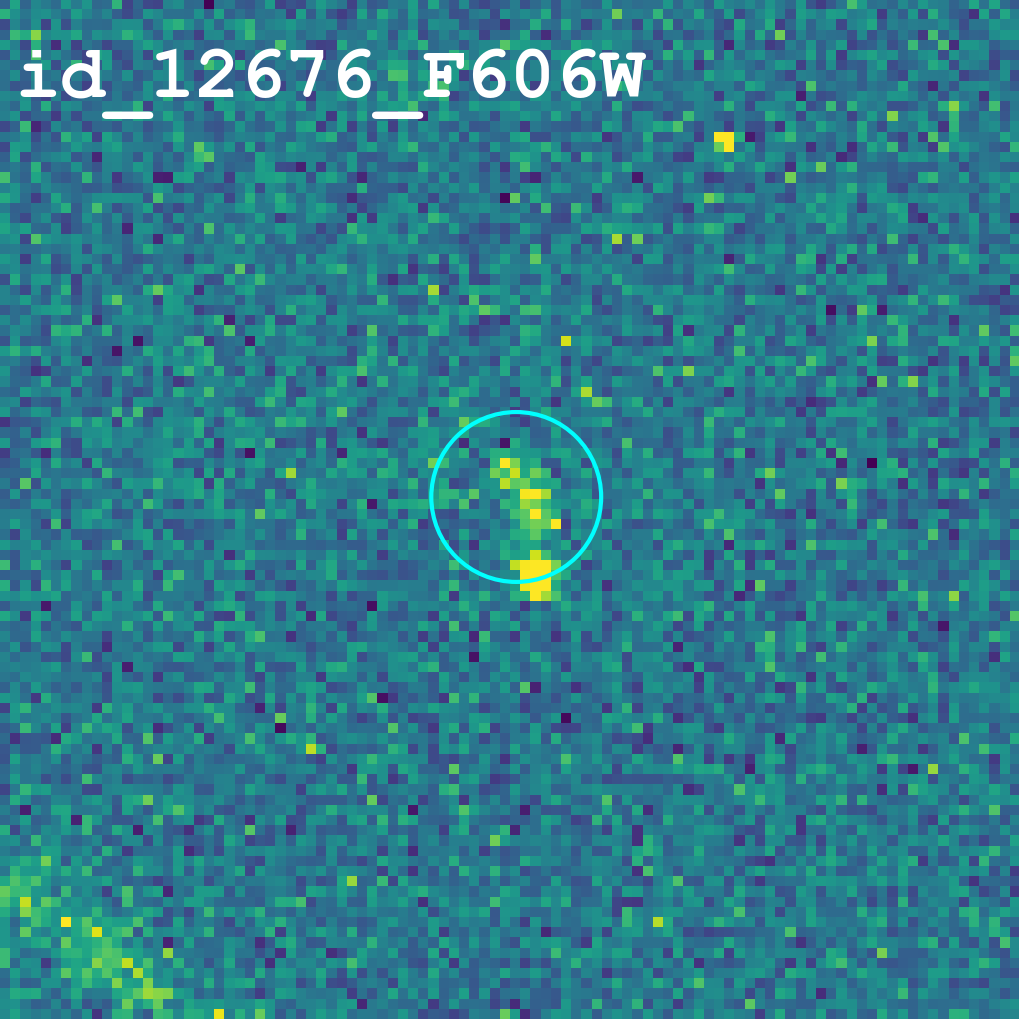}}
    \hspace{-0.01cm}\vspace{-0.05cm}
    \subfloat{\includegraphics[width=.25\textwidth]{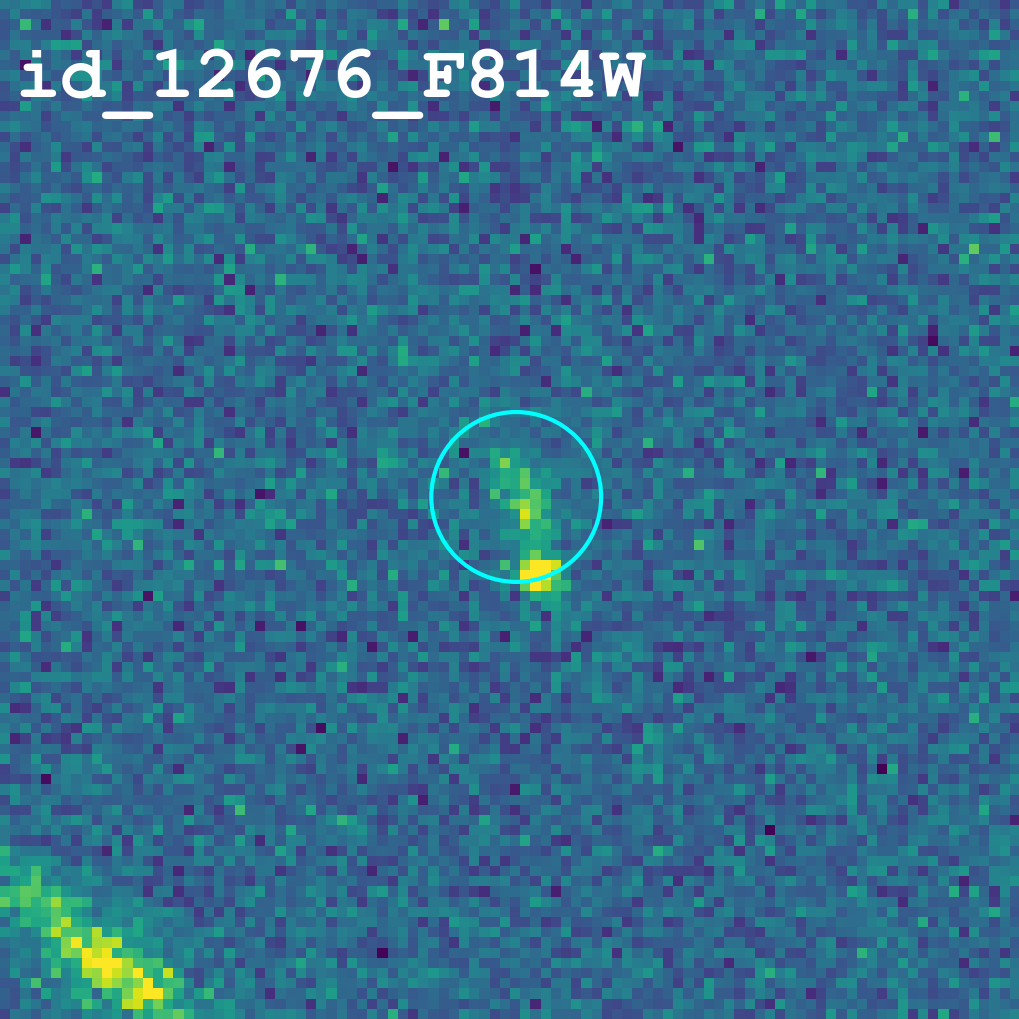}}
    \hspace{-0.01cm}\vspace{-0.05cm}
    \subfloat{\includegraphics[width=.25\textwidth]{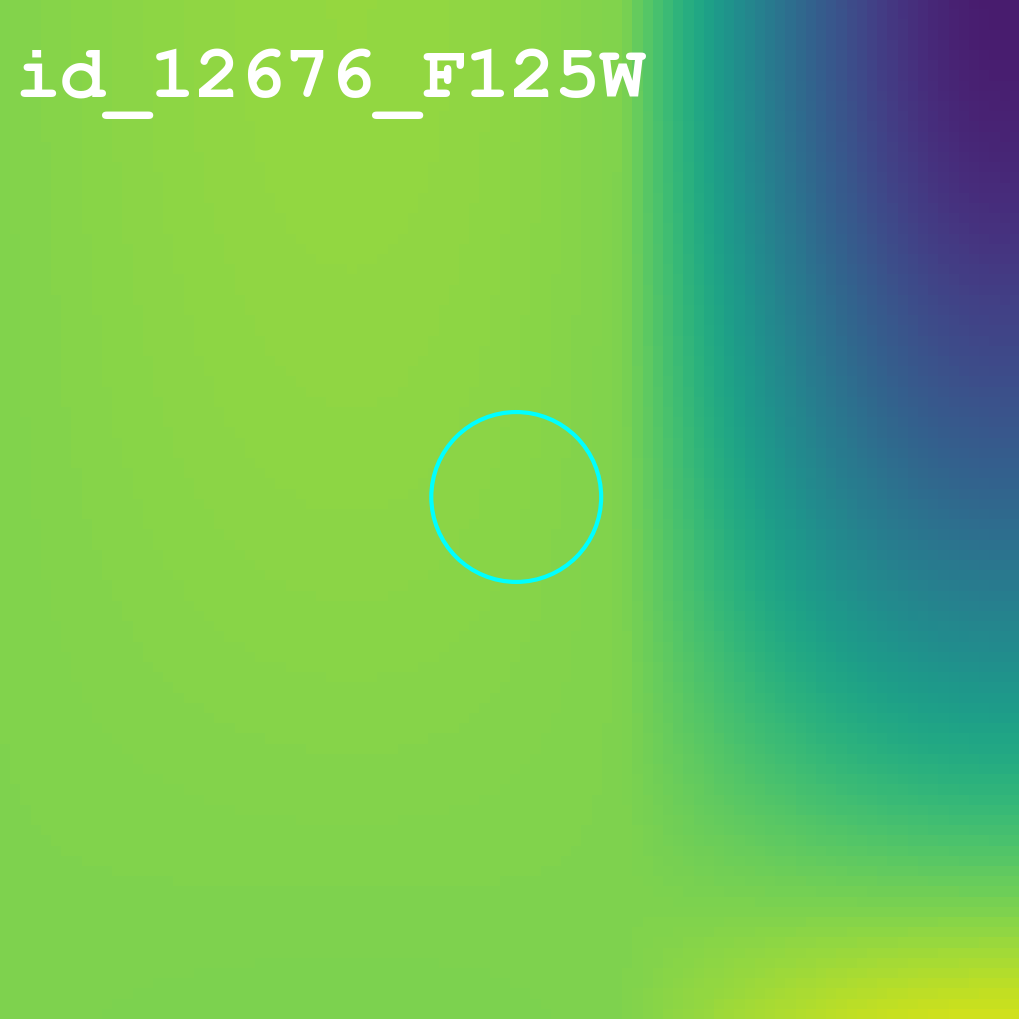}}
    \hspace{-0.01cm}\vspace{-0.05cm}

    \subfloat{\includegraphics[width=.25\textwidth]{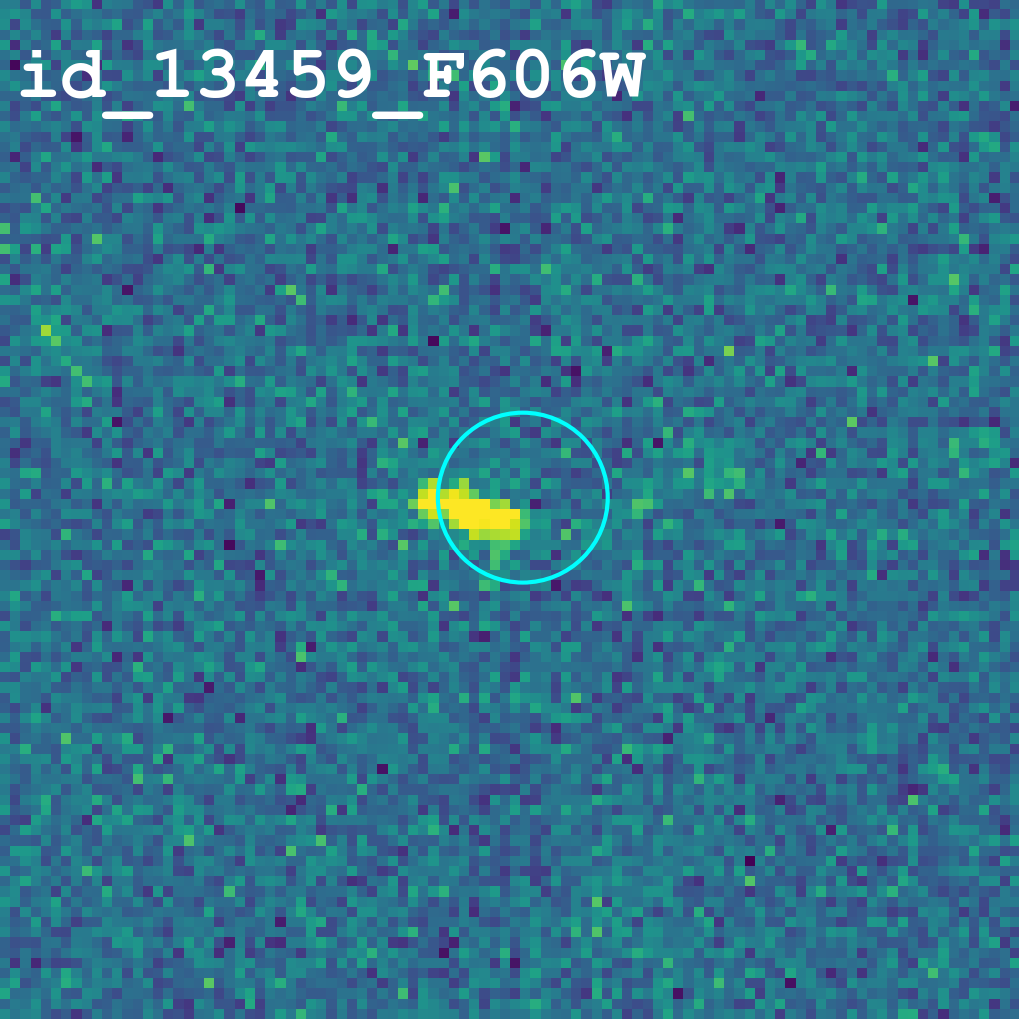}}
    \hspace{-0.05cm}\vspace{-0.05cm}
    \subfloat{\includegraphics[width=.25\textwidth]{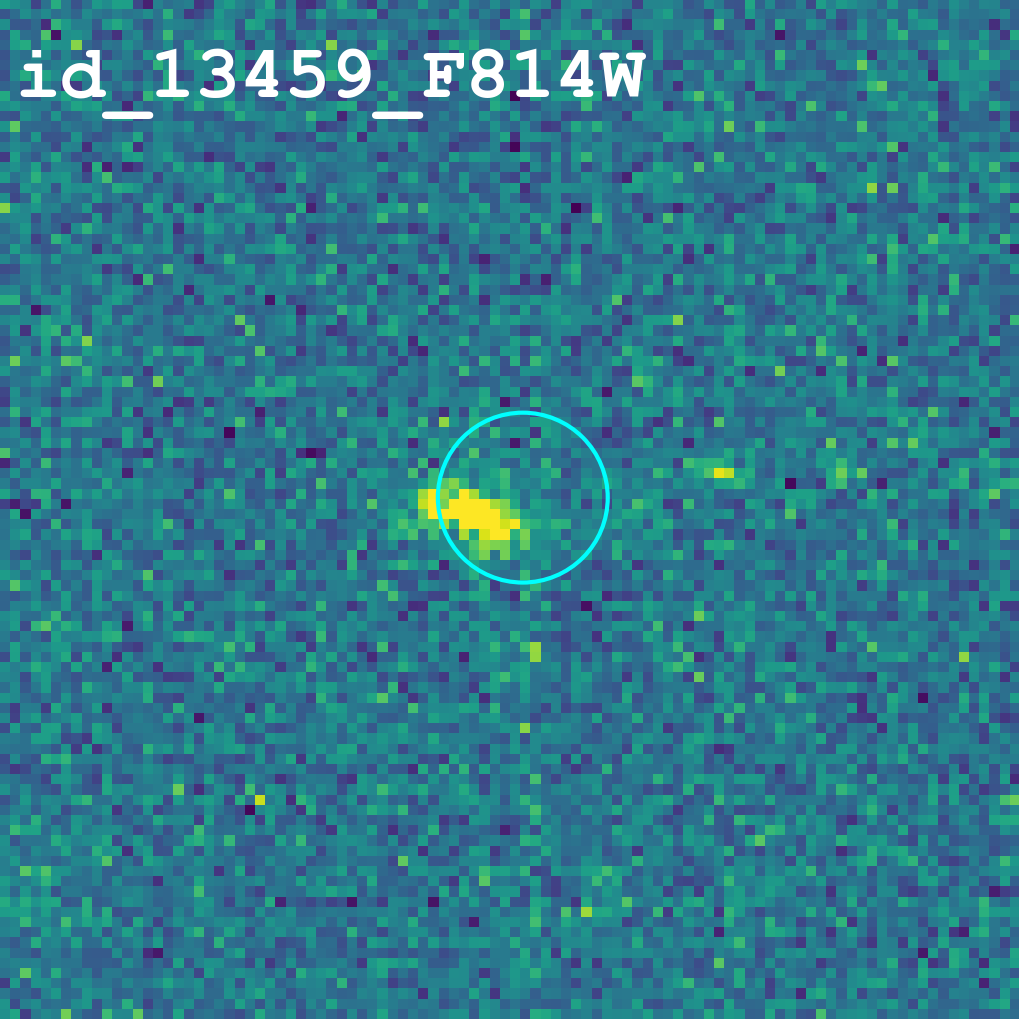}}
    \hspace{-0.05cm}\vspace{-0.05cm}
    \subfloat{\includegraphics[width=.25\textwidth]{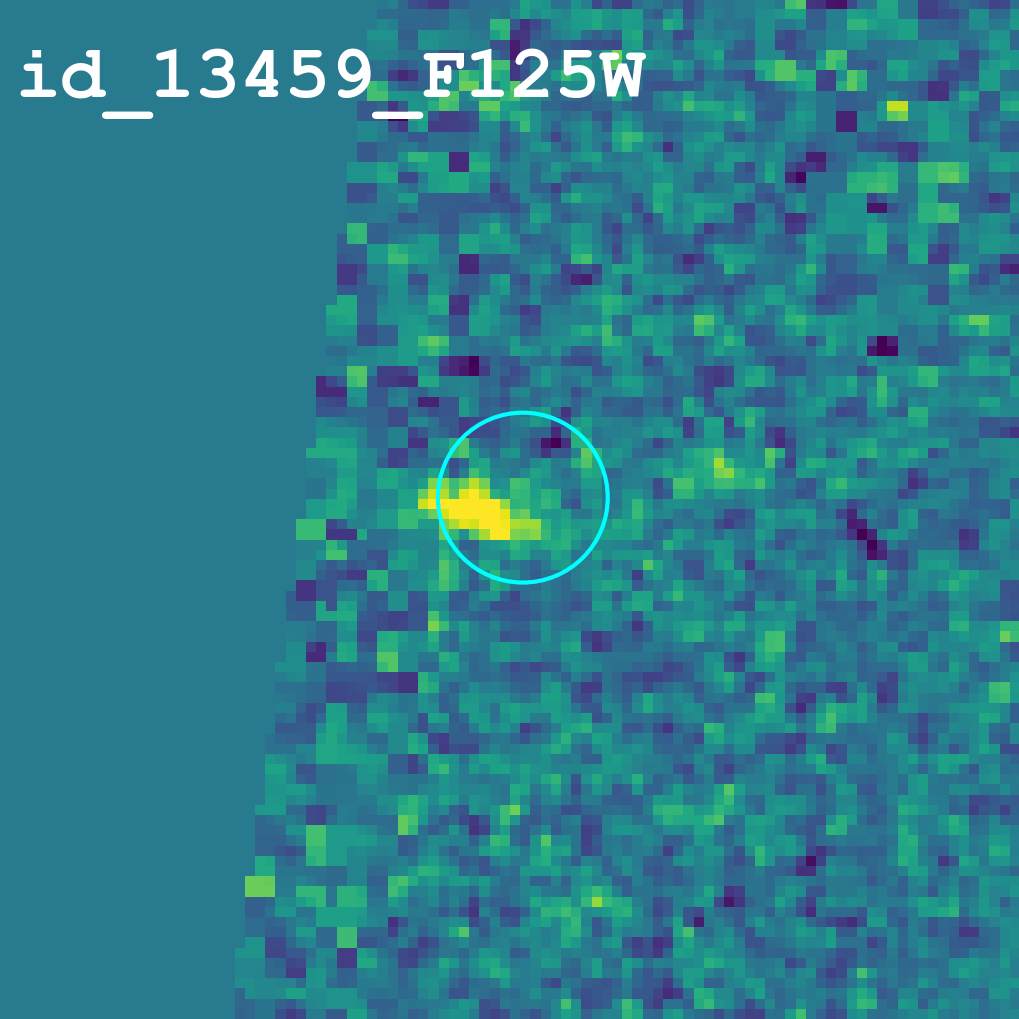}}
    \hspace{-0.05cm}\vspace{-0.05cm}

    \subfloat{\includegraphics[width=.25\textwidth]{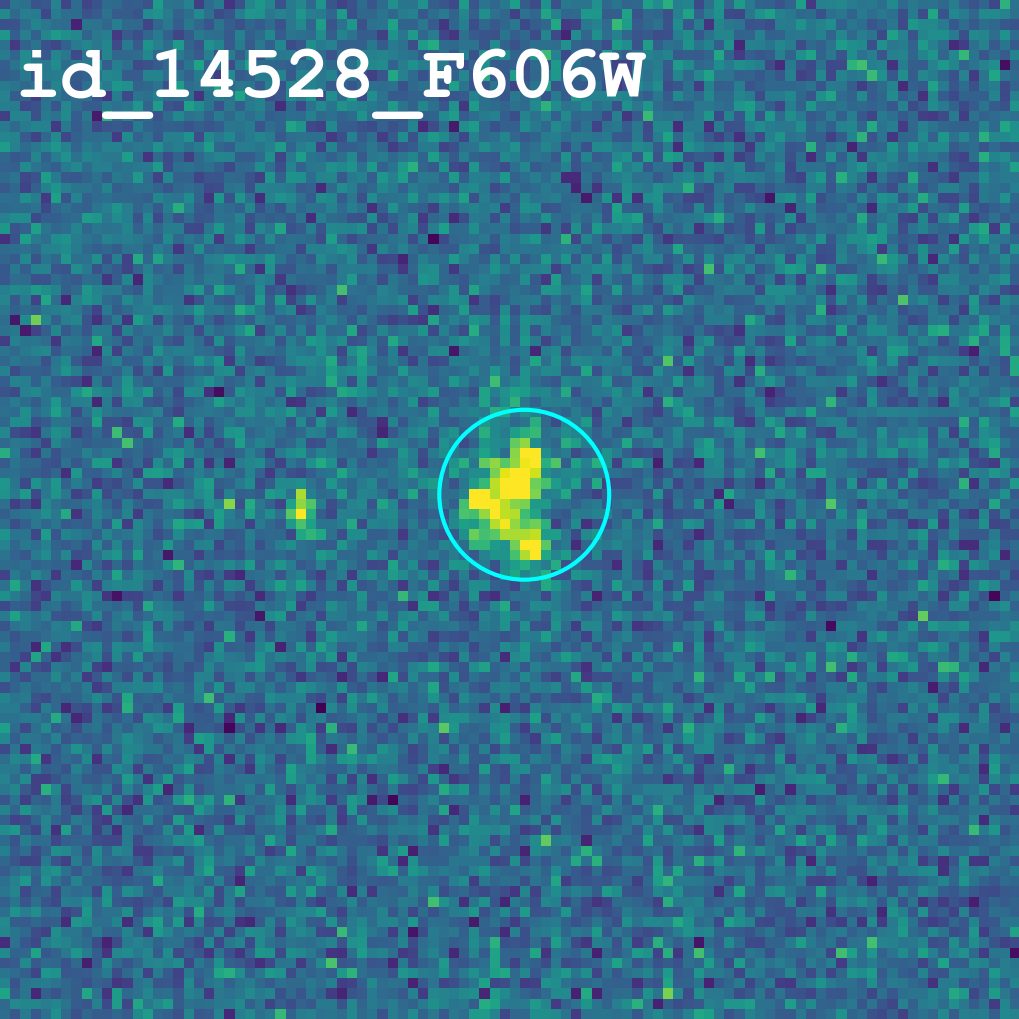}}
    \hspace{-0.01cm}\vspace{-0.05cm}
    \subfloat{\includegraphics[width=.25\textwidth]{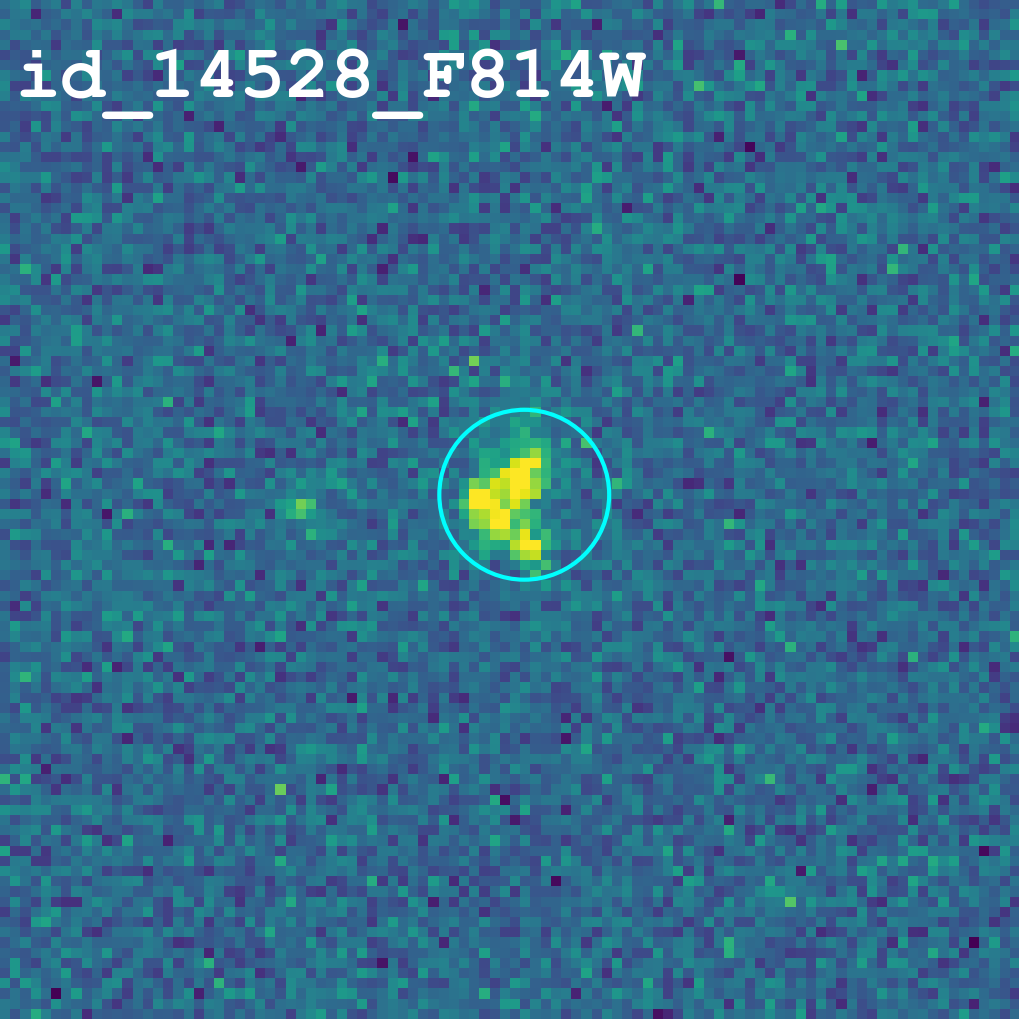}}
    \hspace{-0.01cm}\vspace{-0.05cm}
    \subfloat{\includegraphics[width=.25\textwidth]{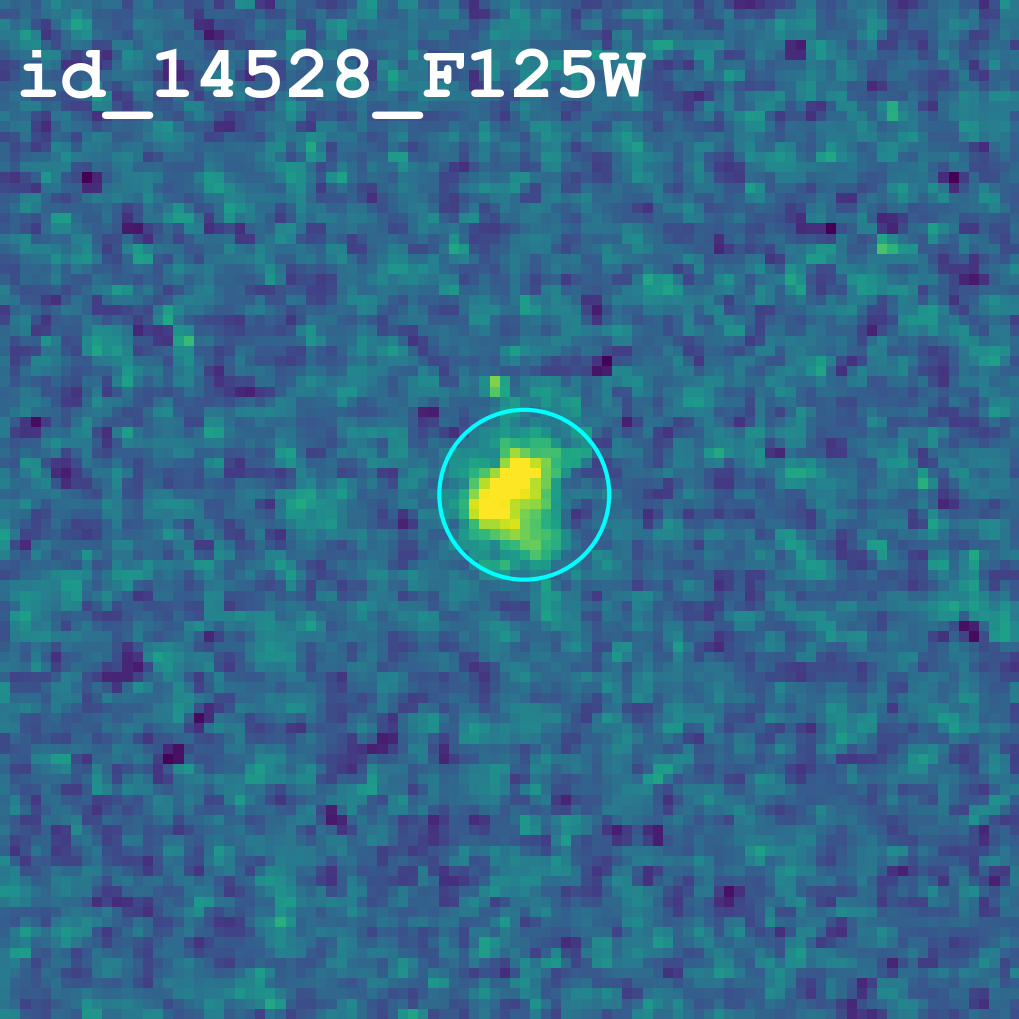}}
    \hspace{-0.01cm}\vspace{-0.05cm}

\caption{The HST imaging in thre bands F606W, F814W and F125W of the 9 LyC candidates presented in this work. Cut outs are $6\times6^{\prime\prime}$ in size. Cyan circle encloses LyC candidate and it is $1^{\prime\prime}$ in diameter. }
\label{A1}
\end{center}
\end{figure*}

\begin{figure*}
\begin{center}

    \subfloat{\includegraphics[width=.25\textwidth]{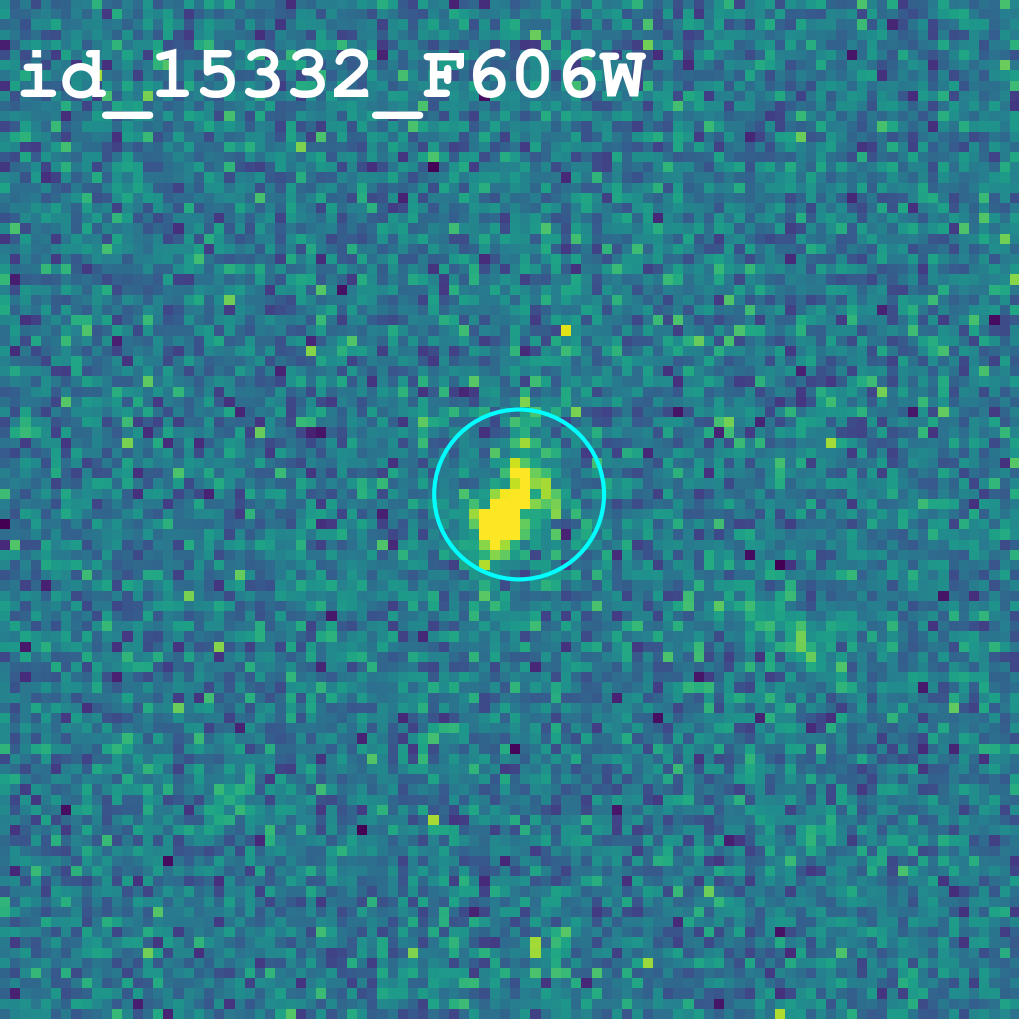}}
    \hspace{-0.01cm}\vspace{-0.05cm}
    \subfloat{\includegraphics[width=.25\textwidth]{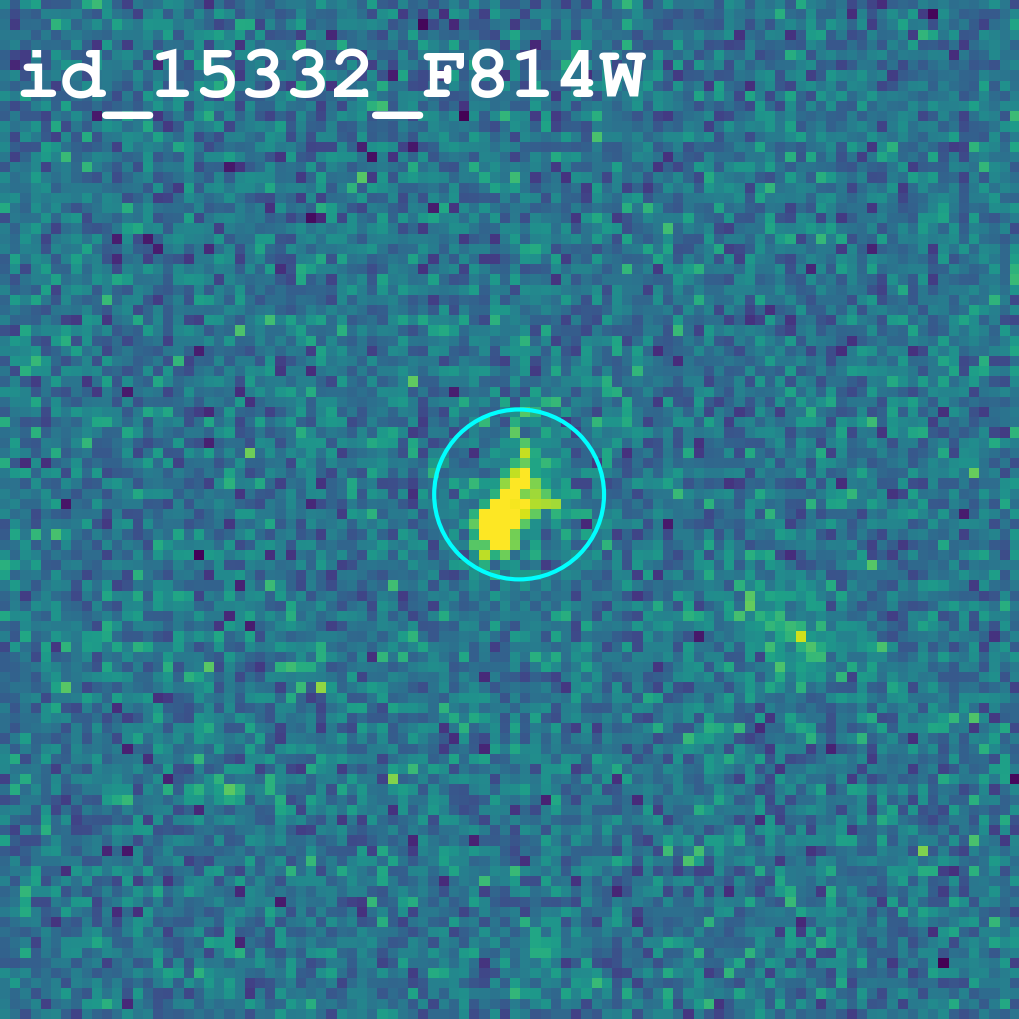}}
    \hspace{-0.01cm}\vspace{-0.05cm}
    \subfloat{\includegraphics[width=.25\textwidth]{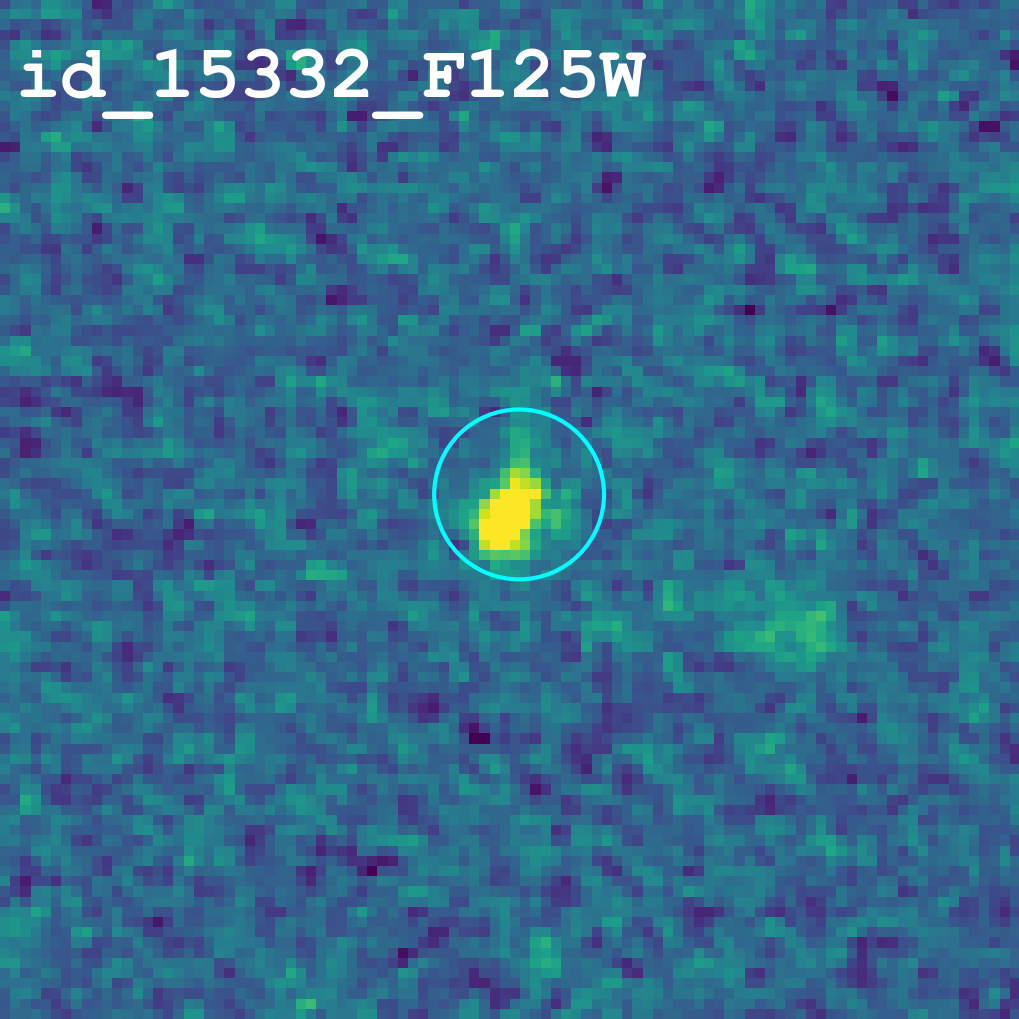}}
    \hspace{-0.01cm}\vspace{-0.05cm}

    \subfloat{\includegraphics[width=.25\textwidth]{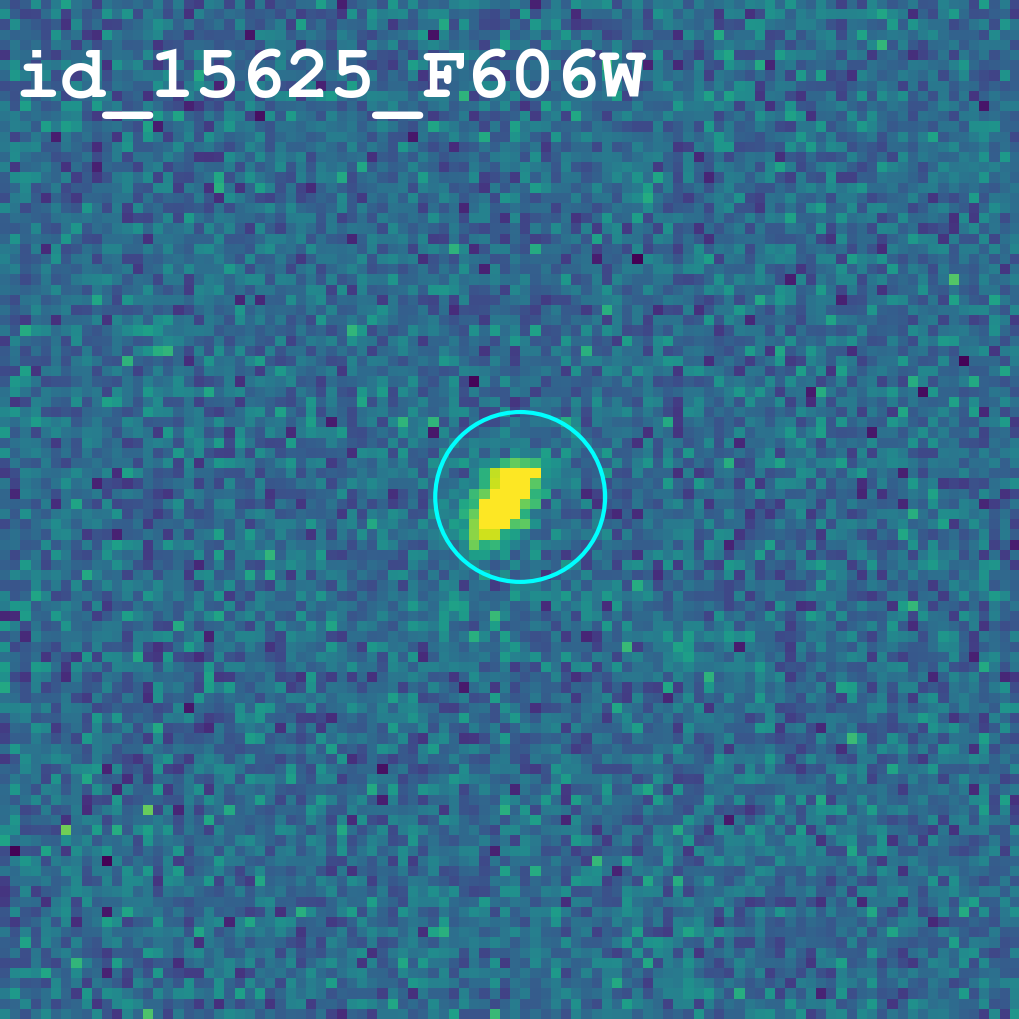}}
    \hspace{-0.01cm}\vspace{-0.05cm}
    \subfloat{\includegraphics[width=.25\textwidth]{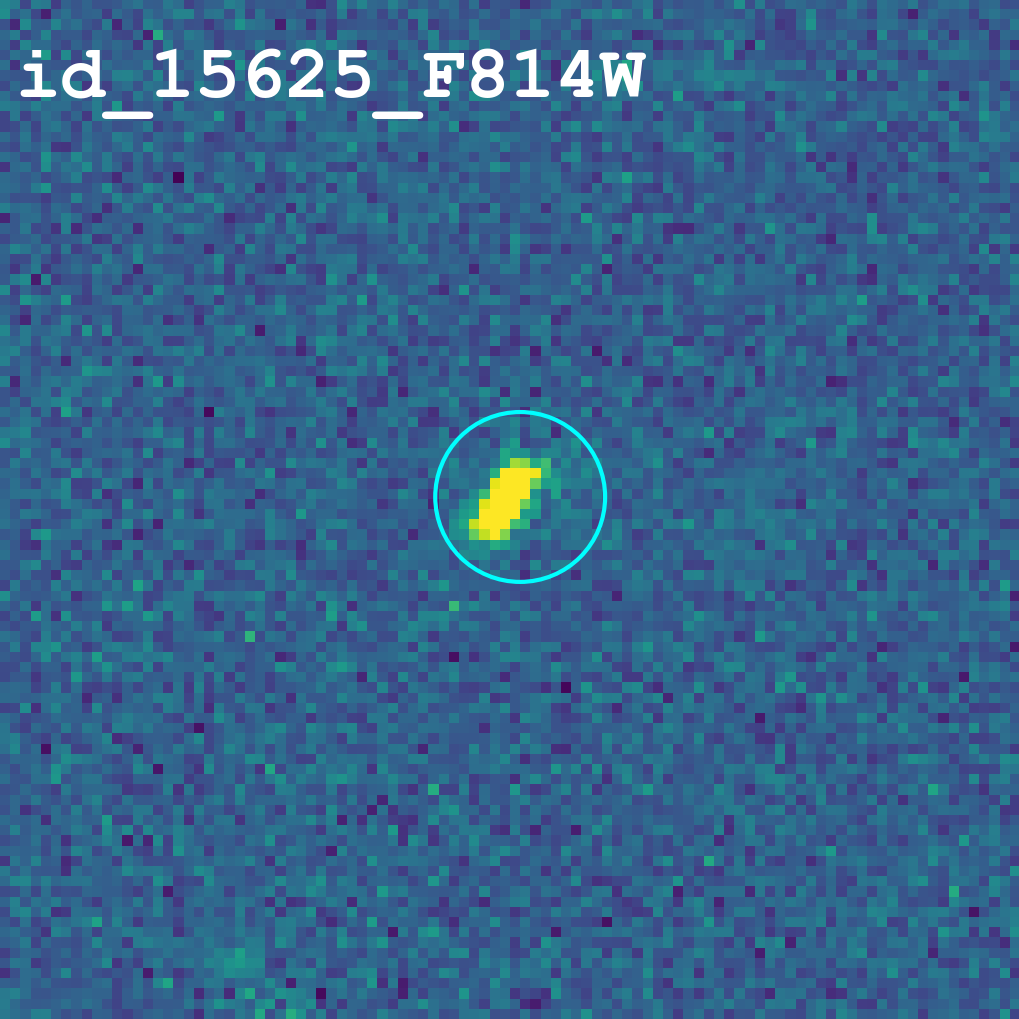}}
    \hspace{-0.01cm}\vspace{-0.05cm}
    \subfloat{\includegraphics[width=.25\textwidth]{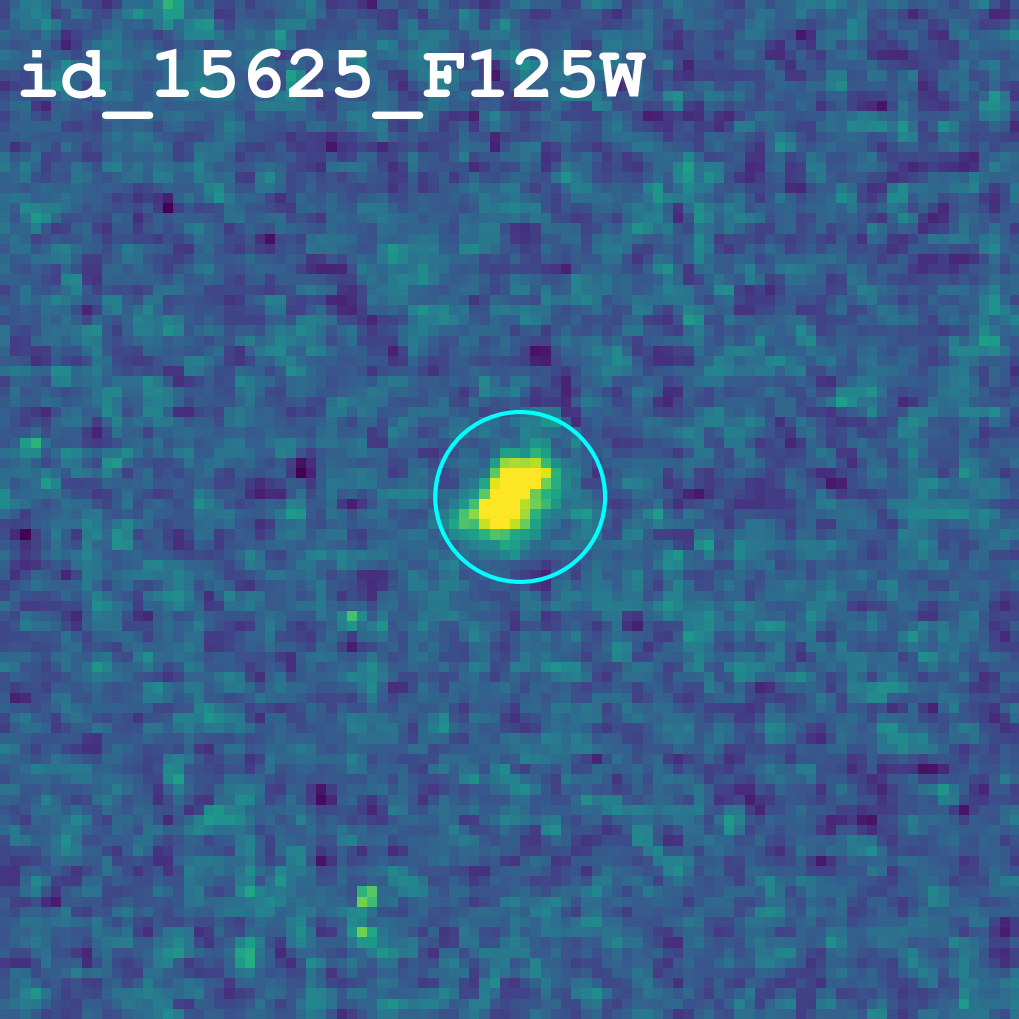}}
    \hspace{-0.01cm}\vspace{-0.05cm}

    \subfloat{\includegraphics[width=.25\textwidth]{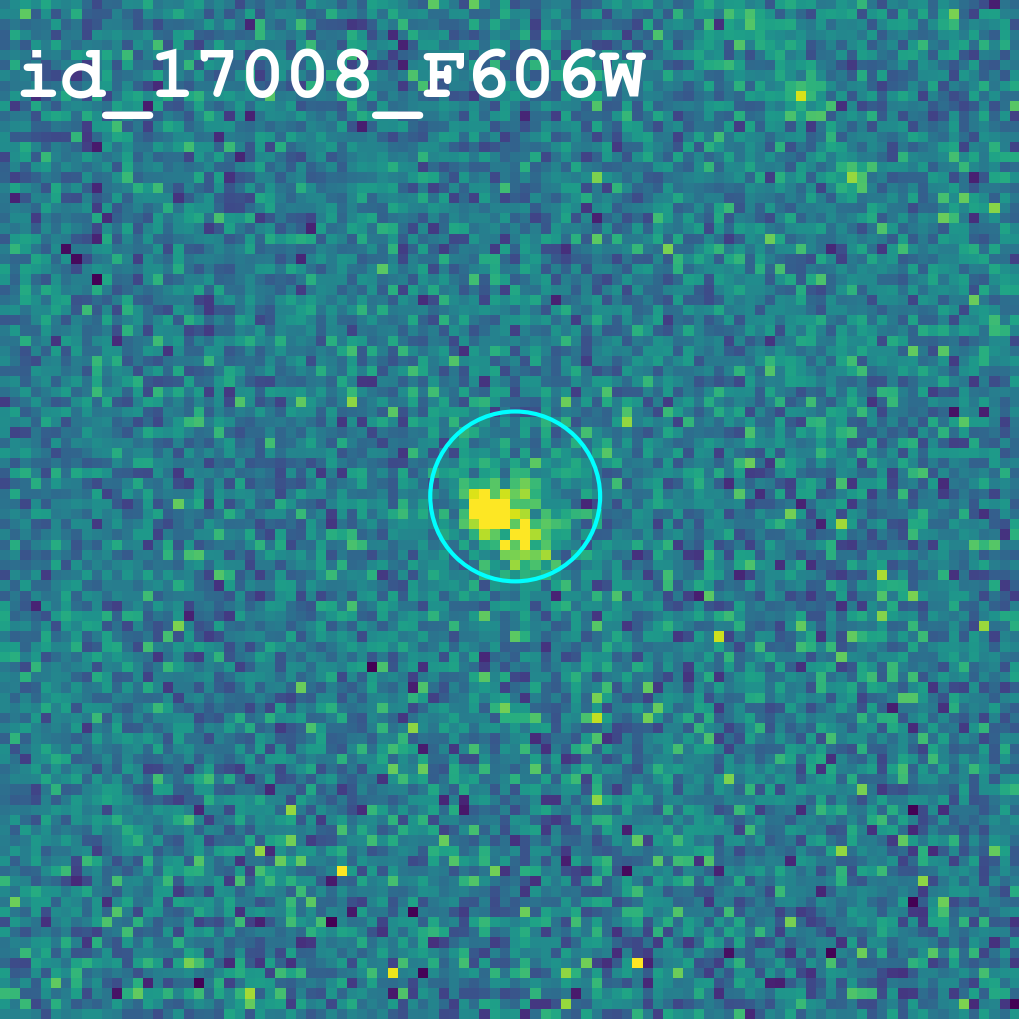}}
    \hspace{-0.01cm}\vspace{-0.05cm}
    \subfloat{\includegraphics[width=.25\textwidth]{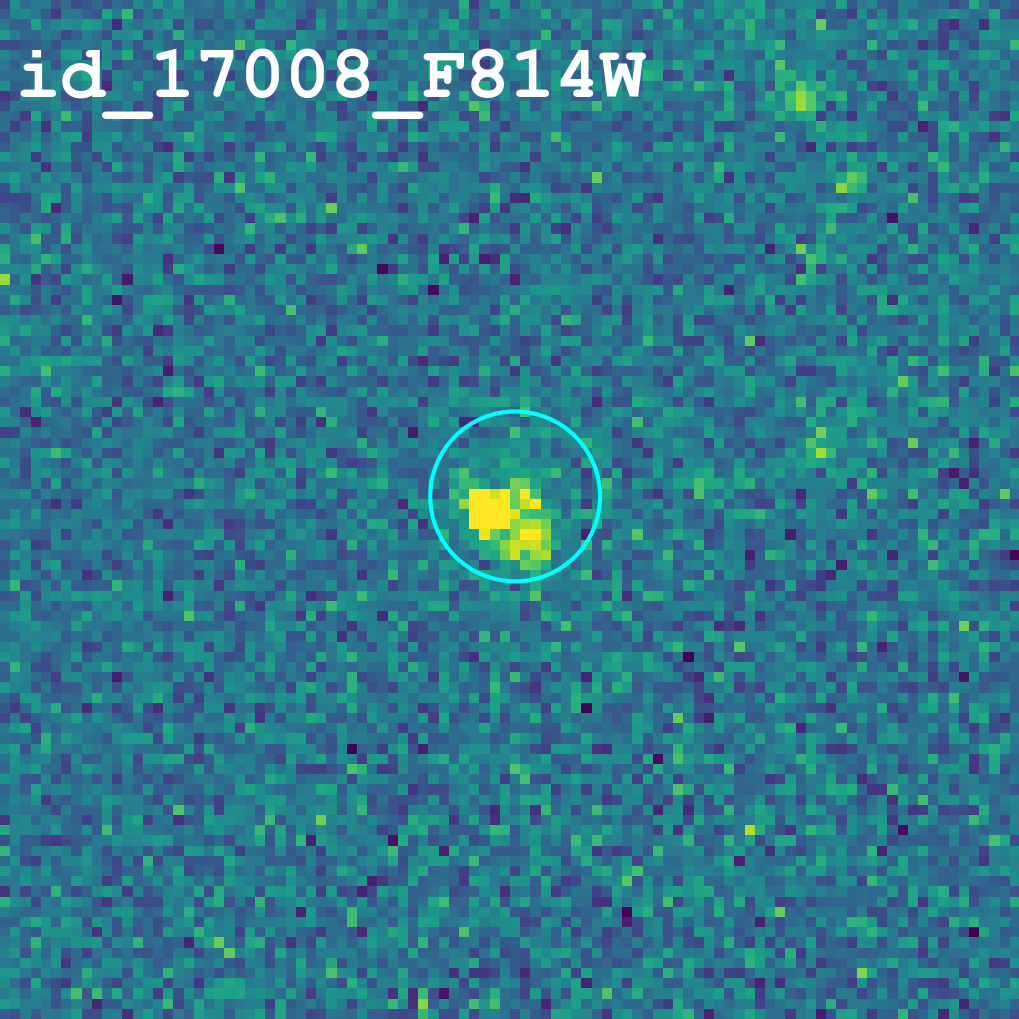}}
    \hspace{-0.01cm}\vspace{-0.05cm}
    \subfloat{\includegraphics[width=.25\textwidth]{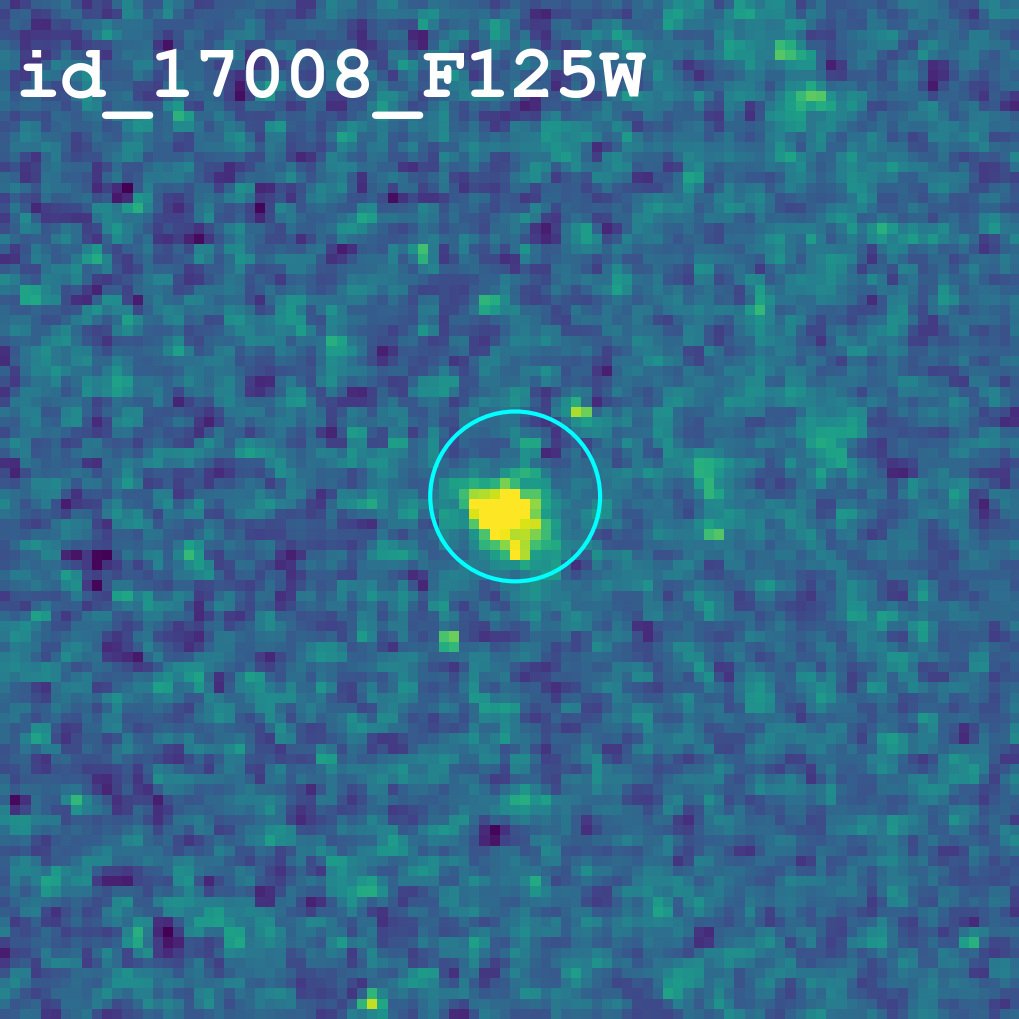}}
    \hspace{-0.01cm}\vspace{-0.05cm}

    \subfloat{\includegraphics[width=.25\textwidth]{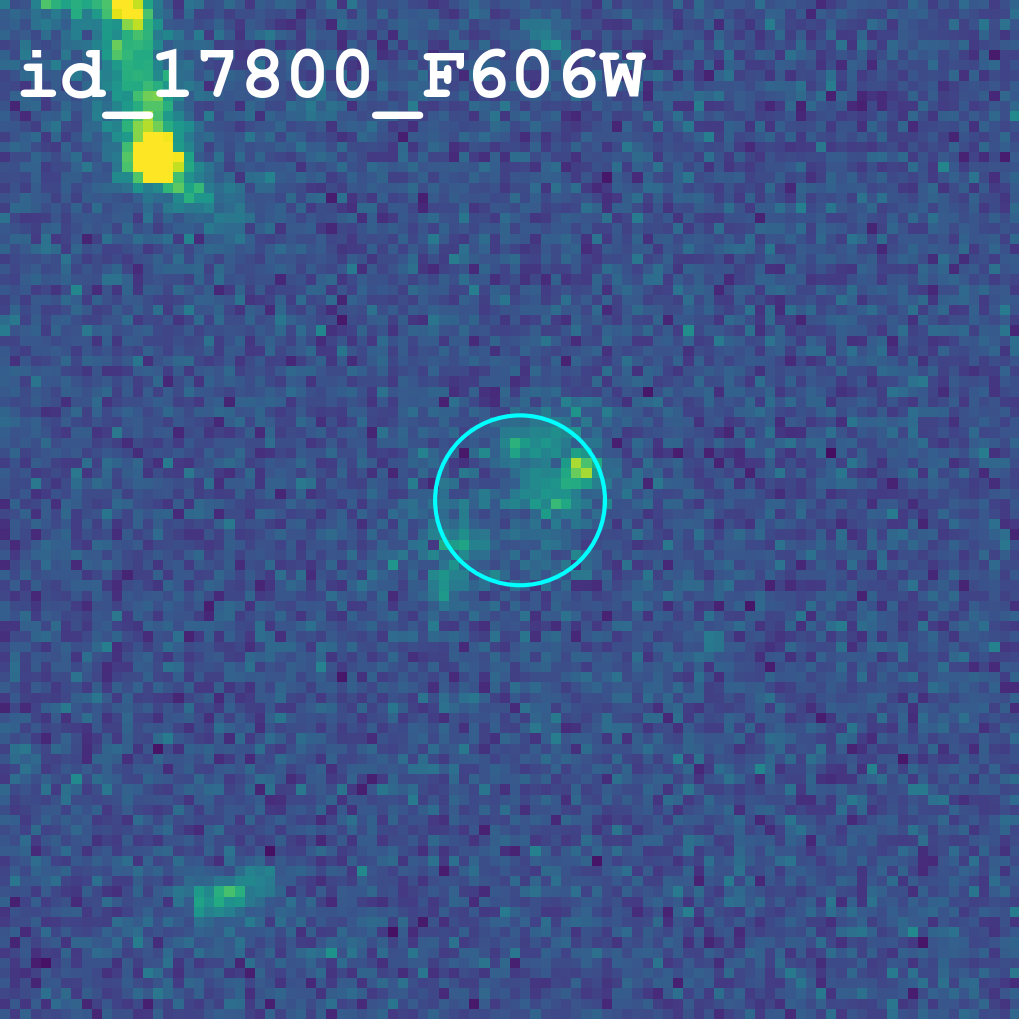}}
    \hspace{-0.01cm}\vspace{-0.05cm}
    \subfloat{\includegraphics[width=.25\textwidth]{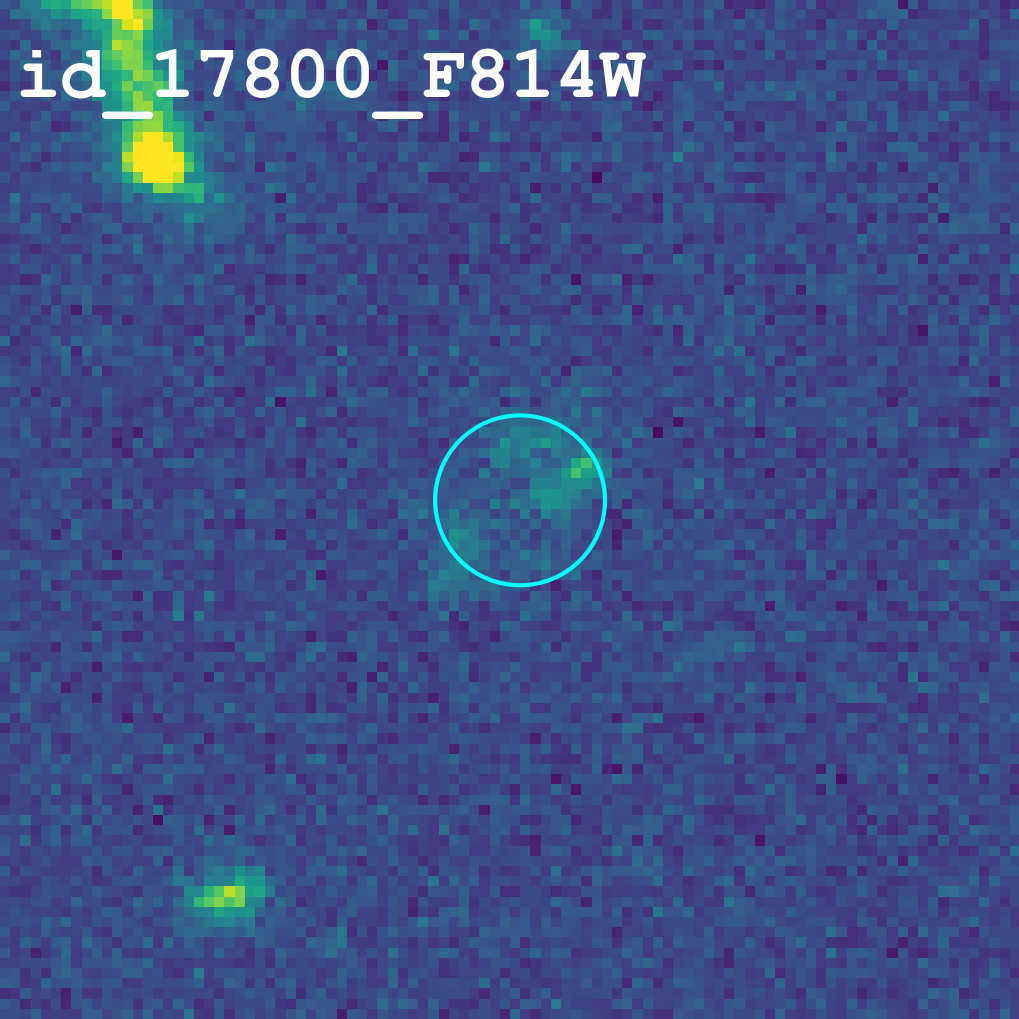}}
    \hspace{-0.01cm}\vspace{-0.05cm}
    \subfloat{\includegraphics[width=.25\textwidth]{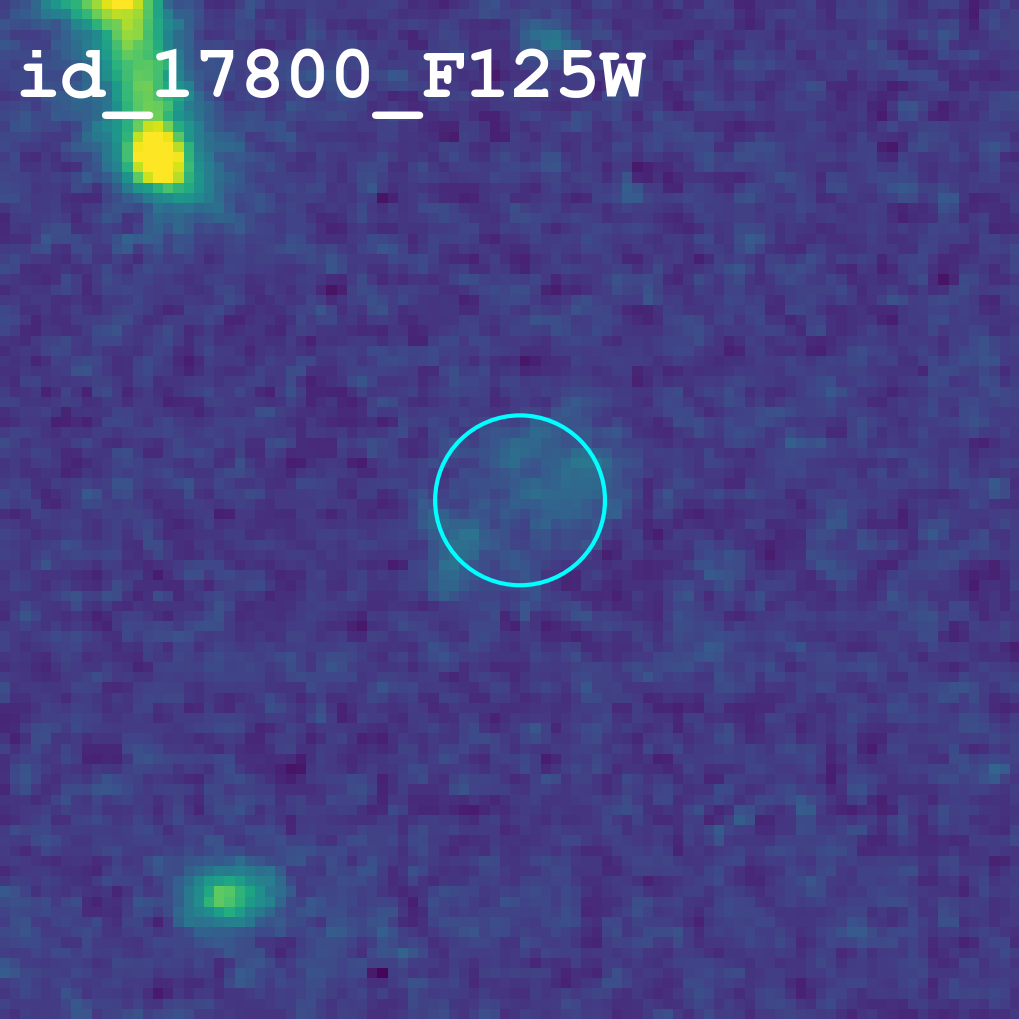}}
    \hspace{-0.01cm}\vspace{-0.05cm}

\caption{Same as \ref{A1}.}
\label{figA2}
\end{center}
\end{figure*}

\section {The 2D Keck LRIS spectra of the 9 presented LyC candidates.}

\begin{figure*}
\begin{center}

    \subfloat{\includegraphics[ width=18cm]{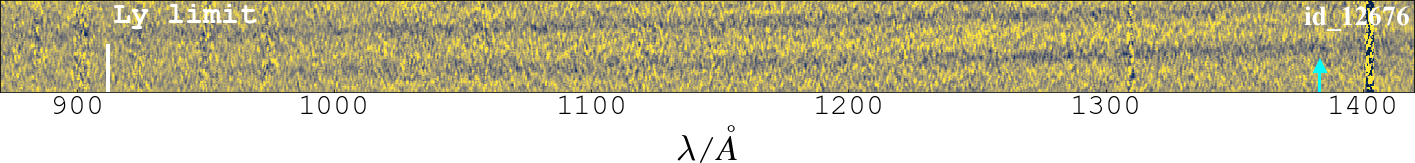}}
    \vspace{0.01cm}
    \subfloat{\includegraphics[ width=18cm]{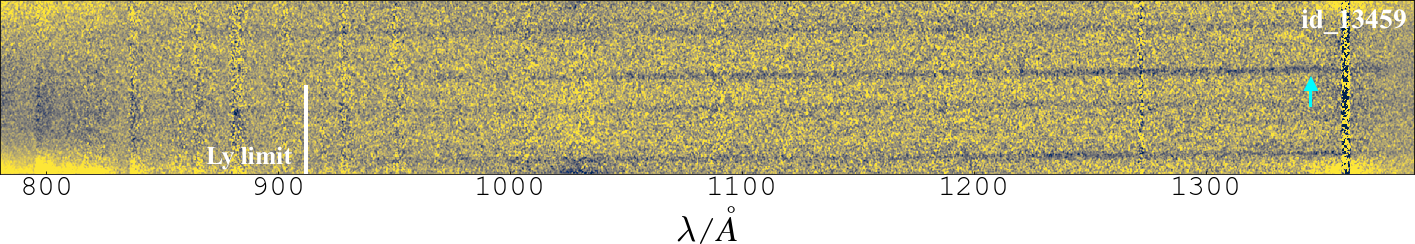}}
    \vspace{0.01cm}
    \subfloat{\includegraphics[ width=18cm]{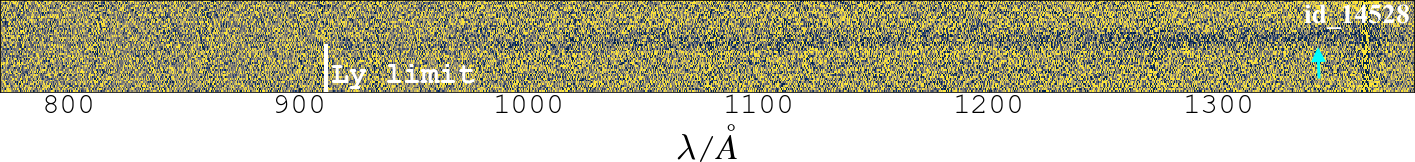}}
    \vspace{0.01cm}
    \subfloat{\includegraphics[ width=18cm]{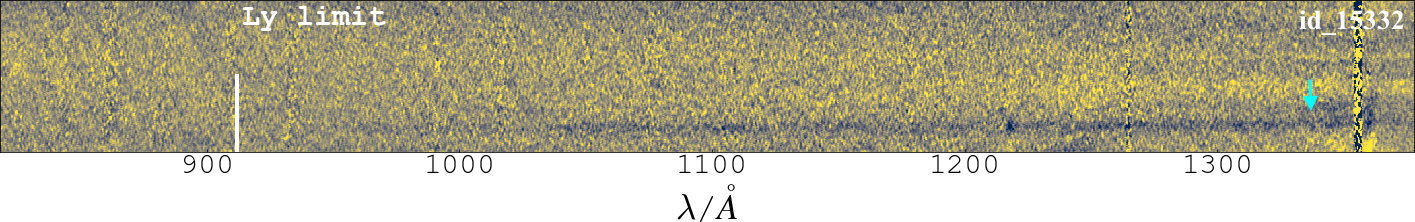}}
    \vspace{0.01cm}
    \subfloat{\includegraphics[ width=18cm]{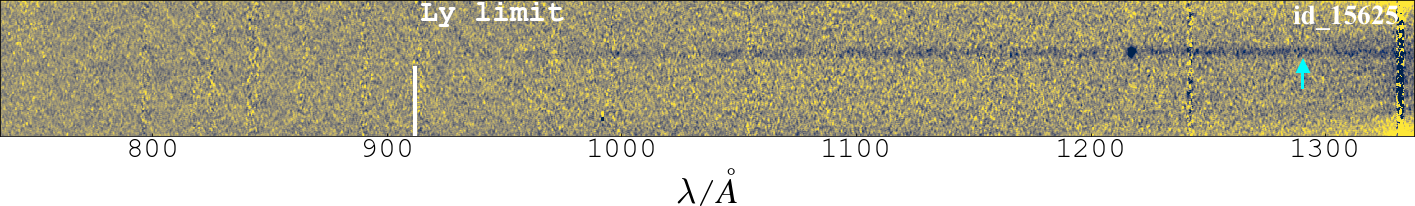}}
    \vspace{0.01cm}
    \subfloat{\includegraphics[ width=18cm]{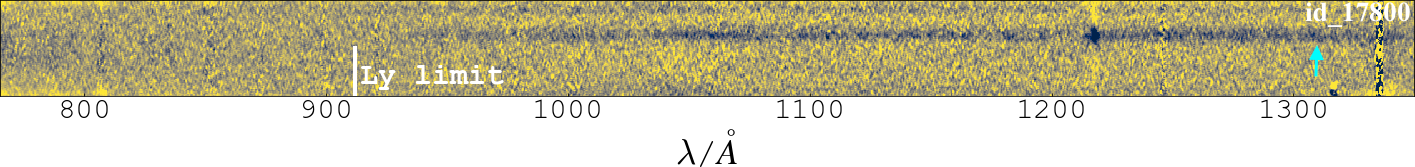}}
    \vspace{0.01cm}
    \subfloat{\includegraphics[ width=18cm]{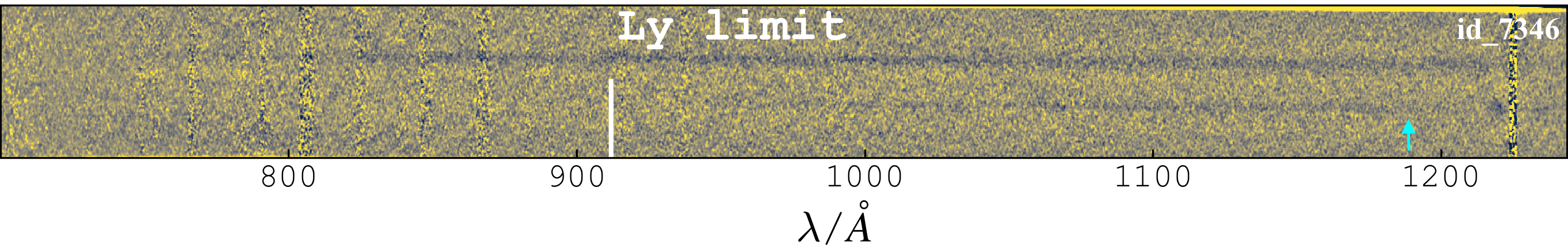}}
    \vspace{0.01cm}
    \subfloat{\includegraphics[ width=18cm]{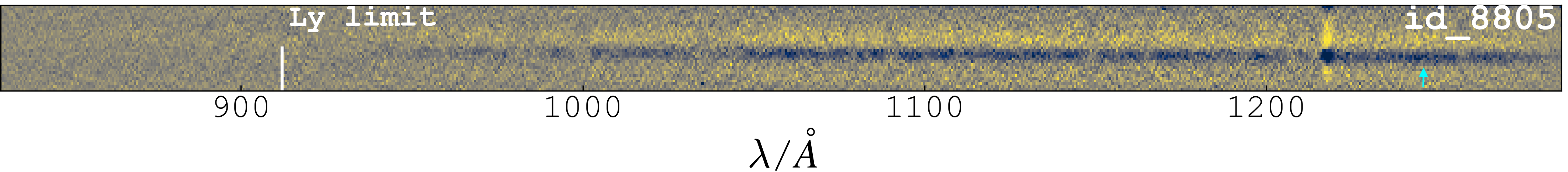}}
    
\caption{The 2D Keck LRIS spectrum of 9 candidates presented in this work. The cyan arrow points into the dispersion trace from LyC candidate, while vertical white line marks the Lyman limit.}
\label{B1}
\end{center}
\end{figure*}

\begin{figure*}
\begin{center}

\subfloat{\includegraphics[ width=18cm]{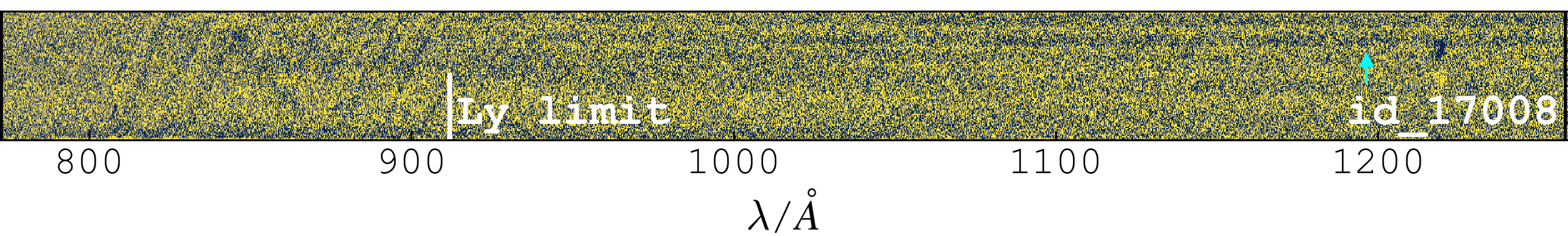}}
\vspace{0.01cm}

\caption{Same as \ref{B1}.}
\label{figB2}
\end{center}
\end{figure*}

\end{appendix}

\bsp	
\label{lastpage}
\end{document}